%% file: main.tex
\documentclass[twoside]{article}

\usepackage[accepted]{aistats2024}

\usepackage[round]{natbib}

\usepackage[blackhypersetup]{shortex} 
\usepackage{xcolor}
\usepackage{algorithm}
\usepackage{algpseudocode}
\input{math-macros.tex}
\begin{document}

\fancyhead[CO]{\small\bfseries Biron-Lattes, Surjanovic, Syed, Campbell, Bouchard-C\^{o}t\'{e}}

\twocolumn[

\aistatstitle{autoMALA: Locally adaptive Metropolis-adjusted Langevin algorithm}

\aistatsaddress{
\setlength\tabcolsep{0pt}
\begin{tabular*}{0.85\linewidth}{@{\extracolsep{\fill}} ccc }
\bf Miguel Biron-Lattes* & \bf Nikola Surjanovic* & \bf Saifuddin Syed \\
University of British Columbia & University of British Columbia & University of Oxford
\end{tabular*} \\[15pt]
\begin{tabular*}{0.65\linewidth}{@{\extracolsep{\fill}} cc }
\bf Trevor Campbell & \bf Alexandre Bouchard-C\^{o}t\'{e} \\
University of British Columbia & University of British Columbia
\end{tabular*}
}
]

\input{abstract}

\input{introduction}
\input{background}
\input{autoMALA}

\input{experiments}
\input{conclusion}

\subsubsection*{Acknowledgements}
ABC and TC acknowledge the support of an NSERC Discovery Grant.
SS acknowledges the support of EPSRC grant EP/R034710/1 CoSines and
NS acknowledges the support of a Vanier Canada Graduate Scholarship. 
We additionally acknowledge use of the ARC Sockeye computing platform from the
University of British Columbia.

\bibliographystyle{apalike}
\bibliography{main.bib}

\input{checklist}

\appendix

\onecolumn
\aistatstitle{Supplementary Materials}

\input{supplement_proofs.tex}

\newpage

\input{supplement_experiments.tex}

\end{document}

%% file: math-macros.tex
\newcommand{\statespace}{\reals^d}
\newcommand{\normtarget}{\pi}
\newcommand{\target}{\gamma}
\newcommand{\hmcspace}{\reals^{2d}}
\newcommand{\hmctarget}{\normtarget_\mathrm{HMC}}
\newcommand{\score}{\grad\log\target}
\newcommand{\involution}{Q_{\mathrm{AM}}}
\newcommand{\amtarget}{\widebar\normtarget}
\newcommand{\amspace}{\mathcal{S}}

\newcommand{\amkern}{K_{\mathrm{AM}}}

\newcommand{\mcS}{\mathcal{S}}

\newcommand{\epsi}{\eps_{\text{init}}}
\newcommand{\exactess}{\text{ESS}(\mu, \sigma)}
\newcommand{\mcT}{\mathcal{T}}

%% file: abstract.tex
\begin{abstract}
Selecting the step size for the Metropolis-adjusted Langevin algorithm (MALA) 
is necessary in order to obtain satisfactory performance. However, finding an
adequate step size for an arbitrary target distribution can be a difficult task
and even the \emph{best} step size can perform poorly in specific
regions of the space when the target distribution is sufficiently complex.
To resolve this issue we introduce \emph{autoMALA}, a new Markov chain Monte Carlo
algorithm based on MALA that automatically sets its step size at each iteration 
based on the local geometry of the target distribution. 
We prove that autoMALA has the correct invariant distribution, despite 
continual automatic adjustments of the step size.
Our experiments demonstrate that autoMALA is competitive with related state-of-the-art 
MCMC methods, in terms of the number of log density evaluations per 
effective sample, and it outperforms state-of-the-art samplers on targets with 
varying geometries. Furthermore, we find that autoMALA tends to find step sizes
comparable to optimally-tuned MALA when a fixed step size suffices for the whole
domain.
\end{abstract}

%% file: introduction.tex
\section{INTRODUCTION}
The Metropolis-adjusted Langevin algorithm (MALA), introduced by 
\cite{rossky1978brownian}, is a well-established 
Markov chain Monte Carlo (MCMC) method for asymptotically obtaining samples 
from a target distribution $\pi$. 
As a gradient-based method, MALA often provides better performance than, for example, 
random-walk Metropolis--Hastings (MH), because gradients direct the sampler to 
high-density regions in the target distribution. 
MALA is based on an approximation to overdamped Langevin dynamics, followed 
by a Metropolis--Hastings correction to account for the approximation.

\begin{figure}[t]
    \includegraphics[width=0.48\textwidth]{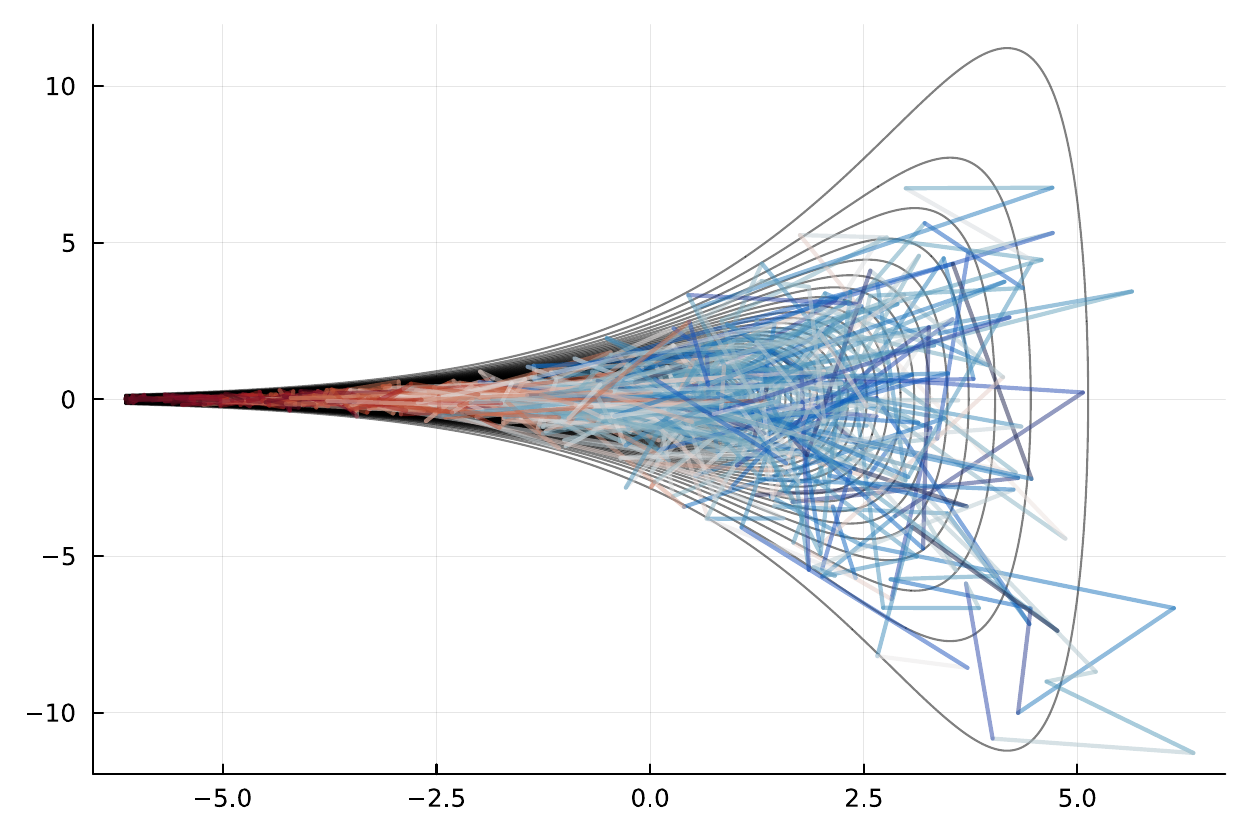}
    \caption{1000 iterations of autoMALA on Neal's funnel. Paths between 
    points are coloured according to the automatically selected step size 
    chosen by autoMALA (red for small steps, blue for large steps).}
    \label{fig:funnel_demo}
\end{figure}

A key challenge in using MALA is that it requires the selection of a step size, $\eps$, 
which controls the level of refinement of the approximation to Langevin dynamics. 
The dynamics are stated as a stochastic differential equation (SDE). 
As $\eps \to 0$, we perform smaller time updates to our SDE approximation, resulting 
in a higher acceptance probability of the proposal. This high acceptance rate 
comes at the expense of slow mixing due to the small step size $\eps$. 
On the other hand, as $\eps \to \infty$, more distant moves are generally proposed,
at the expense of a fall in the acceptance rate, leading again to slow mixing. 
There is therefore a tradeoff between the distance that one proposal can reach 
and the resulting acceptance rates. 
Extensive work \citep{roberts1998optimal,atchade2006adaptive,marshall2012adaptive}
has been carried out to derive optimal step sizes that strike a balance for this tradeoff.

The fact that traditional MALA uses a single fixed 
step size to explore the whole state space makes it inadequate for target
densities whose scales and correlations vary across the state space (e.g., the 
funnel in \cref{fig:funnel_demo}). 
To tackle this fundamental limitation, in this work we introduce \emph{autoMALA},
a new MCMC algorithm based on the Langevin diffusion that dynamically selects
an appropriate step size at every MCMC iteration. 
Crucially, our method ensures that we sample from the correct target distribution 
even in the presence of this dynamic step size selection. 
We conduct experiments that establish the desirable properties 
of autoMALA in terms of the number of time step updates (related to the number of 
density evaluations) required per effective sample when compared to hand-tuned MALA
and NUTS \citep{hoffman2014nuts}.
All proofs and experimental details are provided in the supplement.

\noindent \textbf{Related work.}
There is a large literature devoted to selecting a single optimal step size for MALA. 
\citet{roberts1998optimal} use scaling limits to argue that the step size should
be set to obtain an average acceptance rate of $0.574$. More recently, adaptive
MCMC methods \citep{atchade2005adaptive} have been developed specifically for MALA
\citep{atchade2006adaptive,marshall2012adaptive}, which adjust the step size on the
fly to target the $0.574$ rate while preserving the correct invariant distribution. 
Alternatively, general-purpose tuning procedures for gradient-based MCMC 
\citep[see e.g.][]{coullon2023efficient} tune hyperparameters by directly minimizing
a divergence between the target and the sample-based distribution.
Furthermore, \citet{livingstone2022barker} propose a novel MH algorithm that
exhibits the same dimensional scaling as MALA while being more robust to the choice 
of step size. Still, the fact that these algorithms depend on a fixed step size makes 
them inappropriate for targets with varying local geometry. 

In the Hamiltonian Monte Carlo (HMC) literature there is comparatively more attention on dynamic, automatic selection
of proposal length scale via selection of the number of steps $L$ between momentum refreshments \citep{hoffman2014nuts}. 
However, $n$ steps of doubling the trajectory length $L$ takes compute time 
proportional to $2^n$, whereas $n$ steps of step size doubling in the autoMALA 
algorithm has a cost proportional to $n$. 
Therefore, for target distributions with sufficiently variable geometries, 
dynamic step size selection is preferable to dynamic trajectory length selection,
as we also demonstrate empirically. 
In this vein, several approximate Monte Carlo algorithms---i.e, which do not preserve
the distribution $\normtarget$---based on variable step size integrators of Hamiltonian 
dynamics have been proposed \citep[see e.g.][]{kleppe2022connecting}.
However, our focus is on asymptotically exact MCMC; we therefore exclude 
such approximate algorithms from our analyses.

\cite{kleppe2016adaptive} is one of the few works that tackles $\normtarget$-invariant 
dynamic step size selection. However, its step size selection routine (Algorithm 2
in \citet{kleppe2016adaptive}) does not take into account the direction of the 
proposal and may not terminate in general. \citet{modi2023delayed} also proposes
a dynamic step size algorithm, based on the delayed-rejection method 
\citep[][Sec.\ 5]{tierney1999some}. Because of this, the algorithm in \citet{modi2023delayed}
requires setting \emph{a priori} a maximum number of proposed step sizes, whereas 
autoMALA does not.
Finally, \cite{girolami2011riemann} study MALA and HMC 
with position-dependent preconditioning (a randomized ``mass matrix''), 
which provides an alternative way of making an informed selection 
of the step size. \citet{nishimura2016variable} further extend this idea by employing
explicit variable step size integrators.
However, these methods have a compute cost per leapfrog step that scale superlinearly in the number of dimensions. 
While they provide improved mixing, in medium or high-dimensional problems these 
gains are often not sufficient to counteract exploding cost per step, and therefore 
these methods have limited or no empirical advantages in high-dimensional 
problems \citep{girolami2011riemann}. 
In contrast, the cost of one autoMALA step scales linearly with dimension. 

%% file: background.tex
\section{BACKGROUND}\label{sec:background}

Let $\normtarget$ be a probability distribution of interest on $\statespace$, 
which we assume can be written as 
\[
  \normtarget(x) := \frac{\target(x)}{\int_{\statespace} \target(u) \, \dee u}, 
\]
where $\target$ can be evaluated pointwise and is continuously differentiable.
Consider the following special case of the stochastic differential 
equation known as the \emph{Langevin diffusion} \citep[]{roberts2002langevin}:
\[
  \label{eq:Langevin_dynamics}    
  \dee X_t = \frac{1}{2} C\nabla \log\target(X_t) \dee t + C^{1/2} \dee W_t.
\]
Here, $\{W_t\}_{t\geq0}$ is a Wiener process and $C$ is any user-defined
positive-definite 
matrix. Under certain regularity conditions \citep{roberts2002langevin}, 
the solution $\cbra{X(t)}_{t\geq0}$
of \cref{eq:Langevin_dynamics} is ergodic and has $\normtarget$ as its stationary 
distribution. 
We can build a discrete-time
Markov chain $\{\tilde{x}_n\}_{n\in\nats}$ approximating
$\cbra{X(t)}_{t \geq 0}$ via the Euler--Maruyama discretization, with the update
\[
  \label{eq:EM_discretization}
  \tilde{x}'|\tilde{x} \sim 
    \distNorm\left( \tilde{x} + \frac{h}{2}C\nabla \log\target(\tilde{x}), \, h C \right),
\]
where $h > 0$ denotes a step size. 
This approximation is known as the unadjusted Langevin algorithm (ULA). 
ULA with a fixed step size $h > 0$ does not in general admit $\normtarget$ 
as a stationary distribution. 

MALA is a modification of ULA that restores the desired stationary distribution
by using ULA as a proposal within a
Metropolis--Hastings scheme. Concretely, at any point $x\in\statespace$,
MALA selects the next state $x'$ to be the ULA update $\tilde{x}'|x$ 
with probability
\[
  \min \left\{1, \frac{\target(\tilde{x}')\distNorm\left(x \mid \tilde{x}' + 
    \frac{h}{2}C\nabla \log\target(\tilde{x}'), \, h C\right)}{\target(x)
    \distNorm\left(\tilde{x}' \mid x + \frac{h}{2}C\nabla \log\target(x), \, 
    h C\right)}\right\},
\]
and otherwise sets $x'=x$, where $\distNorm(\cdot | \mu, \, \Sigma)$ is
the density of a $\distNorm(\mu,\Sigma)$ distribution.
This modification makes $\{x_n\}_{n\in\nats}$ an ergodic, $\normtarget$-reversible, 
and $\normtarget$-invariant Markov chain under appropriate conditions. 

We can reframe MALA as
HMC with a single leapfrog step (\cref{eq:leap_frog}) of size $\eps = h^{1/2}$ and 
positive definite mass matrix $M=C^{-1}$ \citep[\S 5.2]{neal2011mcmc}. 
Our exposition of autoMALA exploits this fact. 
We expand the space from $\statespace$ to $\hmcspace$ and augment the target density:
\[
  \label{eq:def_HMC_target}
  \hmctarget(x, p) := \normtarget(x) \cdot \distNorm(p \mid 0, \, M).
\]
Note that $x$ and $p$ are independent, and the  
$x$ marginal is the target distribution of interest $\pi$. The MALA proposal is equivalent to
drawing $p \sim  \distNorm(0,M)$ and then applying a map $L_\eps:\hmcspace\to\hmcspace$ consisting 
of a single leapfrog step of size $\eps > 0$ and a momentum flip: 
\[
  \label{eq:leap_frog}
  p_{\half}'  &= p + \frac{\eps}{2}\score(x) \\
  x'          &= x + \eps M^{-1} p'_{\half} \\
  \check p    &= p'_{\half} + \frac{\eps}{2}\score(x')\\
  p'          &= -\check p.
\]
The map $L_\eps$ is an involution, i.e., $L_\eps = (L_\eps)^{-1}$, and is volume preserving, 
$|\det \nabla L_\eps| = 1$ \citep[see, e.g.,][]{neal2011mcmc}; we use these facts and results from \cite{tierney1998note}
in the analysis of autoMALA. The proposal $(x',p')$ is then accepted with probability 
\[
  \label{eq:alpha_MALA}
  \alpha((x,p),(x', p')) := 
    \min\left\{1, \frac{\target(x') \distNorm(p' \mid 0, M)}{\target(x)\distNorm(p \mid 0, M)}\right\}. 
\]

We finish this section by noting an important motivation for autoMALA: 
it is necessary to determine a step size 
$\eps$ for MALA that appropriately balances the acceptance rate with 
fast exploration of the state space.
However, to account for different length scales of $\pi$ in different regions of state space, this 
tradeoff should be made \emph{at each iteration},
dependent on the current position $(x, p)$ in the state space. 
Our proposed sampler, autoMALA, does exactly this: autoMALA adapts
$\eps$ ``on the fly'' as a function of the current state,
and is carefully constructed to 
ensure  that we still asymptotically obtain samples from the correct target 
distribution $\pi$.

%% file: autoMALA.tex
\section{AUTOMALA}
\label{sec:autoMALA}

\begin{figure*}[!t]
  \centering
  \begin{subfigure}{0.38\textwidth}
    \centering
    \includegraphics[width=\textwidth]{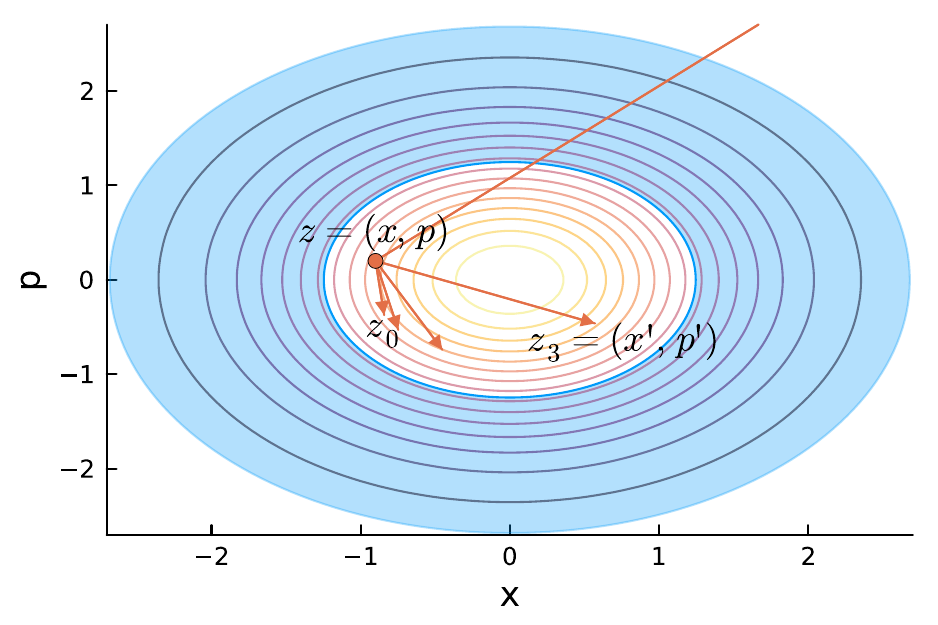}
  \end{subfigure}
  \begin{subfigure}{0.38\textwidth}
    \centering
    \includegraphics[width=\textwidth]{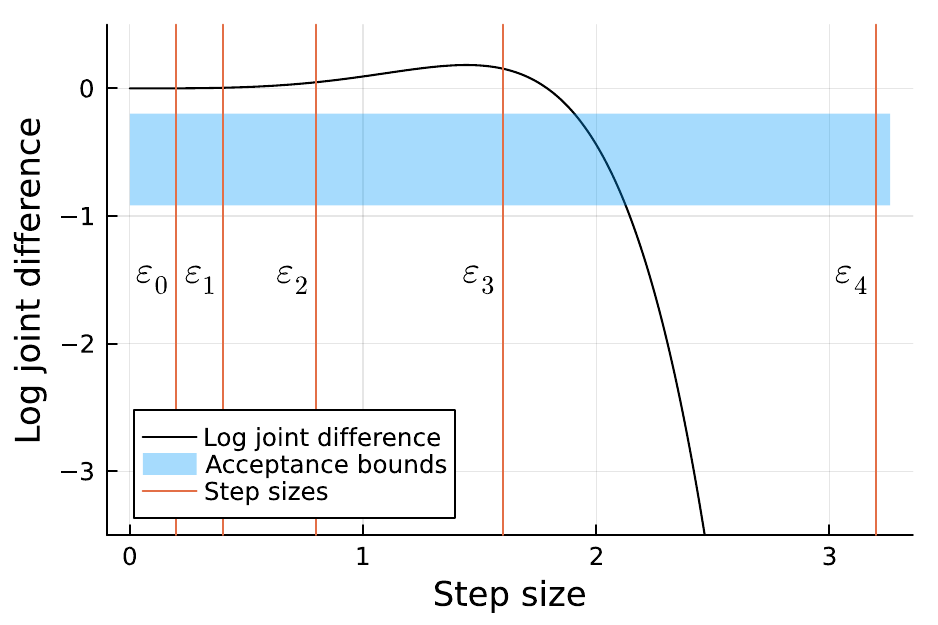}
  \end{subfigure}\\
  \begin{subfigure}{0.38\textwidth}
    \centering
    \includegraphics[width=\textwidth]{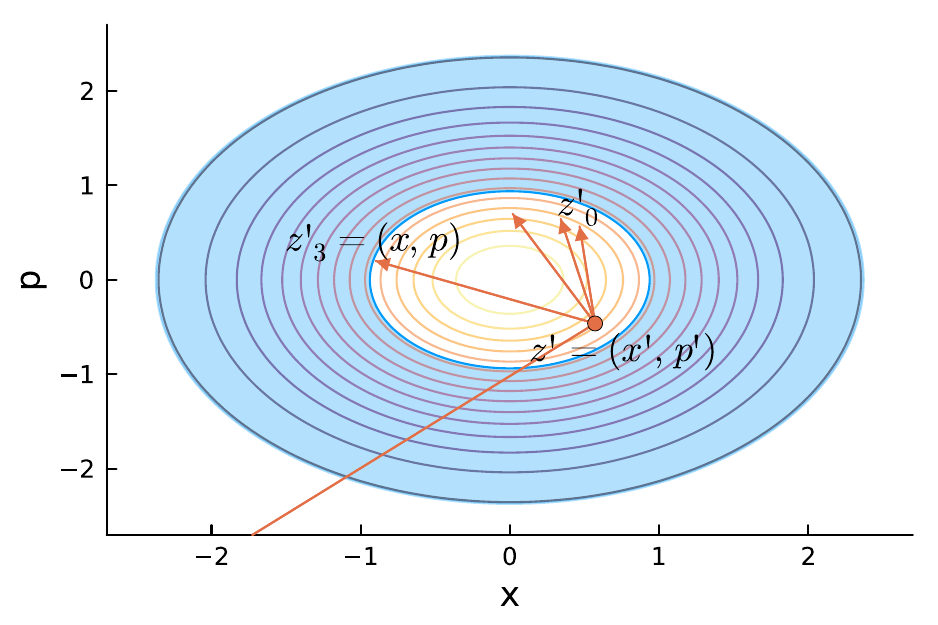}
  \end{subfigure}
  \begin{subfigure}{0.38\textwidth}
    \centering
    \includegraphics[width=\textwidth]{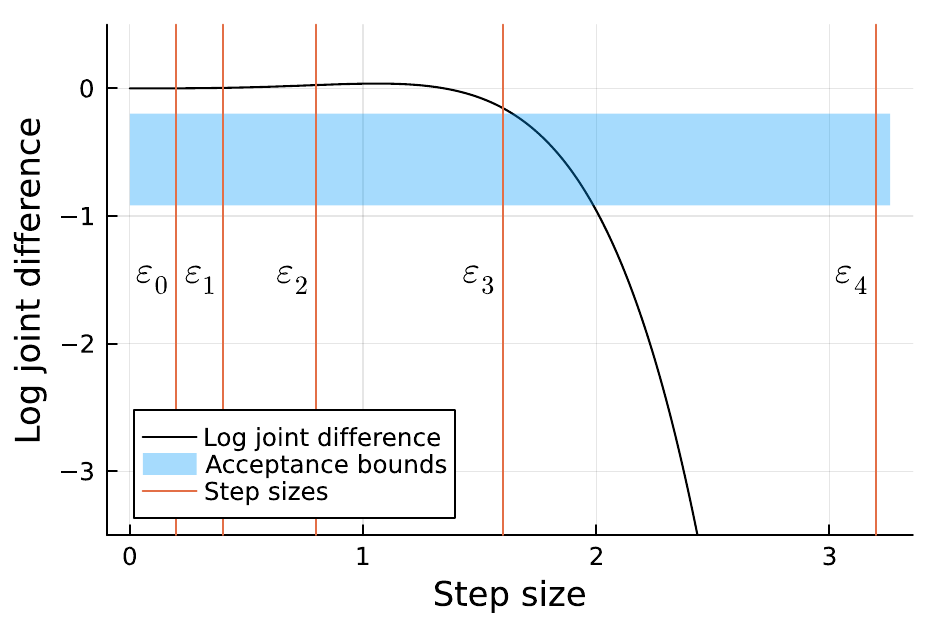}
  \end{subfigure}
  \caption{The autoMALA step size selection procedure on a standard univariate 
  normal distribution. Both figures illustrate the phase space for $(x,p)$. 
  In the figure, $z_j = L_{\epsi \cdot 2^{j}}(x,p)$ and  
  $z'_j = L_{\epsi \cdot 2^{j}}(x', p')$, whereas $z$ and $z'$ 
  denote $(x, p)$ and $(x', p')$, respectively. Similarly, $\eps_j = \epsi \cdot 2^j$.
  \textbf{Top (proposal phase):} Values of $a, b$ are drawn. 
  The initial density ratio is larger than $b$ 
  and the step size is doubled until the first time we  
  exit the acceptance region (orange arrows). The state just before exiting 
  is proposed. 
  \textbf{Bottom (reversibility check):} The same values of $a, b$ are kept, but 
  a new acceptance region is drawn relative to $(x', p')$. 
  The number of doublings is the same in both the forward and 
  reverse directions, and we pass the reversibility check. 
  When the initial step size is too large, a similar approach is taken 
  (with the difference of Line~\ref{line:other} instead of Line~\ref{line:asymmetry} 
  in \cref{alg:step_size_selector}).}
  \label{fig:stepsize_procedure}
\end{figure*}

autoMALA, at a high level, can be thought of as MALA with an automatic step size selection
procedure, followed by corrections to ensure that  $\pi$ is invariant.
One application of the autoMALA kernel consists of:
\begin{enumerate}
  \item \textbf{State augmentation:} Sample a momentum $p\sim \distNorm(0,M)$ 
    and two thresholds $a$ and $b$ uniformly from 
    $\Delta := \{(a,b) \in (0,1)^2: a < b\}$. 
    Let $s = (x, p, a, b)\in \hmcspace \times \Delta$ denote the augmented state.
  \item \textbf{MALA proposal with step size selection:} 
    Start at an initial step size guess, $\epsi > 0$.
    Let $r_{\text{init}}$ be the density ratio of the leapfrog 
    proposal $L_{\epsi}(x, p)$. 
    If $r_\text{init} \in (a,b)$, keep $\epsi$. 
    If $r_\text{init} \leq a$, halve the step size until the 
    ratio is strictly above $a$. 
    Otherwise, if $r_\text{init} \geq b$, double the step size until the ratio
    is strictly less than $b$, and then halve it once. 
    Denote the step size selection function $\eps(s, \epsi)$.
  \item \textbf{Reversibility check:} Let $s' = (x', p', a, b)$ and consider the 
    step size selected from $s'$, namely $\eps' = \eps(s', \epsi)$. 
    If $\eps \neq \eps'$, we remain at $x$. 
    This so-called ``reversibility check''---akin to \citet[Fig. 6]{neal2003slice}---is 
    essential to keeping the target $\pi$-invariant.
  \item \textbf{Metropolis--Hastings:} If the reversibility check passes, 
    apply the MALA Metropolis--Hastings correction, with probability given by \cref{eq:alpha_MALA}. 
\end{enumerate}

The first key feature of autoMALA that distinguishes it from standard MALA is the 
step size selection in Step 2. 
Drawing the acceptance thresholds $a,b$ uniformly generally keeps the acceptance probability
bounded away from the extremes (0 and 1), as desired, without needing to specifically tune their values.
For a given pair of thresholds $a < b$, 
if the initial acceptance ratio is too 
low relative to $a$ ($\epsi$ too large),
we make the steps more conservative by successively halving $\eps$ until the ratio is above $a$ (although not necessarily below $b$).
If instead the initial acceptance ratio is too high relative to $b$ 
($\epsi$ too small), 
we make the steps more aggressive by successively doubling $\eps$ until the ratio is below $b$ (although not necessarily above $a$),
and then \emph{halving the step size once} (Line~\ref{line:asymmetry} in \cref{alg:step_size_selector}). Note that the asymmetry
in these two cases is crucial. Without the final halving, the reversibility check
would always fail in the doubling sub-case. To see this, compare the
value $\ell$ (Line~\ref{line:ell} in \cref{alg:step_size_selector}) for the
last iteration $j^*$ of $\eps(s, \epsi)$ to the value $\ell'$ at iteration $j =
j^*$ of $\eps(s', \epsi)$.  From Line~\ref{line:ell}, $\ell = -\ell'$.
Moreover, since $\ell$ comes from the last iteration, by
Line~\ref{line:logb-terminate}, $\ell < \log(b)$, and since $b \in (0, 1)$,
that implies $\ell < 0$ so $\ell' > 0 > \log(b)$.  Hence $\eps(s', \epsi)$
would generally not terminate at that iteration without the final halving. The same
argument does not apply to the halving sub-case, as it is possible
to have $\ell > \log(a)$ and $-\ell > \log(a)$ simultaneously. 

Another key feature of autoMALA is the reversibility check in Step 3. To understand the basis for this check, 
let $A$ be the region of the augmented state space  
$\mcS = \reals^{2d} \times \Delta$ where the reversibility check succeeds, 
(both the forward and reverse step sizes are the same). 
Then $A$ is precisely the subset on which the autoMALA proposal is an involution. 
Hence, autoMALA preserves $\bar\pi$ on $A$ when combined with the 
Metropolis--Hastings correction
in Step 4 \citep{tierney1998note}. On the complement $A^c$, autoMALA is the 
identity, and so it preserves $\bar\pi$ there, as well.
Because autoMALA never proposes a state in $A^c$ from one in $A$ and vice versa, 
autoMALA keeps $\bar\pi$ invariant on all of $\mcS$.

Finally, in practice, the 
step size selection and reversibility check should operate on the integer exponent $j$ 
of the step size $\epsi \cdot 2^j$, instead of the floating point step size itself, so that the 
reversibility checks do not suffer from floating point errors.
This is the reason why \cref{alg:step_size_selector} returns the integer 
number of doublings as well as the step size.

\begin{algorithm}[t]
	\begin{algorithmic}[1]
    \Require Initial state $x_0$, 
      number of iterations $T$, initial step size $\eps_\text{init}$, 
      preconditioning matrix $\hat{\Sigma}$, number of 
      unadjusted burn-in iterations per round $t_\text{unadj}$
		  (default:  $t_\text{unadj} = 1$)
    \For{$t$ {\bf in} 1, 2, \dots, $T$}
        \State $\eta \gets \distBeta_{01}(1,1, 1/2, 2/3)$
        \LineComment{form random preconditioning matrix}
          \State $(\hat{\Sigma}_\text{AM})_{i,i}^{-1/2}=\eta \hat \Sigma_{i, i}^{-1/2}+(1-\eta)$
        \State $p \gets \distNorm(0_d, \hat{\Sigma}_\text{AM}^{-1})$
          \Comment{sample momentum}
        \State $(u_1, u_2) \gets U([0,1]^2)$
        \LineComment{soft bounds for acceptance ratio}
          \State $(a, b) \gets (\min\cbra{u_1,u_2}, \max\cbra{u_1,u_2})$
        \State $s \gets (x_{t-1}, p, a, b)$ 
        \State $\eps, j \gets \eps(s, \epsi)$ \label{line:forward_step_size}
        \State $s' \gets (L_{\eps}(x_{t-1}, p), a, b)$
          \Comment{proposed state}
        \State $\eps', j' \gets \eps(s', \epsi)$ \label{line:backward_step_size}
        \State $\alpha \gets 1 \wedge \bar{\pi}(s')/\bar{\pi}(s)$
          \Comment{see \cref{eq:pi_augmented}}
        \State $U \gets U[0,1]$
        \If{$t \le t_\text{unadj}$ \textbf{ or }$(j = j'$ \textbf{ and } $U \leq \alpha)$}
            \State $x_t \gets x'$  \Comment{accept}
        \Else 
            \State $x_t \gets x_{t-1}$ \Comment{reject}
        \EndIf
        \State $\eps_t \gets (\eps + \eps')/2$
	\EndFor
    \State \Return $\{(x_t, \eps_t)\}_{t=1}^T$
	\end{algorithmic}
  \caption{{\texttt autoMALA}$(x_0, T, \epsi, \hat{\Sigma}, t_\text{unadj})$}
  \label{alg:autoMALA}
\end{algorithm}

\begin{algorithm}[t]
	\begin{algorithmic}[1]
    \Require state $s = (x, p, a, b)$, initial step size $\eps_\text{init}$.
    \State $\eps \gets \eps_\text{init}$
    \State $s' \gets (L_\eps(x,p), a, b)$
    \State $\ell \gets \log\bar{\pi}(s') - \log\bar{\pi}(s)$
    \State $\delta \gets \ind\{\ell \geq \log(b)\} - \ind\{\ell \leq \log (a)\}$
    \State $j = 0$ \Comment{number of doublings/halvings}
      \label{alg_step:delta}
    \If{$\delta = 0$}
    	\State \Return $\epsi$
    \EndIf
    \While{true} 
        \State $\eps \gets \eps\cdot2^\delta$
        \State $j \gets j + \delta$
        \State $s' \gets (L_\eps(x,p), a, b)$ \label{line:eval}
        \State $\ell \gets \log\bar{\pi}(s') - \log\bar{\pi}(s)$ \label{line:ell}
        \If{$\delta=1$ and $\ell < \log(b)$} \label{line:logb-terminate}
            \State \Return $\eps/2, j-1$ \Comment{See \cref{sec:autoMALA}} \label{line:asymmetry}
        \ElsIf{$\delta=-1$ and $\ell > \log(a)$}
            \State \Return $\eps, j$ \label{line:other}
        \EndIf 
    \EndWhile
	\end{algorithmic}
  \caption{Step size selector $\eps(s, \eps_\text{init})$}
  \label{alg:step_size_selector}
\end{algorithm}

\subsection{Round-based tuning}
The autoMALA sampler described above chooses an appropriate step size $\eps$ at 
any given point $(x, p)$ in the state space. 
The computational cost of the step size selection procedure 
increases logarithmically in the gap between $\eps_\text{init}$ and the selected step $\eps$. 
To decrease this cost, we use a simple round-based approach to tuning $\eps_\text{init}$ (\cref{alg:round_autoMALA}). 
During tuning round $r$ we perform $T_r = 2^r$ iterations of autoMALA. 
In the first tuning round, we use $\eps_\text{init} = 1$.
At the end of the $r^\text{th}$ tuning round, we obtain the average step size used during 
that tuning round; this then becomes the initial step size $\eps_{\text{init}}$ used in the $r+1^{\text{th}}$ tuning round.

Additionally, MALA and autoMALA can perform better with 
preconditioning, i.e.\ a positive definite matrix $C \neq I_d$ in \cref{eq:Langevin_dynamics} 
(equivalently written as a mass matrix $M = C^{-1}$). 
We obtain an appropriate preconditioner as a by-product of our round-based adaptive scheme. 
Our approach is motivated by multivariate normal target distributions with covariance matrix $\Sigma$, where the choice  $C = M^{-1} = \Sigma$ is recommended \citep{neal2011mcmc}, 
but to avoid matrix operations with cost superlinear in $d$, it is 
customary to  
use a diagonal matrix with entry $(i, i)$ given by the marginal variance of 
component $i$, $\hat \Sigma_{i, i} = \widehat \var[x^{(i)}]$. 
To improve the robustness of the diagonal matrix approach---which is prone
to errors even in the Gaussian setting \citep{hird2023quantifying}---at each step 
we perform the random interpolation 
$(\hat{\Sigma}_\text{AM})_{i,i}^{-1/2}=\eta \hat \Sigma_{i, i}^{-1/2}+(1-\eta)$, 
with $\eta$ sampled independently from a zero-one-inflated beta distribution,
denoted as $\distBeta_{01}(\tilde \alpha, \tilde \beta, p, m)$, 
for some values of $\tilde\alpha, \tilde\beta > 0$ and $p,m\in[0,1]$.
Here, $\distBeta_{01}(\tilde \alpha, \tilde \beta, p, m)$ denotes a random variable 
that is distributed according to $\distBern(p)$ with probability $m$ and 
$\distBeta(\tilde \alpha, \tilde \beta)$ with probability $1-m$.
In our experiments, we use $\tilde\alpha = \tilde\beta = 1$, $p=1/2$, and $m=2/3$
so that each of the two endpoints $\{0,1\}$ and the interval $(0,1)$ all have an equal 
chance ($1/3$) of being selected (see \cref{app:preconditioning} for an experimental
validation of this approach). 

Note that our round-based procedure does not introduce additional tuning parameters, and plays well with other round-based 
algorithms such as non-reversible parallel tempering, described in \cite{syed2021nrpt} 
and implemented in \cite{surjanovic2023pigeons}.

\begin{algorithm}[t]
	\begin{algorithmic}[1]
    \Require Initial state $x_0$, number of rounds $R$, number of 
    unadjusted burn-in iterations per round $t_\text{unadj}$
    (default:  $t_\text{unadj} = 1$)
	\State $\eps_\text{init} = 1$
	\State $\hat{\Sigma} \gets I_d$
    \For{$r$ {\bf in} 1, 2, \dots, $R$}
      \State $T \gets 2^r$       
      \State $\cbra{(x_t, \eps_t)}_{t=1}^T \gets 
        \texttt{autoMALA}(x_0, T, \eps_\text{init}, \hat{\Sigma}, t_\text{unadj})$       
      \State $\eps_\text{init} \gets T^{-1} \sum_{t=1}^T \eps_t$
      \State $x_0 \gets x_T$ 
      \State $\hat{\Sigma} \gets \text{diag}\left(
        \widehat{\var}[x_t^{(1)}]_{t=1}^T, \ldots, \widehat{\var}[x_t^{(d)}]_{t=1}^T \right)$
  	\EndFor
    \State \Return $\{x_t\}_{t=1}^T$
	\end{algorithmic}
  \caption{Round-based autoMALA}
  \label{alg:round_autoMALA}
\end{algorithm}

\subsection{Unadjusted burn-in}

Empirically, the reversibility check succeeds with 
high probability at stationarity in all cases investigated. However, we 
have also observed situations where an arbitrary initialization
yields a near-zero success probability. 
As a result, we skip the reversibility check and the Metropolis--Hasting rejection step for 
a constant number of iterations $t_\text{unadj} = 1$ at the beginning of each round. 
Empirically we find that this step helps avoid the sampler getting stuck at a bad initial point. 

\subsection{Theoretical results}

In this section we establish that
the step size selection algorithm given by \cref{alg:step_size_selector} terminates 
almost surely (\cref{thm:termination}) and that 
autoMALA has the correct invariant distribution $\bar{\pi}$ (\cref{thm:invariance}). 
The proofs of these theoretical results can be found in the supplementary material.
In what follows, we introduce some regularity conditions on the target distribution 
$\pi$. For a vector $x \in \reals^d$, let $\abs{x}$ be its Euclidean norm. 

\bassump{(Smoothness)}
\label{assump:smoothness}
$\pi$ is twice continuously differentiable on $\reals^d$.
\eassump

\bassump{(Tails)} 
\label{assump:tail}
  $\lim_{\abs{x} \to \infty} \pi(x) = 0$.  
\eassump

To analyze the autoMALA algorithm, it will be useful to define the augmented density
\[
\label{eq:pi_augmented}
\amtarget(s) := 2 \normtarget(x) \cdot \distNorm(p \mid 0,M) \cdot \ind_\Delta(a,b),
\]
where $\ind_\Delta(\cdot)$ is the indicator for the set $\Delta$. 

Our first result confirms that the step size selection 
(\cref{alg:step_size_selector}) terminates almost surely. 
For $s \in \mcS$ and $\epsi > 0$, we define $\tau(s, \epsi) \geq 1$ to be 
the number of iterations of the while loop in \cref{alg:step_size_selector}.

\bthm{(Step size selector termination)}
\label{thm:termination}
Let $\epsi > 0$ and suppose that $\pi$ satisfies \cref{assump:smoothness}
and \cref{assump:tail}. Then, $\tau(s, \epsi) < \infty$ $\bar \pi$-\as
\ethm

We now formally state the $\pi$-invariance property of autoMALA.
Suppressing $\eps_{\text{init}}$ in the notation, 
define $L_\eps(s) = (L_\eps(x, p), a, b)$,  $T(s) = L_{\eps(s)}(s)$ and the region $A \subset \amspace$ where the 
reversibility check succeeds,  $A = \{s \in \amspace : \eps(s) = \eps \circ T(s)\}$. 
Define the deterministic proposal 
\[ \involution(s) = T(s)\ind_A(s) + s\ind_{A^c}(s), \]
from which we construct the autoMALA kernel
\[ 
K_\textrm{AM}(s, \dee s' ) &= (1 - \alpha(s))\delta_{s}(\dee s') +  \alpha(s) \delta_{\involution(s)}(\dee s') \\
\alpha(s) &= \min \left\{ 1, \frac{\bar \pi(\involution(s))}{\bar \pi(s)} \right\}.
\]

\begin{figure*}[t!]
  \centering
  \begin{subfigure}{0.32\textwidth}
    \centering
    \includegraphics[width=\textwidth]{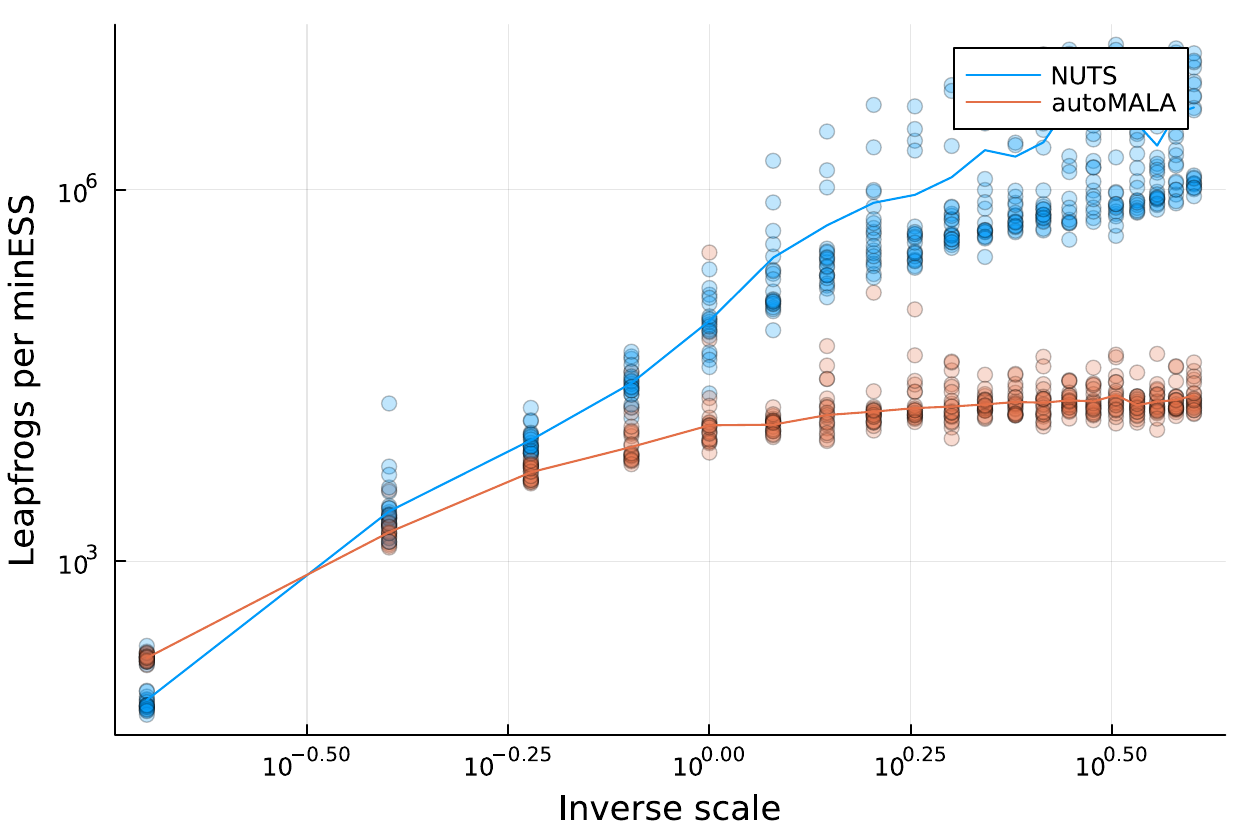}
  \end{subfigure}
  \begin{subfigure}{0.32\textwidth}
    \centering
    \includegraphics[width=\textwidth]{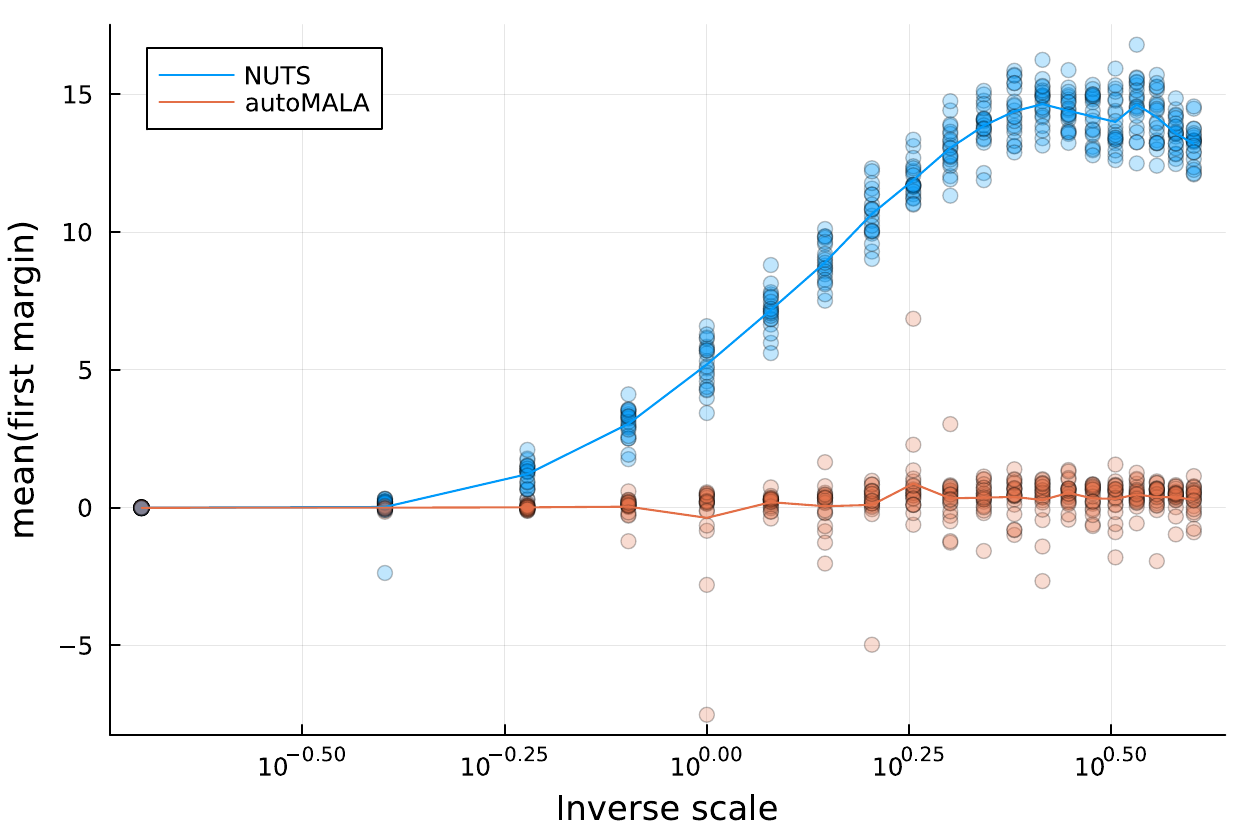}
  \end{subfigure}
  \begin{subfigure}{0.32\textwidth}
    \centering
    \includegraphics[width=\textwidth]{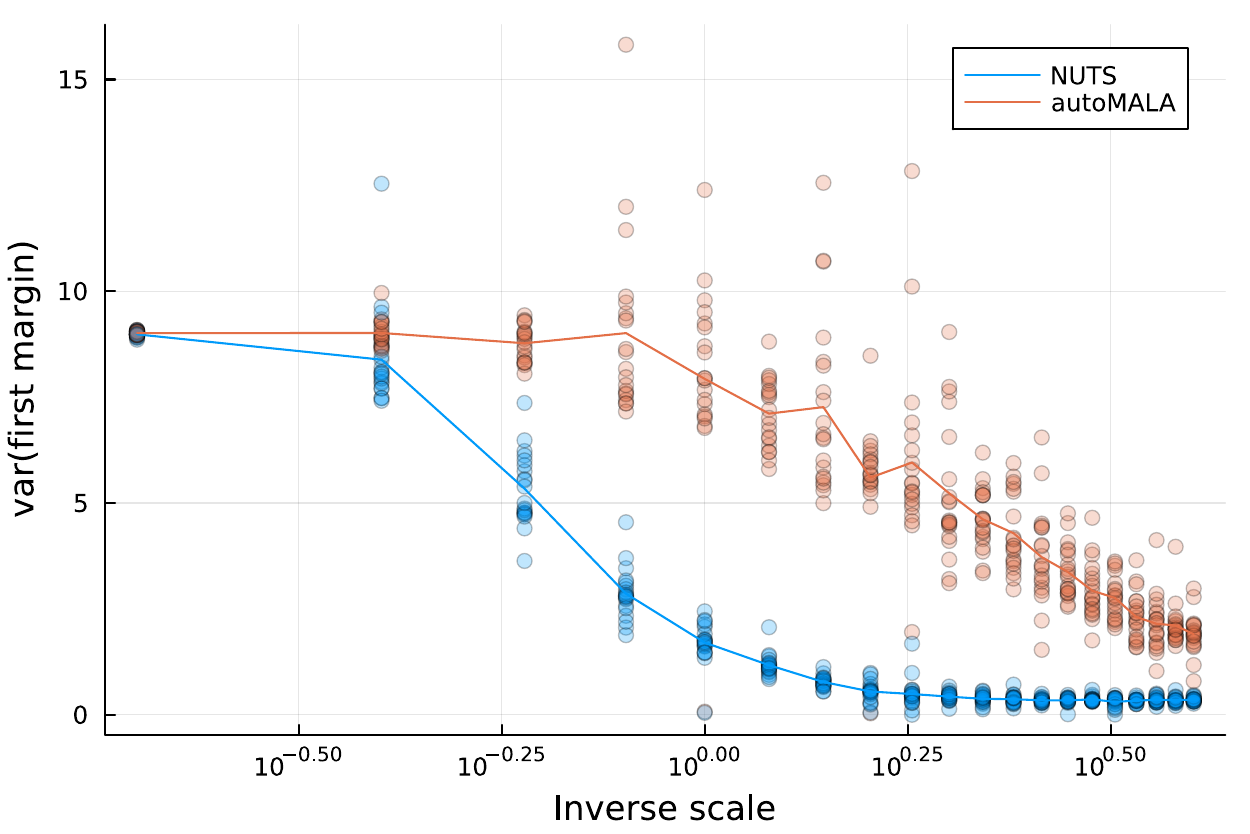}
  \end{subfigure}
  \caption{
    Comparison of autoMALA and NUTS on Neal's funnel with a varying scale parameter
    for 20 different seeds.
    Bold curves indicate averages across seeds. 
    Geometries increase in difficulty from left to right.
    \textbf{Left to right:}
    Number of leapfrog evaluations per 1000 minESS (lower is better), 
    estimated mean of the first marginal (exact mean is 0),
    estimated variance of the first marginal (exact variance is 9).}
  \label{fig:varying_local_geometry}
\end{figure*}

\bthm{(Invariance)}
\label{thm:invariance}
Under \cref{assump:smoothness} and \cref{assump:tail},
for any measurable $B \subset \amspace$, 
\[  \int \bar \pi(\dee s) \amkern(s, B) = \bar \pi(B).  \]
\ethm

Notice that \cref{alg:autoMALA} is a deterministic composition 
of $\amkern$ with a block Gibbs update on $p, a, b$, and 
therefore is $\bar\pi$-invariant as a corollary of \cref{thm:invariance}.

Regarding irreducibility, it does not appear straightforward to show 
that autoMALA can visit any $x$ state after only a \emph{single step}, unlike MALA.
However, we conjecture that autoMALA can still visit any $x$ state after multiple 
steps under reasonable conditions, which we leave as an open problem. 
Additionally, one can easily guarantee irreducibility---if desired---by mixing autoMALA
with another kernel known to be irreducible. For example, using the mixture 
$\lambda K_\text{AM} + (1-\lambda) K_\text{MALA}$, where $K_\text{MALA}$ is a 
MALA transition kernel and $0 < \lambda < 1$, leads to an irreducible sampler. Of
course, $\lambda$ should be chosen close to one in order to retain the adaptive 
benefits of autoMALA.

%% file: experiments.tex
\section{EXPERIMENTS}
\label{sec:experiments}
In this section, we present experiments that investigate
the performance of autoMALA on targets with varying geometry
and increasing dimension, as well as the convergence behaviour
of the initial step size. We compare autoMALA against
the locally adaptive sampler NUTS \citep{hoffman2014nuts}, 
as well as standard MALA, which is a non-adaptive method.
We refer readers to the supplementary material for a complete specification of 
all experimental details.

We use the effective sample size (ESS) \citep{flegal_markov_2008}
to capture the statistical efficiency of Markov chains.
For synthetic distributions with known marginals, 
we complement the ESS 
with: comparison of the estimated means and variances to their known values;
one-sample Kolmogorov-Smirnov test statistics;
and a more reliable estimator of ESS, labelled $\exactess$, that takes into account 
the known target moments. (See the supplement for details on $\exactess$.)
We combine the traditional ESS and $\exactess$ by examining the
statistic $\text{minESS} := \min \cbra{\text{ESS}, \exactess}$. 

\paragraph{Software and reproducibility} autoMALA is 
available as part of an open-source Julia package, \href{https://pigeons.run/dev/}{\texttt{Pigeons.jl}}.
The package can use targets specified as Stan models,
Julia functions, and \texttt{Turing.jl} models. The code for 
the experiments is available at \url{https://github.com/Julia-Tempering/autoMALA-mev}.
Experiments were performed on Intel i7 CPUs and the ARC Sockeye computer 
cluster at the University of British Columbia.

\begin{figure}[t!]
  \centering 
  \includegraphics[width=0.48\textwidth]{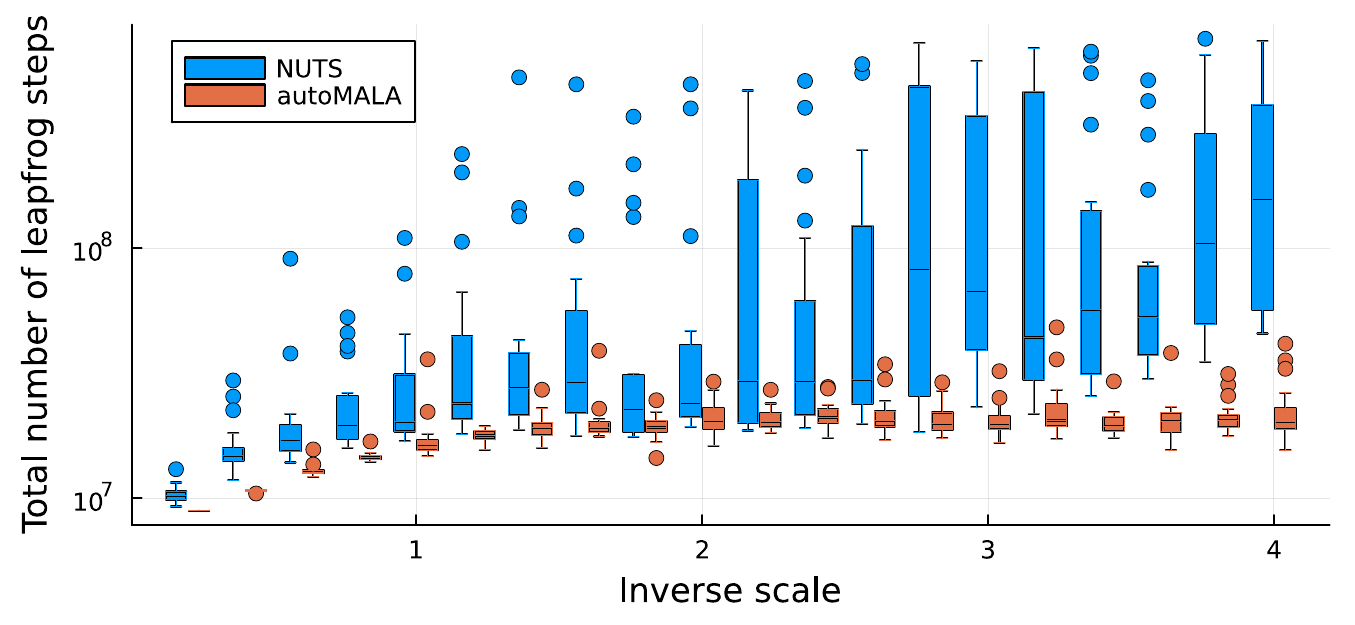}
  \caption{
     Total number of leapfrog steps used by autoMALA and NUTS for each scale parameter 
     in Neal's funnel.
  }
  \label{fig:funnel_scale_leapfrogs}
\end{figure}

\begin{figure*}[!t]
  \centering
  \begin{subfigure}{0.32\textwidth}
    \centering
    \includegraphics[width=\textwidth]{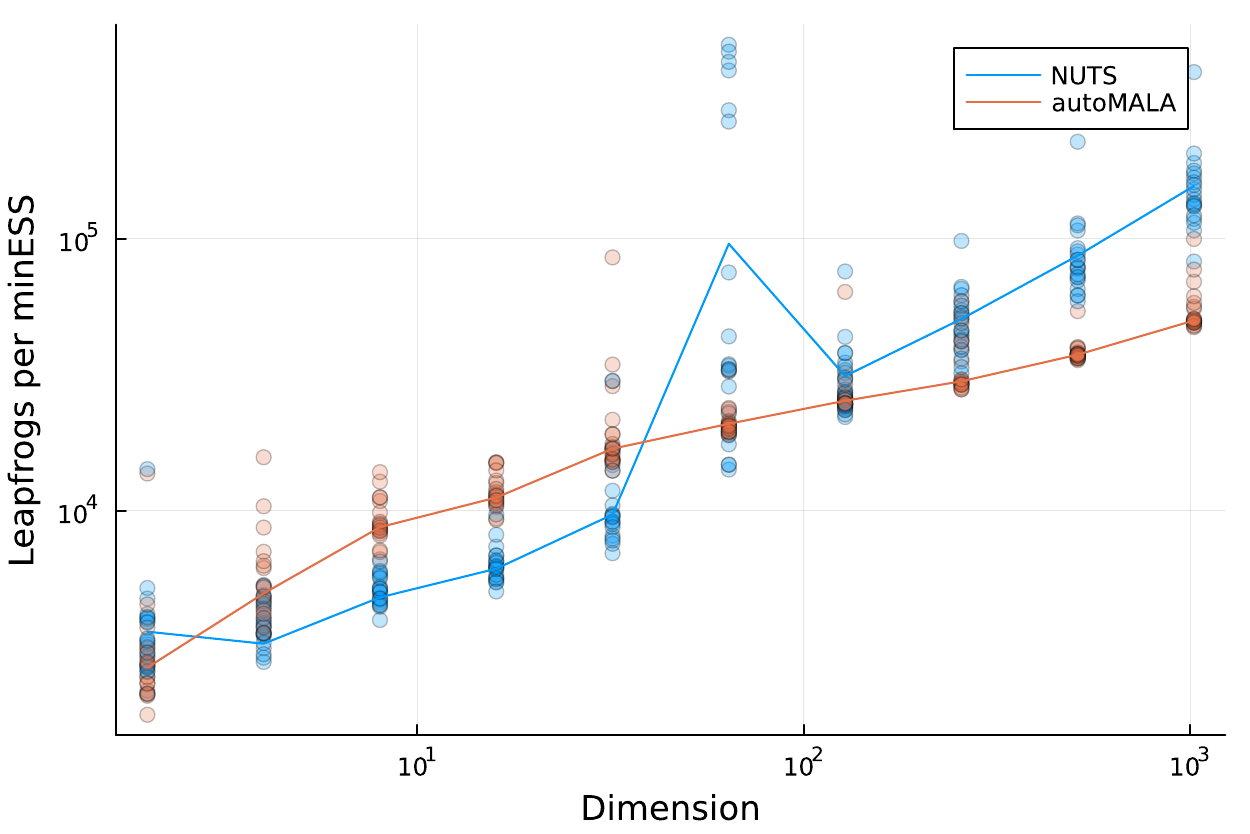}
  \end{subfigure}
  \begin{subfigure}{0.32\textwidth}
    \centering
    \includegraphics[width=\textwidth]{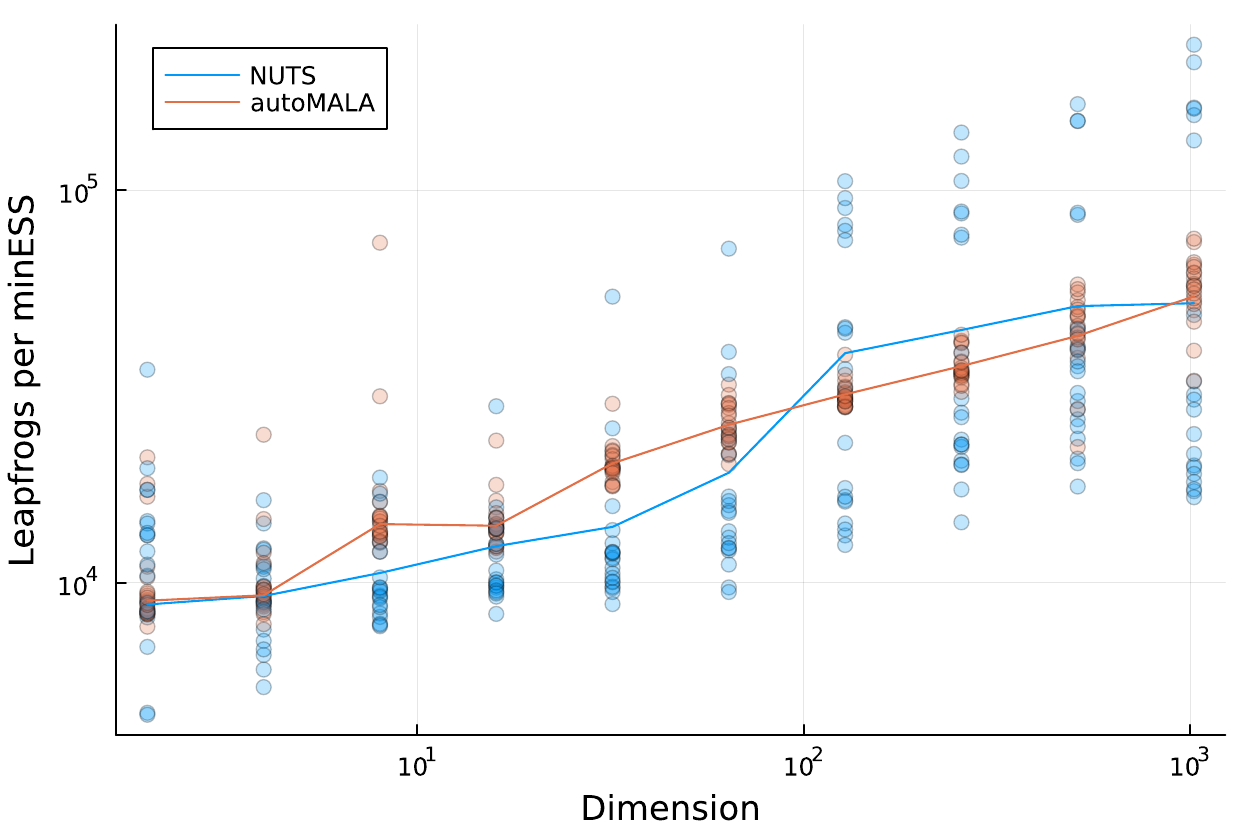}
  \end{subfigure}
  \begin{subfigure}{0.32\textwidth}
    \centering
    \includegraphics[width=\textwidth]{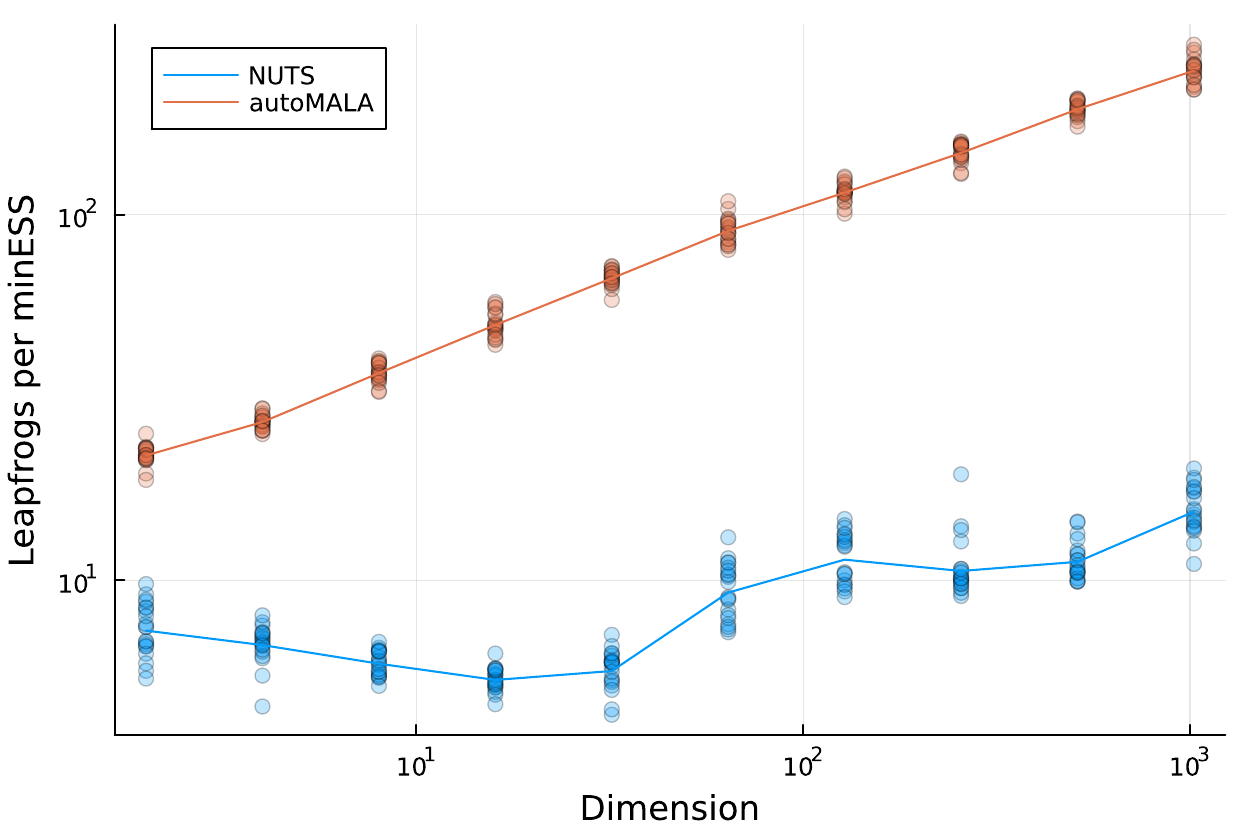}
  \end{subfigure}
  \caption{
    Number of leapfrog evaluations per 1000 minESS (lower is better) 
    on targets with increasing dimension.
    Each point represents results from a separate seed for the experiment. 
    Bold curves indicate averages across seeds. 
    \textbf{Left to right:} Neal's funnel, banana distribution, and multivariate 
    normal distribution.}
  \label{fig:high_dimensional}
\end{figure*}

\subsection{Varying local geometry}

We first investigate the performance of autoMALA on targets with varying local
geometry. There are two target distributions that we consider for this first
synthetic example: Neal's funnel and a banana distribution.
Both targets contain a scale parameter that we can tune to make the
distributions more difficult to sample from (greater variation in local
geometry); difficulty increases as the scale parameter approaches zero.
We compare autoMALA to NUTS: NUTS automatically adapts the number of leapfrog
steps, but fixes a single step size after an initial warmup. In cases like the funnel, 
where the target distribution requires both very large and very small steps,
the NUTS step size will be too large in the narrow part of the funnel;
varying the number of leapfrog steps will not resolve the issue. 
Furthermore, note again that $n$ steps of doubling the trajectory length 
for NUTS takes compute time proportional to $2^n$, whereas $n$ steps of step 
size doubling in the autoMALA algorithm has a cost proportional to $n$. 
Therefore, for distributions with variable local geometries, 
we expect dynamic step size selection (autoMALA) to have 
a lower per-cost step compared to 
trajectory length selection (NUTS), and also less running time 
variability.

In \cref{fig:varying_local_geometry} we see that autoMALA outperforms NUTS 
in terms of the number of leapfrog evaluations per $\text{minESS}$ when applied to Neal's funnel.
We also see that autoMALA provides significantly more accurate estimates of the mean
and variance on the first marginal (i.e., the ``difficult'' marginal direction
along the elongated funnel). Moreover, 
autoMALA provides improved statistical performance despite using significantly
fewer leapfrog steps, as shown in \cref{fig:funnel_scale_leapfrogs},
as well as providing a smaller variability in the number of leapfrog steps.
This makes autoMALA more suitable for distributed MCMC algorithms with synchronization,
such as parallel tempering, where it is desirable for the runtime 
across machines to be approximately equal 
\citep{syed2021nrpt,surjanovic2022parallel,surjanovic2023pigeons}.
Similar results hold for the scaled banana distribution (see supplement), 
although there is less variation in geometry and so the 
two samplers are more comparable.

As a cautionary point, note that standard estimates of the ESS can be potentially
misleading in these problems; the $\exactess$ is more reliable. 
In particular, before reaching stationarity---which can take a long time for difficult
problems---it is possible for the ESS estimate to be very high, even though 
the obtained samples do not resemble the target distribution. 
For instance, even though draws from NUTS are not 
representative of the target distribution, 
standard ESS estimates can still be very high (see the supplement). 
When target marginal means and variances are known, which is typical in synthetic problems,
we recommend using the $\exactess$ for evaluation, which can detect poor mixing
via a misestimated mean and variance.

\subsection{Dimensional scaling}

\cref{fig:high_dimensional} shows an investigation of the scaling properties of autoMALA
as the dimension increases for the funnel, banana, and normal distributions. 
We again compare to NUTS, which is known for its favourable scaling properties on
\iid high-dimensional target distributions \citep{beskos2013optimal}, in terms of 
the number of leapfrog evaluations
per effective sample. The results for the normal target agree with this theory,
with NUTS performing better. In contrast, autoMALA is highly competitive in the two 
targets with extreme variation in local geometry, achieving the same efficiency
values and scaling law as NUTS.

\subsection{Step size convergence}
\label{sec:stepsize_convergence}
\begin{figure}[t]
  \centering
  \begin{subfigure}{0.48\textwidth}
    \centering
    \includegraphics[width=\textwidth]{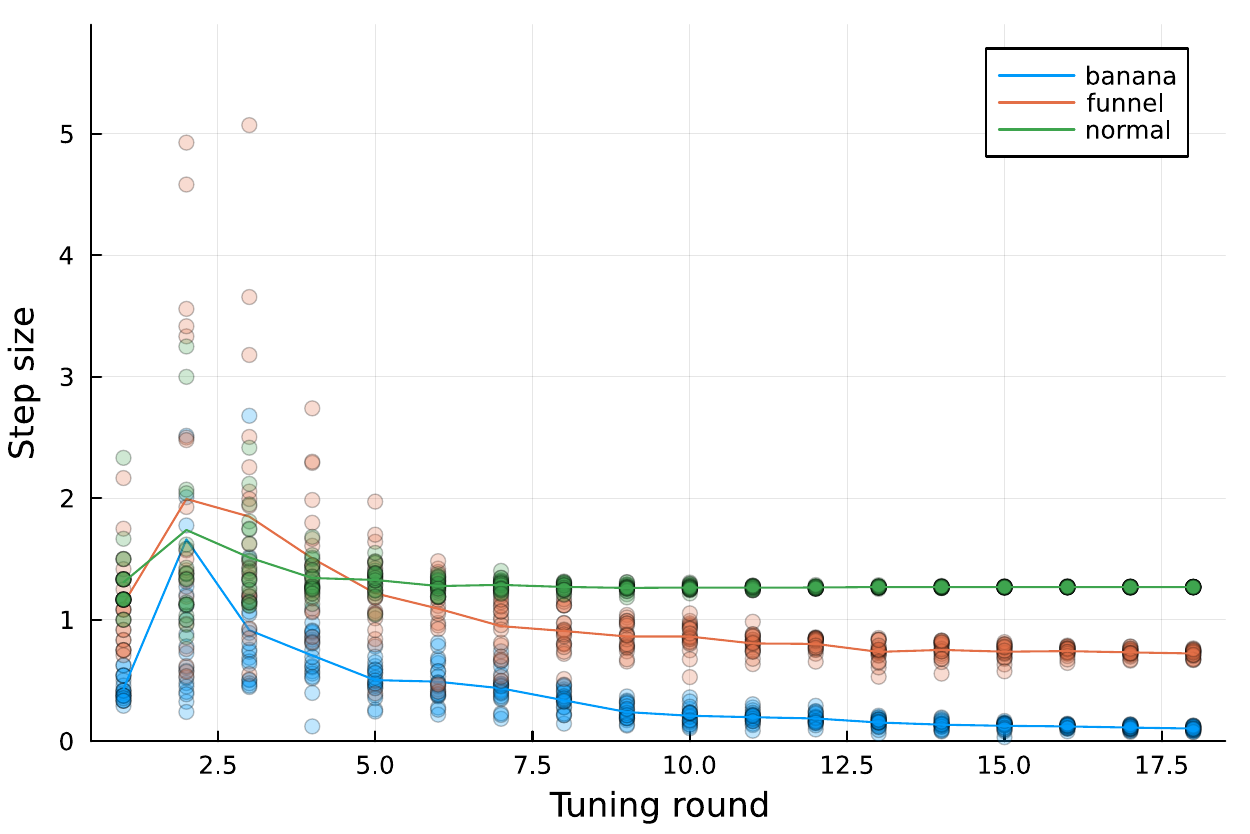}
  \end{subfigure}
  \caption{autoMALA initial step sizes as a function of the tuning round for $d=2$. 
    Each new tuning round corresponds to a doubled number of samples used to 
    estimate the initial step size. Results for higher-dimensional targets 
    are provided in the supplement.}
  \label{fig:stepsize_convergence}
\end{figure}

Round-based autoMALA with a doubling of the number of MCMC iterations at each round 
(\cref{alg:round_autoMALA}) should converge to a reasonable choice of 
initial step size, $\epsi$. 
\cref{fig:stepsize_convergence} shows the chosen autoMALA default step size 
as a function of the tuning round for the three synthetic targets (banana, funnel, normal). 
In general, we see that the initial step size guess 
converges as the tuning rounds proceed. 
We note that generally a greater number of tuning rounds is needed for the step size to converge 
when autoMALA is applied to higher-dimensional distributions, due to a greater 
degree of Monte Carlo estimate variability, as explained in the supplement.

\subsection{Comparison to non-adaptive algorithms}
\label{sec:comparison_to_nonadaptive}

\begin{figure}[t]
  \centering
  \includegraphics[width=0.48\textwidth]{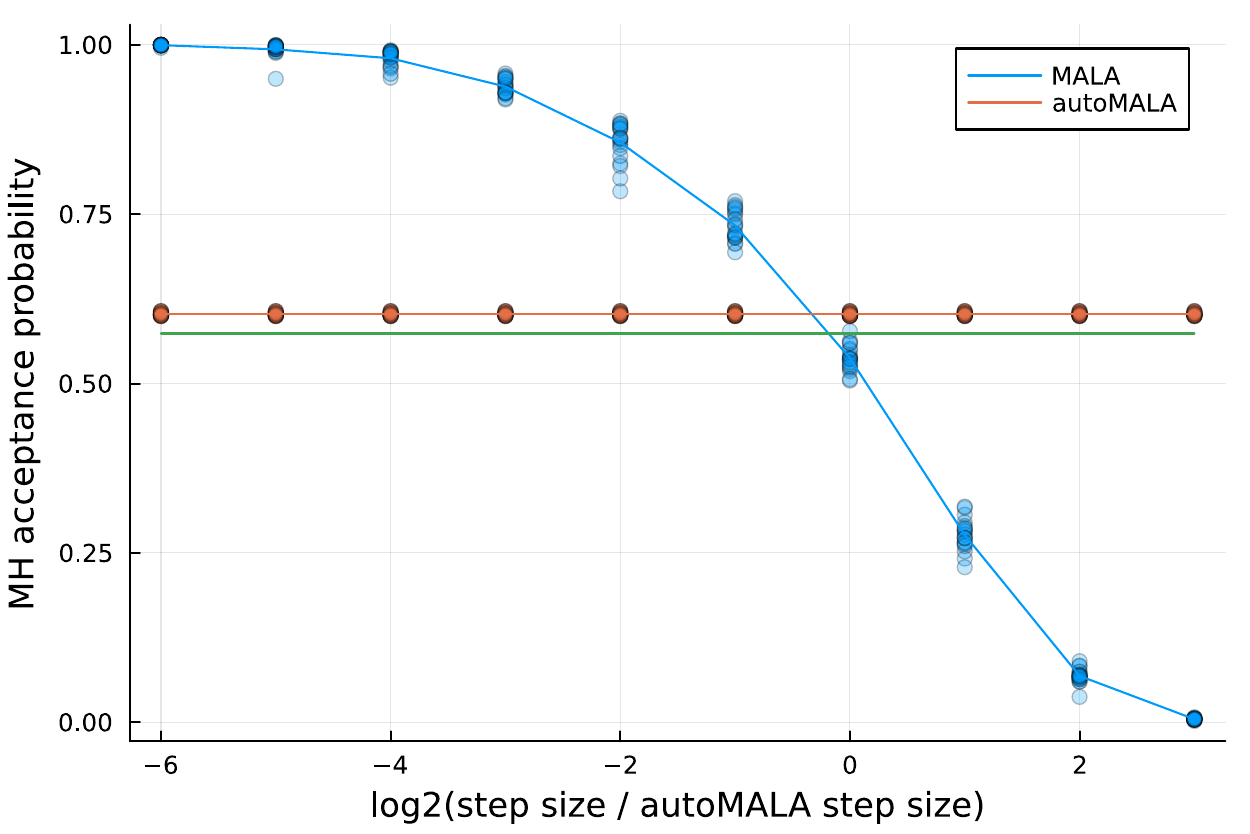}
  \caption{Metropolis--Hastings acceptance probability for autoMALA and MALA 
    as a function of the initial step size for Neal's funnel ($d=2$).
    The green horizontal line indicates the theoretical optimal 0.574 acceptance 
    probability for MALA.
    Step sizes are selected relative to the optimal autoMALA step size.}
  \label{fig:MALA_stepsize_acceptance}
\end{figure}

It is of interest to compare autoMALA to its non-adaptive predecessor MALA. To
this end, we first perform a long run of autoMALA and retrieve its final step size $\eps_\text{final}$.
We then do a grid search for a MALA step size targeting an acceptance 
probability of 0.574, a value known to be optimal for various types of targets 
\citep{roberts1998optimal}. 
\cref{fig:MALA_stepsize_acceptance} shows that $\eps_\text{final}$ is a good proxy
for the optimal MALA step size 
with respect to the ideal Metropolis--Hastings acceptance probability.

We additionally assess the robustness (and lack thereof) of round-based autoMALA and MALA 
with respect to the initial step size. We find that even with very poor choices 
for the initial step size with autoMALA, we are able to converge to a reasonable step 
size that yields adequate acceptance probabilities. This is not the case for 
MALA, where the acceptance probabilities depend very heavily on the selected 
step size (see supplement).

\begin{figure}[!t]
  \centering 
  \includegraphics[width=0.48\textwidth]{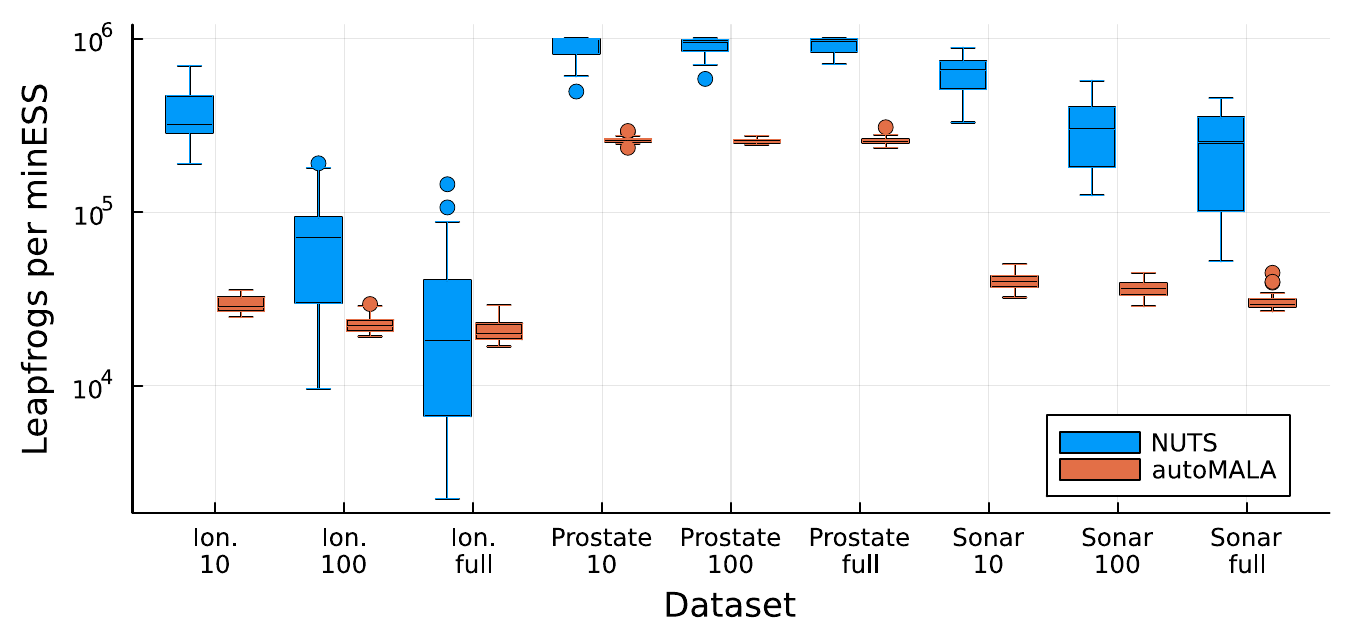}
  \caption{
     Number of leapfrog evaluations per 1000 minESS (lower is better) for the 
     horseshoe variable selection model on three datasets with varying random 
     subsampling.
  }
  \label{fig:boxplots-nleap_to_min_ess-model-horseshoe}
\end{figure}

\subsection{Real data experiments}
\label{sec:real_data}

Finally, we compare autoMALA and NUTS on three joint variable selection and 
binary classification tasks: a dataset of 
radar returns from the ionosphere \citep{igillito1989ionosphere}; data of sonar
signals to distinguish metal from rock objects \citep{sejnowski_connectionist}, 
and the prostate cancer dataset from \citet{piironen2018iterative}.
For each dataset we use a Bayesian logistic regression model with a horseshoe prior to 
induce sparsity on the weights of the predictors \citep{carvalho2009handling}.
When the number of observations is 
low, the horseshoe prior creates varying geometries: 
the inclusion probabilities are non-degenerate, so as the sampler runs, the number 
of ``active'' variables---i.e., the effective dimensionality---changes, and so does 
the appropriate step size to use.

To investigate the small-data regime, we form two additional versions
of each of the datasets by randomly sub-sampling $10$ and $100$ observations---for
a total of $9$ possible combinations. We then run autoMALA and NUTS on all 
for $2^{14}$ iterations, and repeat this process $20$ times. Since there is no
parameter in the model with known distribution, we compute minESS by 
taking the minimum over two different ESS estimators---batch means and 
auto-covariance estimation---across all parameters.
\cref{fig:boxplots-nleap_to_min_ess-model-horseshoe} reports the results of this
experiment, showing that autoMALA generally outperforms NUTS.
As anticipated, the difference in performance is higher in the low data 
regime due to varying geometries in the target distributions.

%% file: conclusion.tex
\section{CONCLUSION}
\label{sec:conclusion}
This work introduced a new MCMC method, autoMALA, that 
can be thought of as the Metropolis-adjusted Langevin algorithm
with a step size that is automatically selected \emph{at each iteration}
to adapt to local target distribution geometry.
We proved that autoMALA preserves the correct stationary distribution despite
its adaptivity.
Further, we developed a round-based tuning procedure that automatically
finds a reasonable initial step size guess and leapfrog step preconditioner matrix.
In our experiments we observed that autoMALA performs 
well on various target distributions and outperforms NUTS when applied to target distributions 
with varying local geometry.

There are several directions for future research. As mentioned in \cref{sec:autoMALA},
a proof of the irreducibility of autoMALA is left for future work. We posit that
a two-step analysis of the Markov kernel---similar to the one carried out for NUTS in 
\citet[Thm.\ 8]{durmus2023convergence}---should be enough to prove irreducibility.
Additionally, it would be interesting to analyze if there are better choices than
the uniform distribution over $\Delta$ for the pair $(a,b)$. Alternatively, a
non-reversible update strategy for the pair $(a,b)$---as done in 
\citet{neal2020uniform} for the Metropolis acceptance decision---might introduce
desirable persistence in the behavior of autoMALA. 
Finally, we note that the proposed automatic step size selection procedure 
can be generalized and extended to other involution-based samplers whose performance
is critically dependent on one or a few hyperparameters.

%% file: checklist.tex
\section*{Checklist}

 \begin{enumerate}

 \item For all models and algorithms presented, check if you include:
 \begin{enumerate}
   \item A clear description of the mathematical setting, assumptions, 
   algorithm, and/or model. [\textbf{Yes}/No/Not Applicable]
   \item An analysis of the properties and complexity (time, space, sample size) 
   of any algorithm. [\textbf{Yes}/No/Not Applicable]
   \item (Optional) Anonymized source code, with specification of all dependencies, 
   including external libraries. [\textbf{Yes}/No/Not Applicable]
 \end{enumerate}

 \item For any theoretical claim, check if you include:
 \begin{enumerate}
   \item Statements of the full set of assumptions of all theoretical results. 
   [\textbf{Yes}/No/Not Applicable]
   \item Complete proofs of all theoretical results. [\textbf{Yes}/No/Not Applicable]
   \item Clear explanations of any assumptions. [\textbf{Yes}/No/Not Applicable]     
 \end{enumerate}

 \item For all figures and tables that present empirical results, check if you include:
 \begin{enumerate}
   \item The code, data, and instructions needed to reproduce the main experimental 
   results (either in the supplemental material or as a URL). [\textbf{Yes}/No/Not Applicable]
   \item All the training details (e.g., data splits, hyperparameters, how they were 
   chosen). [\textbf{Yes}/No/Not Applicable]
         \item A clear definition of the specific measure or statistics and error 
         bars (e.g., with respect to the random seed after running experiments 
         multiple times). [\textbf{Yes}/No/Not Applicable]
         \item A description of the computing infrastructure used. (e.g., type of 
         GPUs, internal cluster, or cloud provider). [\textbf{Yes}/No/Not Applicable]
 \end{enumerate}

 \item If you are using existing assets (e.g., code, data, models) or 
 curating/releasing new assets, check if you include:
 \begin{enumerate}
   \item Citations of the creator if your work uses existing assets. 
   [\textbf{Yes}/No/Not Applicable]
   \item The license information of the assets, if applicable. [Yes/No/\textbf{Not Applicable}]
   \item New assets either in the supplemental material or as a URL, if applicable. 
   [\textbf{Yes}/No/Not Applicable]
   \item Information about consent from data providers/curators. 
   [Yes/No/\textbf{Not Applicable}]
   \item Discussion of sensible content if applicable, e.g., personally 
   identifiable information or offensive content. [Yes/No/\textbf{Not Applicable}]
 \end{enumerate}

 \item If you used crowdsourcing or conducted research with human subjects, 
 check if you include:
 \begin{enumerate}
   \item The full text of instructions given to participants and screenshots. 
   [Yes/No/\textbf{Not Applicable}]
   \item Descriptions of potential participant risks, with links to Institutional 
   Review Board (IRB) approvals if applicable. [Yes/No/\textbf{Not Applicable}]
   \item The estimated hourly wage paid to participants and the total amount 
   spent on participant compensation. [Yes/No/\textbf{Not Applicable}]
 \end{enumerate}

 \end{enumerate}

%% file: supplement_proofs.tex
\section{PROOFS}

\bprfof{\cref{thm:termination}}
We split the proof into two cases depending on the value of $\delta$ in Line
\ref{alg_step:delta} of \cref{alg:step_size_selector}: $\delta = -1$ and $\delta = 1$.
From this point on, fix any $s = (x, p, a, b) \in \mcS$ and $\epsi > 0$.

If $\delta = -1$, our initial step size was too large and we begin our step size 
halving procedure. Our claim is that there exists an $0 < \eps' < \epsi$ such that 
if $s' = (L_{\eps'}(x, p), a, b)$, then 
$\ell(\eps') = \log\bar\pi(s') - \log\bar\pi(s) > \log(a)$. 
Once we show this, it follows that $\tau(s, \epsi) < \infty$ by our use of the 
step size halving procedure. 
To see that this claim holds, observe that by combining the leapfrog steps in \cref{eq:leap_frog}, 
we have for any $\eps > 0$ that 
\[
  L_\eps(x, p)
  = (\tilde x(\eps), \tilde p(\eps)), 
\] 
where 
\[
  \label{eq:leapfrog_expanded}
  \tilde x(\eps) &= x + \eps M^{-1} p + \frac{\eps^2}{2} M^{-1} \nabla \log \gamma(x) \\
  \tilde p(\eps) &= -\left(p + \frac{\eps}{2} \nabla \log \gamma(x) + 
    \frac{\eps}{2} \nabla \log \gamma(\tilde x(\eps))\right).
\]
From the continuous differentiability of $\pi$ (and hence $\log\gamma$), 
it follows that $\tilde x(\eps) \to x$ and $\tilde p(\eps) \to -p$ as $\eps \to 0$.
This implies that $\ell(\eps) \to 0$ as $\eps \to 0$. 
We therefore require that $\log(a) < 0$ to ensure that $\tau(s, \epsi) < \infty$.

If $\delta = 1$, this means that the initial step size was too small and that 
we begin our step size doubling procedure.
We claim that there exists an $\eps' > \epsi$ such that if 
$s' = (L_{\eps'}(x,p), a, b)$, then 
$\ell(\eps') = \log \bar\pi(s') - \log\bar\pi(s) < \log(b)$. 
Provided that $\log(b) > -\infty$,
it suffices to prove that as $\eps \to \infty$, we have 
$\log\bar\pi(\tilde x(\eps), \tilde p(\eps), a, b) \to -\infty$.
Using the expansion of the leapfrog step given by \cref{eq:leapfrog_expanded}, 
observe that $\abs{\tilde x(\eps)} \to \infty$ as $\eps \to \infty$, 
provided that either $p \neq 0$ or $\nabla \log\gamma(x) \neq 0$.
Then, $\pi(\tilde x(\eps)) \to 0$ as $\eps \to \infty$, by the assumption that 
$\pi(x) \to 0$ as $\abs{x} \to \infty$.
We conclude that $\bar\pi(\tilde s) \to 0$ as $\eps \to \infty$, as well, 
because $0 \leq \bar\pi(\tilde s) \leq C \cdot \pi(\tilde x)$ for some $C > 0$. 

Combining these two cases, we have that $\tau(s, \epsi) < \infty$ 
provided that $s$ satisfies the following 
conditions: $\log(a) < 0$, $\log(b) > -\infty$, and $p \neq 0$. 
These conditions hold with probability one under $\bar\pi$, thereby completing the proof.
\eprfof

\bprfof{\cref{thm:invariance}}
We use \citet[Theorem 2]{tierney1998note}, more specifically, Corollary 2 
(``deterministic proposal'').
From \citet[Theorem 2]{tierney1998note}, standard calculations (reviewed, e.g., in \cite{geyer2003mhg}), 
establish that sufficient conditions for invariance of $\amkern$ are:  (1) that $\involution$ is an 
involution, i.e., $\involution = \involution^{-1}$, and (2) that the set of points 
$s$ where the change of variable formula \citep[Thm. 2.47]{folland_real_1999} applies has $\bar \pi$ probability one, 
and at those points, the absolute determinant of the Jacobian of $\involution$ is one. 
We establish (1) and (2) in Lemmas \ref{lem:invo} and \ref{lem:jacobian} respectively. 
\eprfof

\blem
\label{lem:invo} 
The mapping $\involution$ is an involution, $\involution = \involution^{-1}$. 
\elem 

\bprf
We split the argument into two sub-cases, either $s \in A$ 
(introduced in the main text), or $s \notin A$. If $s \notin A$, $\involution$ is 
equal to the identity, so the involution property holds. Suppose now $s \in A$. 
We first show that for $s \in A$, $T\circ T(s) = s$. 
We have $T\circ T(s)= L_{\eps\circ T(s)}(T(s))$ by the definition of $T$. 
Next, since $s \in A$, $L_{\eps\circ T(s)}(T(s)) = L_{\eps(s)}(T(s))$, again by definition. 
Applying again the definition of $T$, $L_{\eps(s)}(T(s)) = L_{\eps(s)} \circ L_{\eps(s)}(s)$. 
Now, using the fact that the leapfrog is time-reversible \citep[\S 2.3]{neal2011mcmc}, 
a synonym for the involution property in this context, we have 
$L_{\eps(s)} \circ L_{\eps(s)}(s) = s$. 
Finally, if $s \in A$, then by the above argument, $\eps \circ T \circ T(s) = \eps(s)$. Since 
$s \in A$, then $\eps(s) = \eps\circ T(s)$, hence $T(s) \in A$. 
This allows us to complete the argument: for $s \in A$,  $\involution \circ \involution(s) =  \involution \circ T(s)$, and using $T(s)\in A$, $\involution \circ T(s) = T\circ T(s) = s$.
\eprf

\blem
\label{lem:Q_is_differentiable}\label{lem:jacobian} 
Under the conditions of \cref{thm:invariance}, there exists an open set $G \subset \amspace$ such that:
\begin{enumerate}
  \item $\bar\pi(G) = 1$,
  \item $\involution$ is continuously differentiable on $G$ with $| \det \nabla \involution | = 1$. 
\end{enumerate}
\elem 

\bprfof{\cref{lem:Q_is_differentiable}} 
We start by identifying a ``bad'' set $B$ of potential discontinuities. 
We will then show it is $\bar \pi$-null, 
and finally, use its complement as a building block for the ``good'' set $G$ 
satisfying the differentiability conditions from point 2.\ of the above statement. 
 
{\bf Construction of the bad set.} For any given point $s = (z, a, b) \in \mcS$, where $z = (x,p)$, 
note that the set of states that one autoMALA step could visit is countable (by step, we mean one iteration of the for loop in \cref{alg:autoMALA} of the main text; by visit, we mean evaluation of the density at a point, the creation of state for which such evaluations occur are at Lines \ref{line:forward_step_size}--\ref{line:backward_step_size} of \cref{alg:autoMALA}, which in turn call \cref{alg:step_size_selector}, where 
evaluations occur at Line~\ref{line:eval}).
Define the \emph{trace of autoMALA} as the countable set 
\[
  \mcT_s = \cbra{\Phi_0(s), \Phi_1(s), \ldots},
\]
where $\Phi_i(z)$ is the $i^\text{th}$ point visited by autoMALA in Lines 
\ref{line:forward_step_size} or \ref{line:backward_step_size} of \cref{alg:autoMALA}. 
(Any ordering suffices for our purposes.)
By inspection of the algorithm,  each $\Phi_i(s)$ is of the form 
$\Phi_i(s) = (L_{\eps_i}(z), a, b)$ (forward pass) or $(L_{\eps_i} \circ L_{\eps'_i}(z), a, b)$ 
(reversibility check) for some $\eps_i, \eps'_i > 0$.
Define the set where we have a finite number of such evaluations as 
$F = \cbra{s : \abs{\mcT_s} < \infty}$.
By \cref{thm:termination}, we have $\bar\pi(F) = 1$.
We also define a superset to the trace,  
the \emph{potential trace of autoMALA} $\bar{\mcT}_s$:
\[
	\mcT_z &= \{ L_{\epsi 2^i}(z) : i \in  \ints\}, \\
	\bar \mcT_s &= \mcT_z \cup \left( \bigcup_{\tilde z \in \mcT_z} \mcT_{\tilde z}  \right) \times \{a\} \times \{b\},	
\]
constructed so that the mappings contained in $\bar \mcT_s$ depend only on $z$ (and not on $a, b$, in contrast to $\Phi_i(s)$), while also having  
$\mcT_s \subset \bar\mcT_s$. 
Since the potential trace $\bar \mcT_s$ is a countable union of countable sets, it is countable, so we index it as: 
\[
  \bar{\mcT}_s = \cbra{\bar{\Phi}_0(z), \bar{\Phi}_1(z), \ldots}  \times \{a\} \times \{b\}.  
\]

We now define a collection of bad points, $B$, that identify possible sources of 
discontinuity of $\involution$:
\[
  B = \cbra{s  = (z, a, b)  : \ell_{ij}(s) \in \cbra{\log a, \log b}, \text{for some} \, i \neq j, \, i,j \leq \abs{\mcT_s}}, \qquad
  \ell_{ij}(s) = \log \bar\pi(\Phi_i(s)) - \log \bar\pi(\Phi_j(s)).
\]
Also, define 
\[
  \bar B &= \cbra{s = (z, a, b) : \bar\ell_{ij}(z) \in \cbra{\log a, \log b}, \, i \neq j}, \qquad
  \bar\ell_{ij}(z) = \log \bar\pi(\bar\Phi_i(z)) - \log \bar\pi(\bar\Phi_j(z)), \\
  &= \bigcup_{i \neq j} \cbra{s : \bar\ell_{ij}(z) \in \cbra{\log a, \log b}}
\]

{\bf The bad set is null. }Note that $B \subset \bar B$.
We argue that $\bar\pi(B) = 0$ by showing that $\bar\pi(\bar B) = 0$. 
To see this, note that by Tonelli's theorem, 
\[
  \bar\pi(\bar B) &= 2 \int \pi(\dee z) \int_\Delta \ind_{\bar B}(s) \, \dee a \, \dee b \\
  &\le 2 \int \pi(\dee z)  \sum_{i, j} \pr(\bar \ell_{i,j}(z) \in \{\log A, \log B\}).
\]
where $(A, B) \sim \text{Unif}(\Delta)$. Since $A, B$ are non-atomic random variables, i.e.\ for all 
$c \in [0, 1]$, $\pr(A = c) = \pr(B = c) = 0$, we have
$ \pr(\bar \ell_{i,j}(z) \in \{\log A, \log B\}) = 0$ and hence $\bar \pi(\bar B) = 0$. 

{\bf Construction of the good set.} From here, set $G = F \cap B^c$. 
 Note that  from \cref{thm:termination}, $\bar \pi(F) = 1$ and hence $\bar\pi(G) = 1$. 

{\bf Showing that the good set is open and satisfies 2.} In the following, 
we use a re-parameterization $(u,v) = (\log a, \log b)$.  
For any $s = (x, p, u, v) \in G$ we would like to show that there exists a 
$\delta > 0$ such that 
the differentiability statement 2.\ of the result hold in a ball of radius $\delta$, 
denoted $N_\delta(s) = \{\tilde s \in \amspace : \| s - \tilde s \| < \delta \}$. 
For $s \in G$ we have that $\min\cbra{\abs{\ell_{ij} - u}, \abs{\ell_{ij} - v}} > 0$ for all 
$i, j \in \cbra{1, 2, \ldots, \abs{\mcT_s}}$ (otherwise if this minimum would be zero, we would have $s \in B$, contradicting $s \in G$ since $G = F \cap B^c$).
Now, set 
\[
  \delta_\ell = \min_{i,j \leq \abs{\mcT_s}} \cbra{\abs{\ell_{ij} - u}, \abs{\ell_{ij} - v}} > 0.
\]  
By the continuity of $\log\bar\pi \circ \Phi_i$ for all $i \in \cbra{1, 2, \ldots, \abs{\mcT_s}}$, 
we have that there exists a $\delta_i > 0$ such that for all 
$\tilde s \in N_{\delta_i}(s)$ we have 
\[
  \abs{\log \bar\pi(\Phi_i(\tilde s)) - \log\bar\pi(\Phi_i(s))} < \frac{\delta_\ell}{3}.
\]
Then, taking $\delta = \min\cbra{\delta_1, \ldots, \delta_{\abs{\mcT_s}}, \delta_\ell/3}$,
we have that inside $N_\delta(s)$ all branching decisions made by the autoMALA algorithm are identical and 
hence for all $\tilde s \in N_\delta(s)$ we have that 
either  $\{s, \tilde s\} \subset A$ or $\{s, \tilde s\} \subset A^c$. Also, $\eps(\tilde s) = \eps(s)$
and hence $\eps$ is constant on $N_\delta(s)$.

Next, we verify that $\involution(\tilde s)$ is 
continuously differentiable in the ball $N_\delta(s)$. As noted above, $N_\delta(s) \subset A$ or  $N_\delta(s) \subset A^c$, so we consider these two sub-cases in turn. If  $\tilde s \in N_\delta(s) \subset A^c$, 
$\involution(\tilde s) = \tilde s$ for all $\tilde s$, which is differentiable and has $| \det \nabla \involution (\tilde s) | = 1$. 
Otherwise, if $\tilde s \in N_\delta(s) \subset A$, we have 
$\involution(\tilde s) = L_{\eps(\tilde s)}(\tilde s) = L_{\eps(s)}(\tilde s)$. 
Because the step size is constant in this neighbourhood, and by the differentiability 
of the leapfrog operator under the assumption that $\pi$ is twice continuously 
differentiable (Assumption 3.1), we have that $\involution(\tilde s)$ is also continuously differentiable in this case.
Since $\eps(\cdot)|_{N_\delta(s)} \equiv \eps_0$ for some $\eps_0$, 
$\involution|_{N_\delta(s)} = L_{\eps_0}(\cdot)|_{N_\delta(s)}$, 
and hence we obtain in this sub-case as well that $| \det \nabla \involution (\tilde s) | = 1$, 
this time from standard properties of the leap-frog operator reviewed in the main text. 
\eprfof

%% file: supplement_experiments.tex
\section{ADDITIONAL EXPERIMENTS}

Below we offer additional details about our experiments and provide the full 
set of figures produced for each of the experiments.
Unless otherwise stated, we counted the number of leapfrogs in the warmup and final phases, 
but only retained samples for computing the ESS and other statistics on the final phase.

\subsection{Synthetic data and models} 
We lay out the synthetic data and models used in \cref{sec:experiments}. 

The $d$-dimensional Neal's funnel with scale parameter $\beta > 0$ and $d \geq 2$ 
is given by 
\[
  \label{eq:neals_funnel}    
  X_1 \sim \distNorm(0, \, 9), \qquad 
  X_2, \ldots, X_d \mid X_1 = x_1 \stackrel{iid}{\sim} \distNorm(0, \, \exp(x_1/\beta)).
\]
Note that we write $\distNorm(0, \sigma^2)$ to denote a normal random variable with 
variance $\sigma^2$.

The $d$-dimensional banana distribution with scale parameter $\beta > 0$ and $d \geq 2$ 
is given by 
\[
  \label{eq:banana}    
  X_1 \sim \distNorm(0, \, 10), \qquad 
  X_2, \ldots, X_d \mid X_1 = x_1 \stackrel{iid}{\sim} \distNorm(x_1^2, \, \beta^2/10).
\]

The $d$-dimensional normal distribution is in all cases given by 
\[
  \label{eq:normal}
  X_1, \ldots, X_d \stackrel{iid}{\sim} \distNorm(0, \, 1).  
\]

\subsection{Varying local geometry}
We carried out experiments comparing autoMALA to NUTS on Neal's funnel and the banana 
distribution with varying scale parameters. As $\beta \to 0$, the sampling problem 
increases in difficulty. \cref{fig:funnel_scale_all,fig:banana_scale_all,%
fig:banana_scale_leapfrogs} show various metrics used to assess autoMALA and NUTS 
for the two targets (a subset of the results for the funnel were already presented 
in \cref{sec:experiments}).
For the funnel distribution we considered values of the scale parameter 
$\beta \in \cbra{1/0.2, 1/0.4, 1/0.6, \ldots, 1/4.0}$.
For the banana distribution we used values of 
$\beta \in \cbra{2^{13}, 2^{12}, \ldots, 2^{-6}}$. 
We used 20 different seeds and $2^{20}$ samples for each seed to estimate statistics with 
$2^{20}$ samples to warm up and tune NUTS. 
We used the same number of samples
for autoMALA in the banana case, but used only $19$ rounds for the funnel in order to
match the overall computational effort of NUTS (see \cref{fig:funnel_scale_leapfrogs}).

\begin{figure*}[!t]
  \centering
    \begin{subfigure}{0.32\textwidth}
        \centering 
        \includegraphics[width=\textwidth]{deliverables/AM_funnel_scale/scaling-leapfrog_min-model-funnel_scale.pdf}
    \end{subfigure}
    \begin{subfigure}{0.32\textwidth}
        \centering 
        \includegraphics[width=\textwidth]{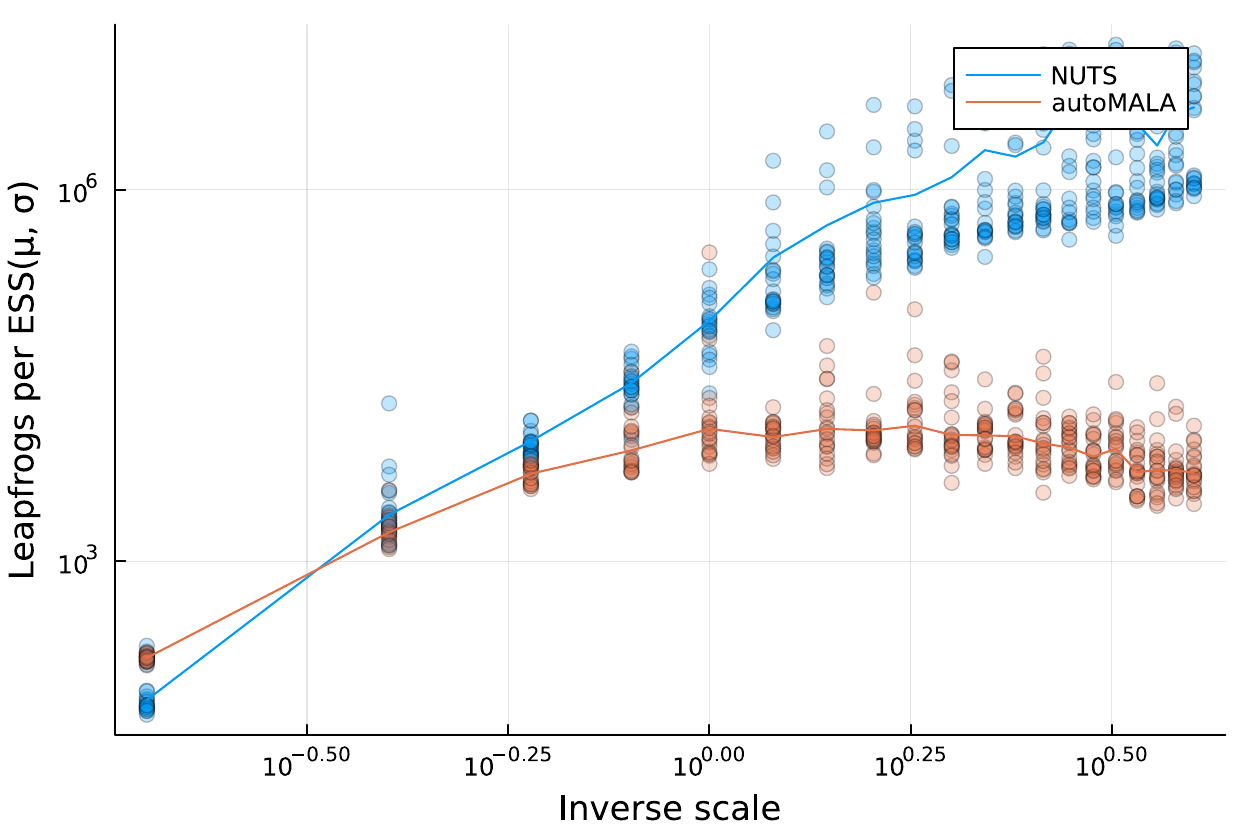}
    \end{subfigure}
    \begin{subfigure}{0.32\textwidth}
        \centering 
        \includegraphics[width=\textwidth]{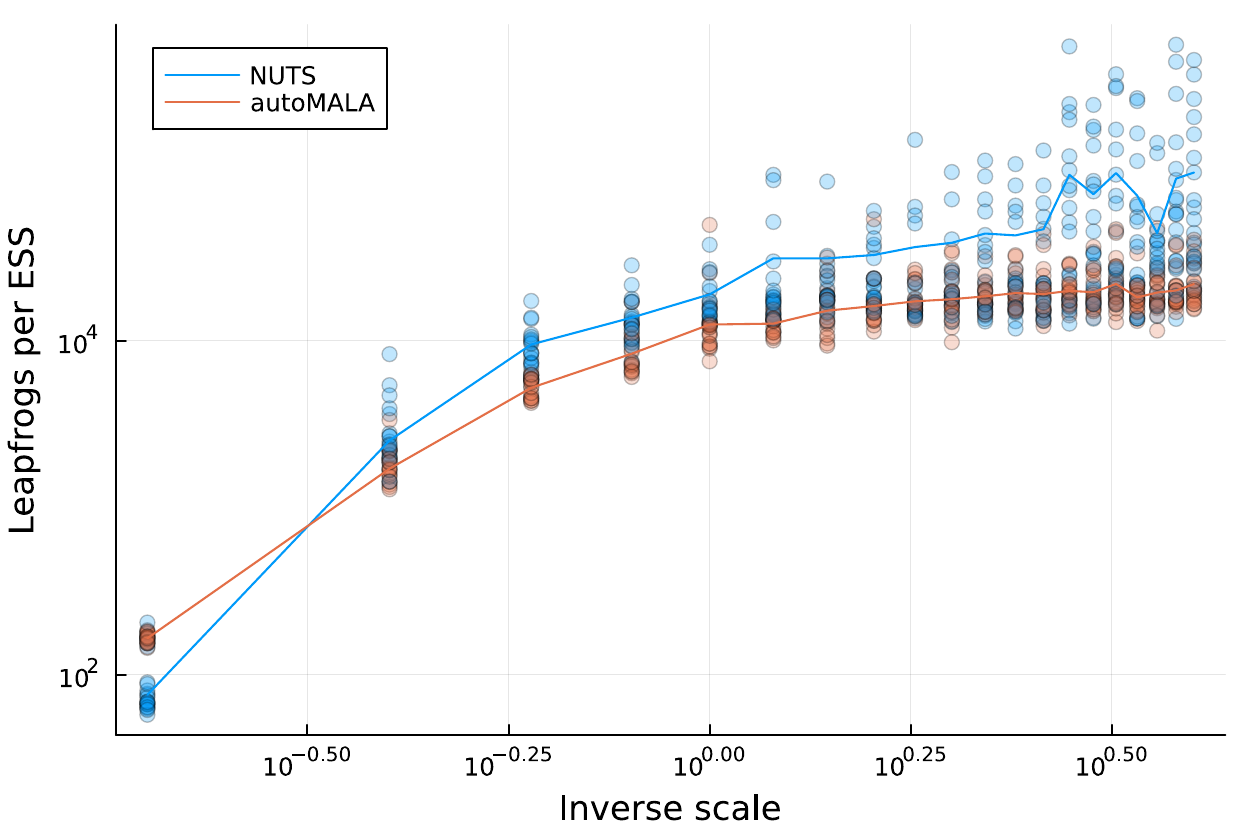} 
    \end{subfigure}
    \begin{subfigure}{0.32\textwidth}
        \centering 
        \includegraphics[width=\textwidth]{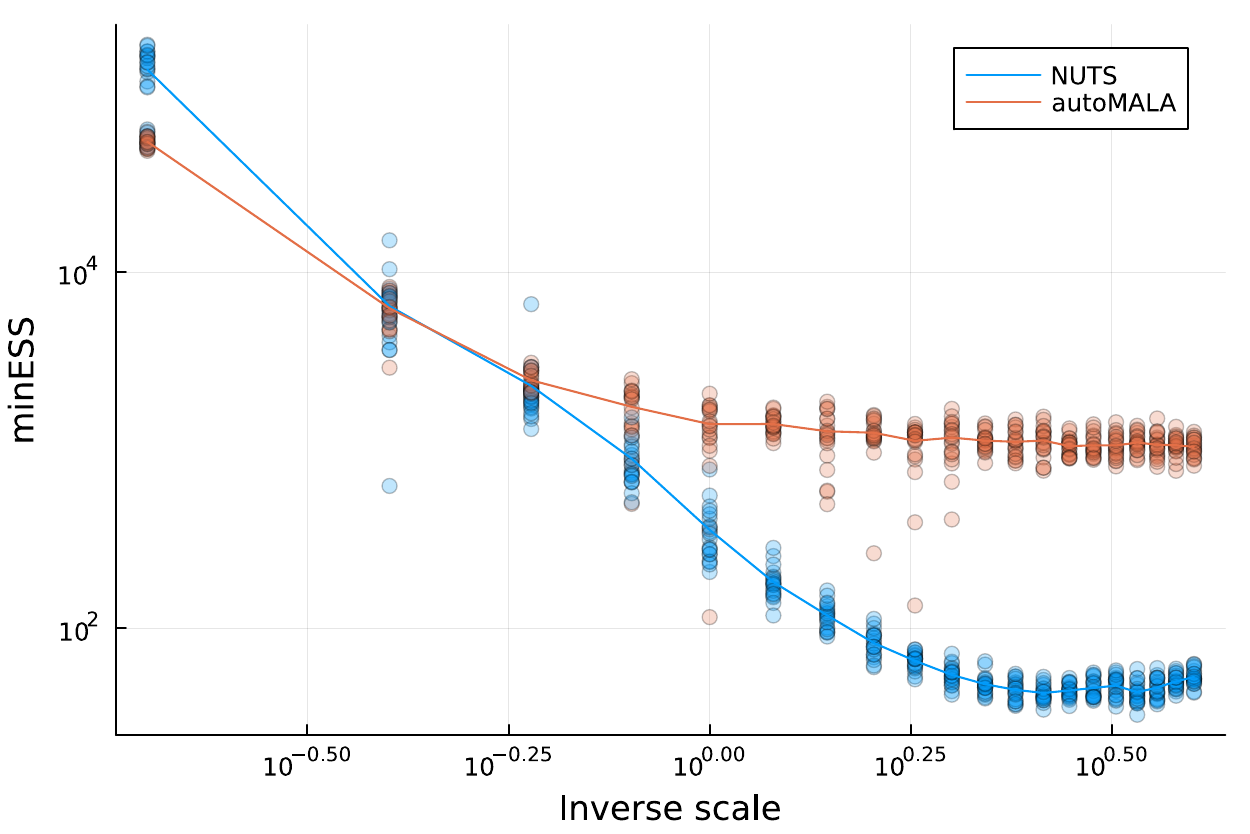}
    \end{subfigure}
    \begin{subfigure}{0.32\textwidth}
        \centering 
        \includegraphics[width=\textwidth]{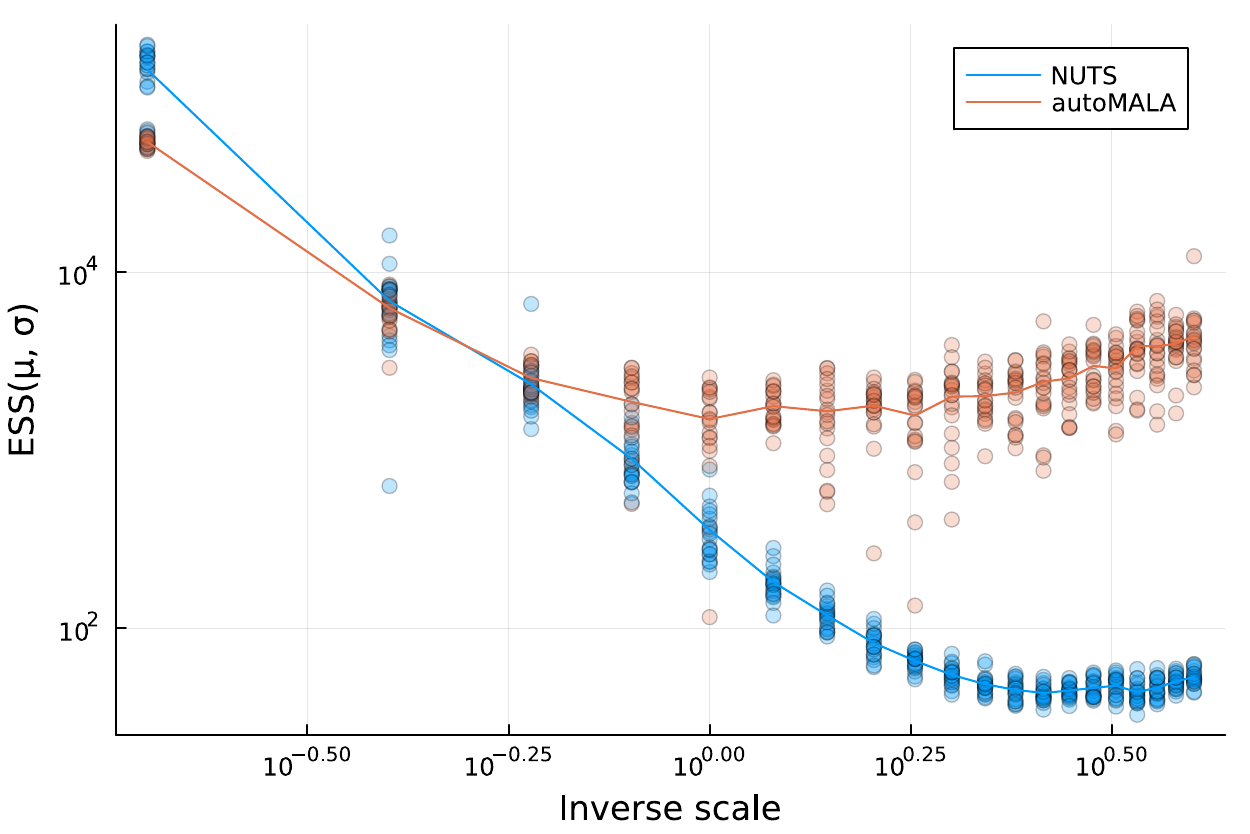}
    \end{subfigure}
    \begin{subfigure}{0.32\textwidth}
        \centering 
        \includegraphics[width=\textwidth]{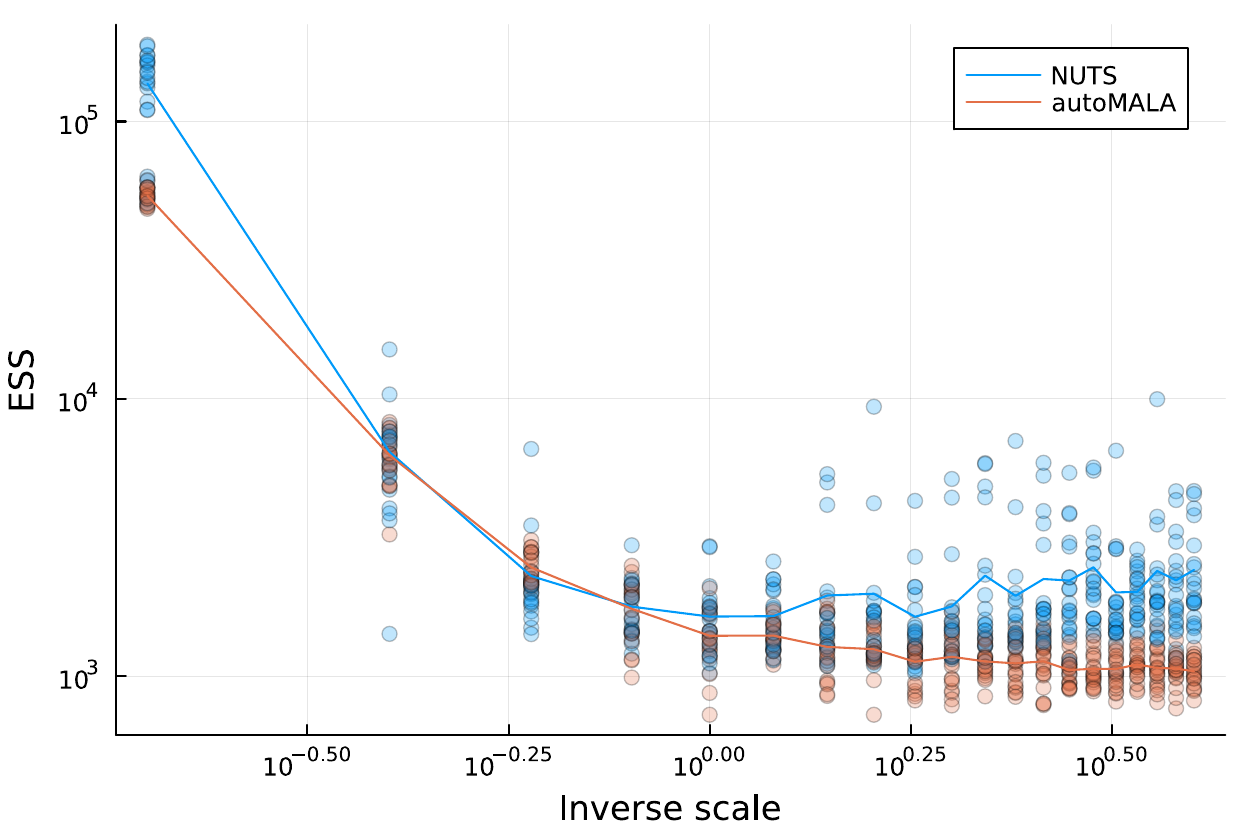}         
    \end{subfigure}
    \begin{subfigure}{0.32\textwidth}
        \centering 
        \includegraphics[width=\textwidth]{deliverables/AM_funnel_scale/scaling-margin1_mean-model-funnel_scale.pdf}
    \end{subfigure}
    \begin{subfigure}{0.32\textwidth}
        \centering 
        \includegraphics[width=\textwidth]{deliverables/AM_funnel_scale/scaling-margin1_var-model-funnel_scale.pdf}
    \end{subfigure}
    \begin{subfigure}{0.32\textwidth}
        \centering 
        \includegraphics[width=\textwidth]{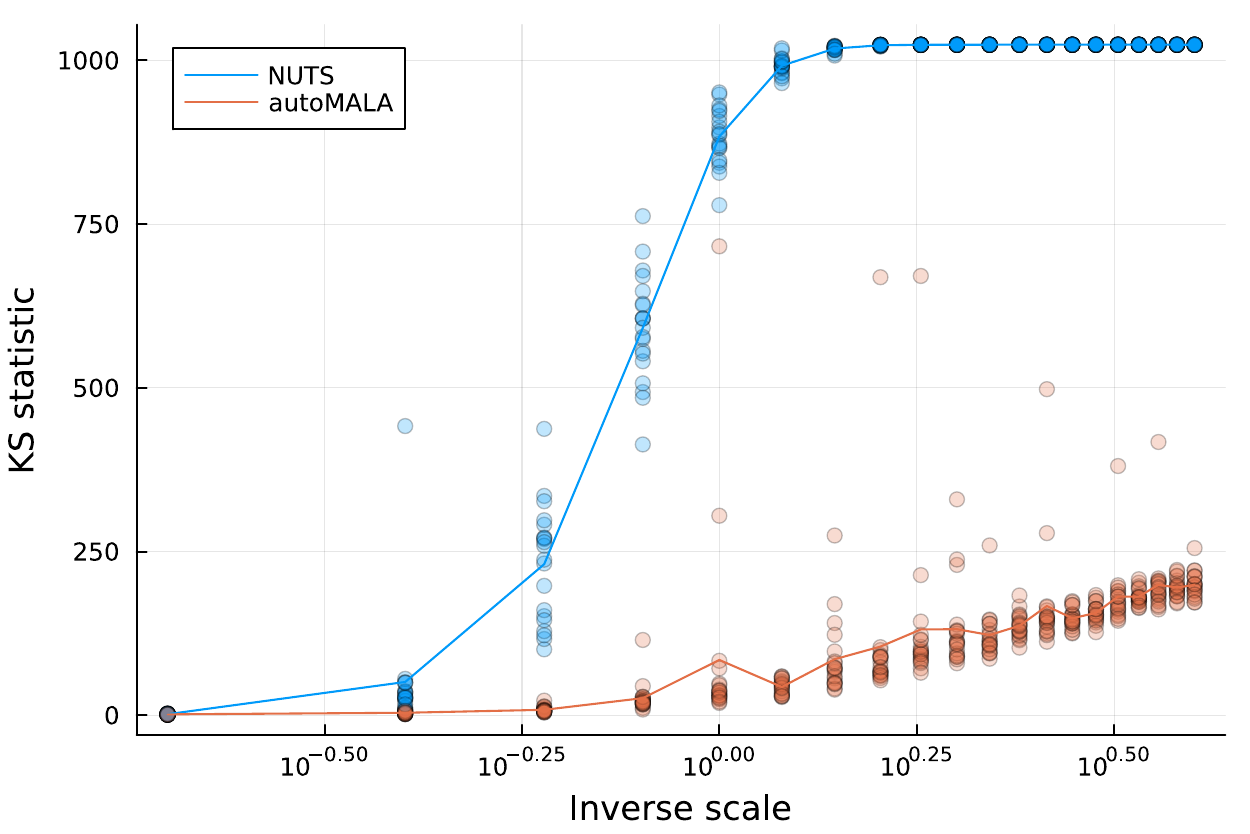}
    \end{subfigure}
    \caption{
        Various autoMALA and NUTS metrics for the scaled funnel experiments. 
        \textbf{Top row:} Number of leapfrog steps per 1000 effective samples. 
        From left to right we consider the minESS, $\exactess$, and regular ESS. 
        \textbf{Middle row:} minESS, $\exactess$, and regular ESS. 
        \textbf{Bottom row:} mean, variance, and Kolomogorov-Smirnov test statistic 
        for the known first marginal of the distribution. 
    }
    \label{fig:funnel_scale_all}
\end{figure*}

\begin{figure*}[!t]
  \centering
    \begin{subfigure}{0.32\textwidth}
        \centering 
        \includegraphics[width=\textwidth]{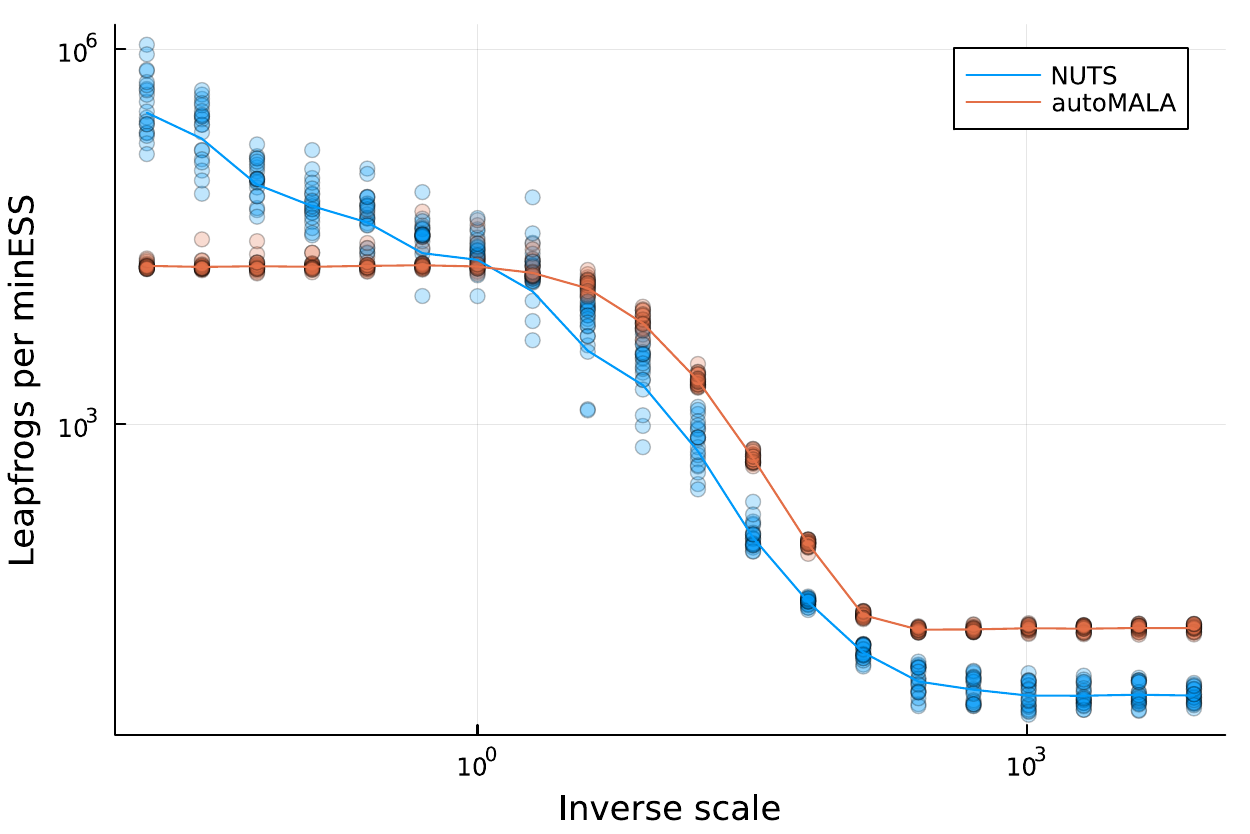}
    \end{subfigure}
    \begin{subfigure}{0.32\textwidth}
        \centering 
        \includegraphics[width=\textwidth]{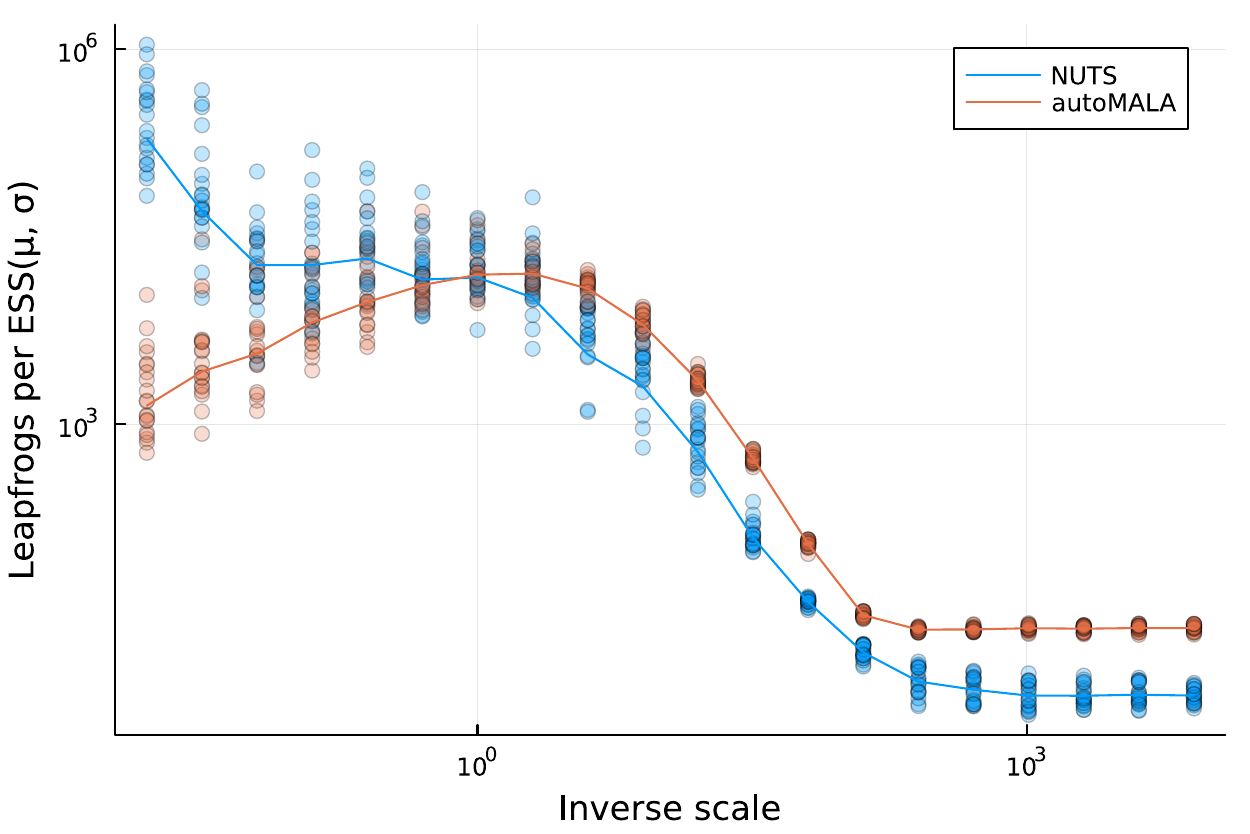}
    \end{subfigure}
    \begin{subfigure}{0.32\textwidth}
        \centering 
        \includegraphics[width=\textwidth]{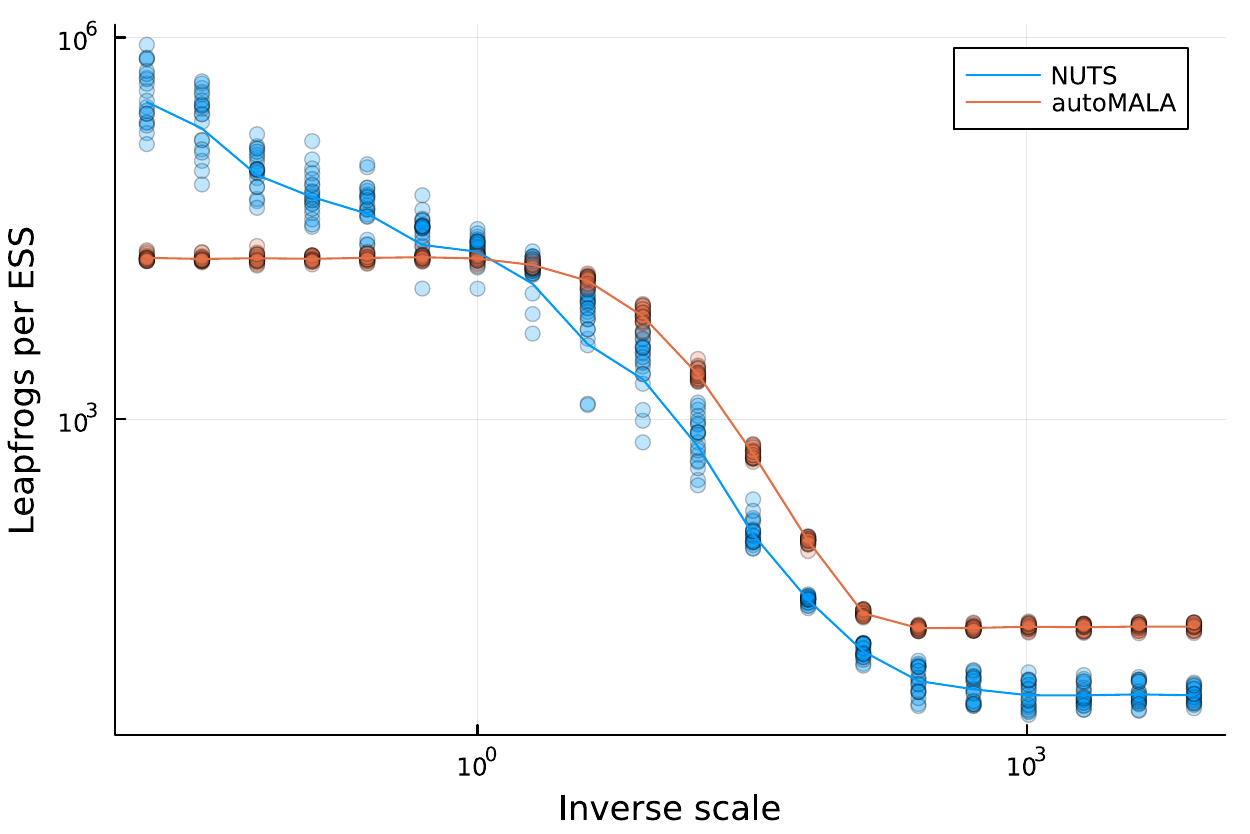} 
    \end{subfigure}
    \begin{subfigure}{0.32\textwidth}
        \centering 
        \includegraphics[width=\textwidth]{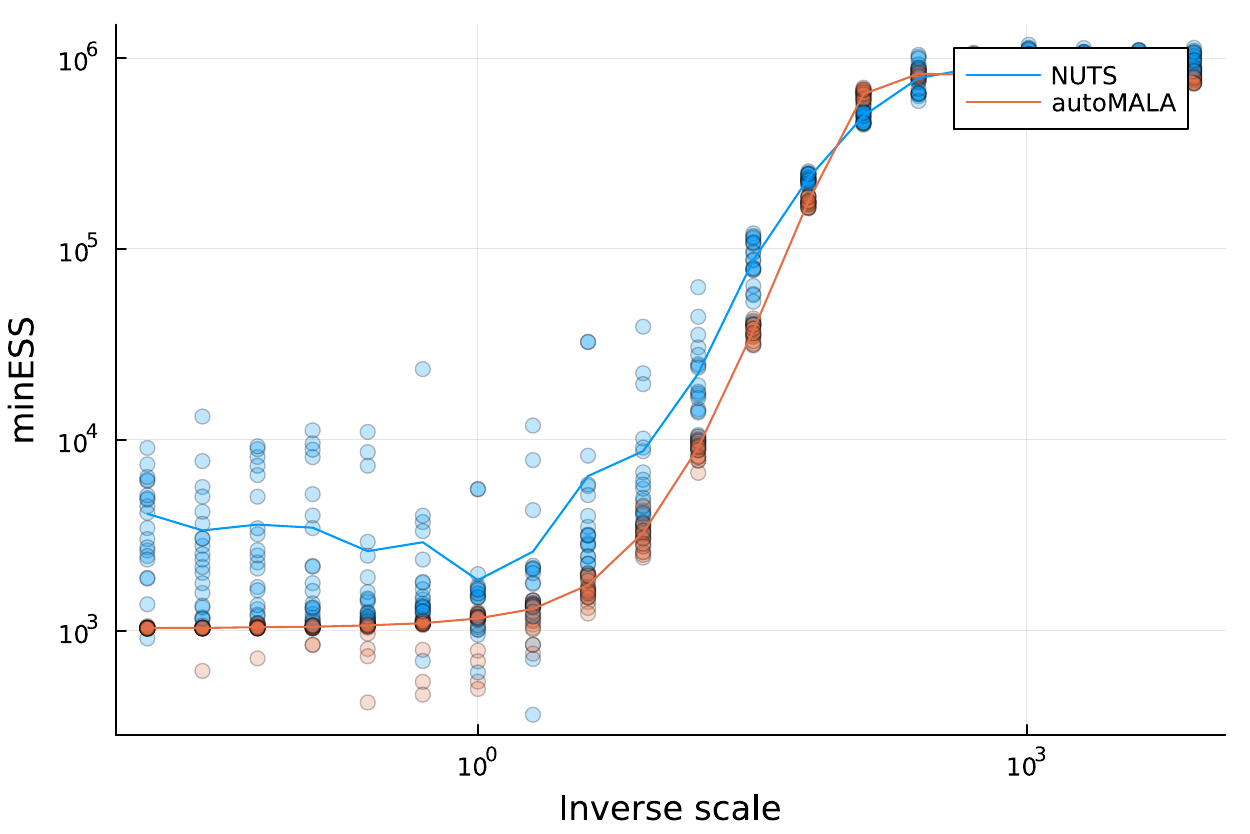}
    \end{subfigure}
    \begin{subfigure}{0.32\textwidth}
        \centering 
        \includegraphics[width=\textwidth]{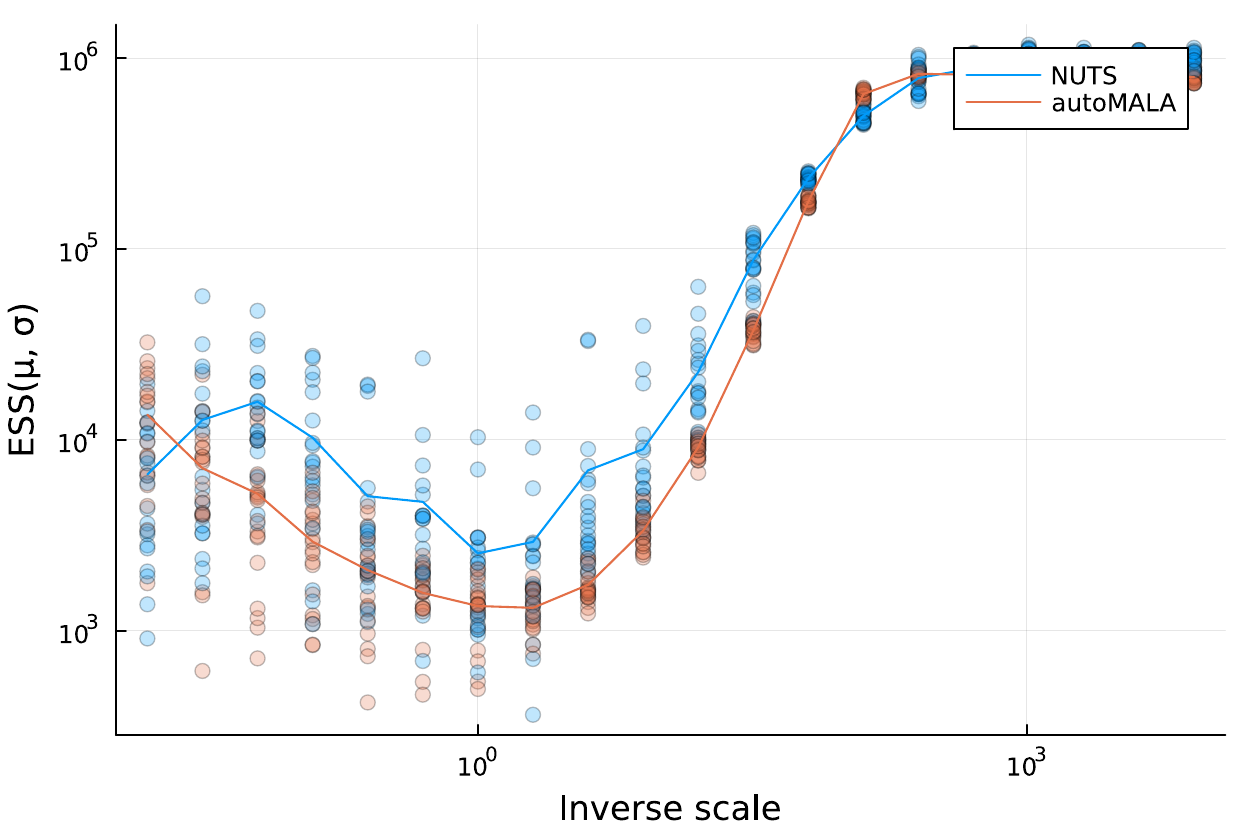}
    \end{subfigure}
    \begin{subfigure}{0.32\textwidth}
        \centering 
        \includegraphics[width=\textwidth]{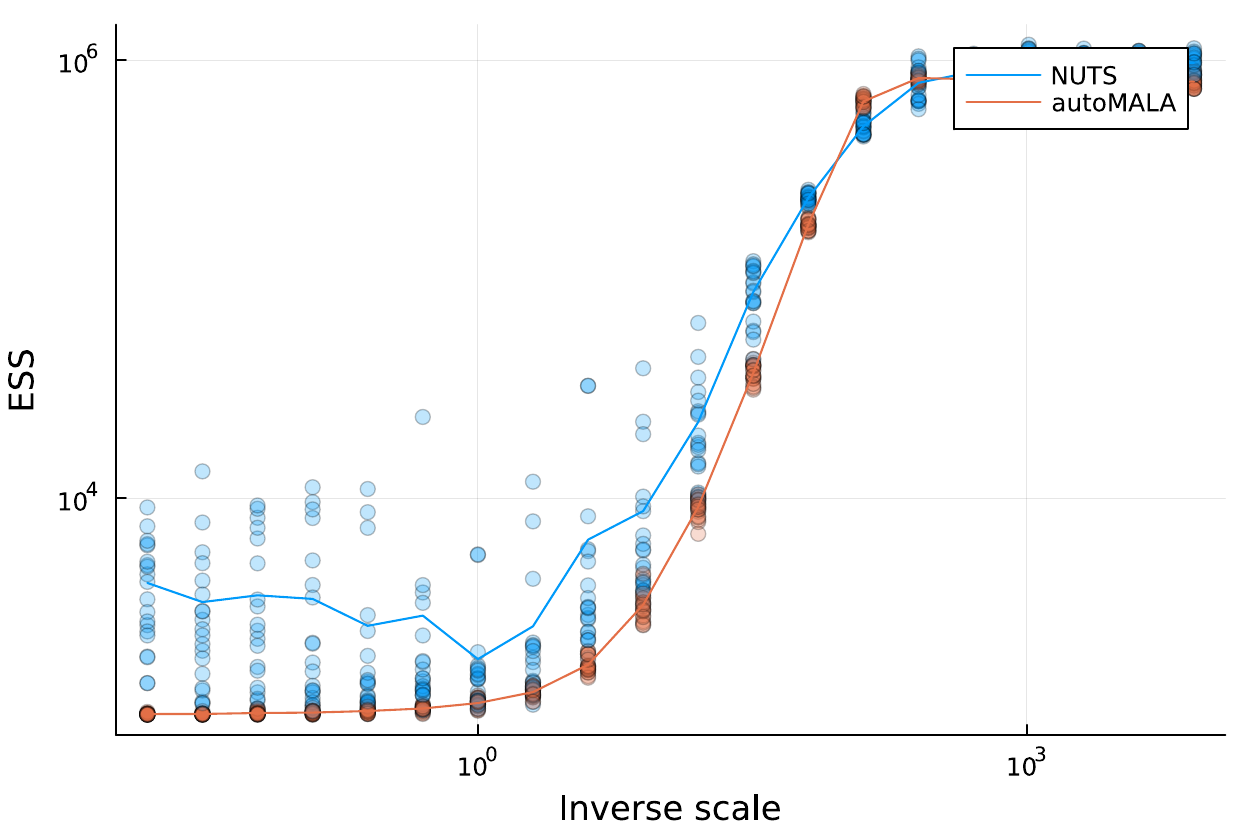}         
    \end{subfigure}
    \begin{subfigure}{0.32\textwidth}
        \centering 
        \includegraphics[width=\textwidth]{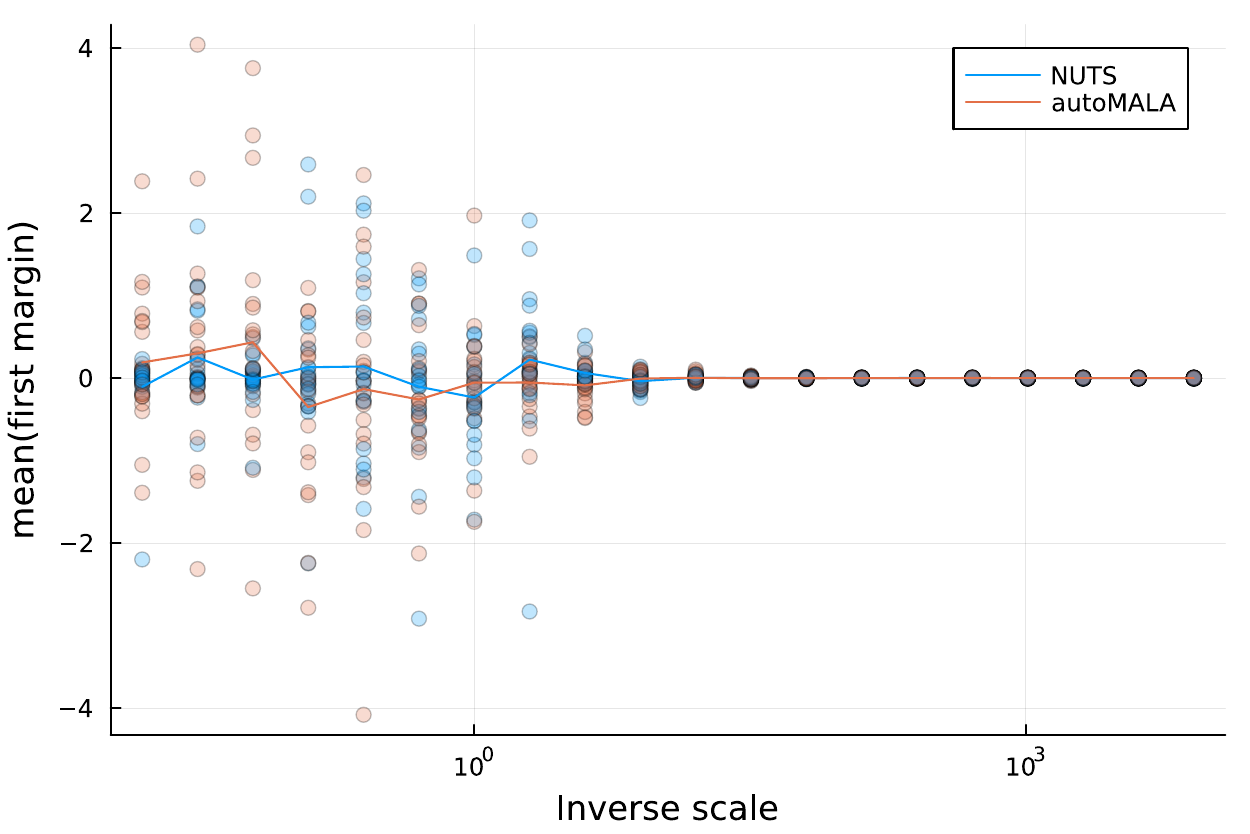}
    \end{subfigure}
    \begin{subfigure}{0.32\textwidth}
        \centering 
        \includegraphics[width=\textwidth]{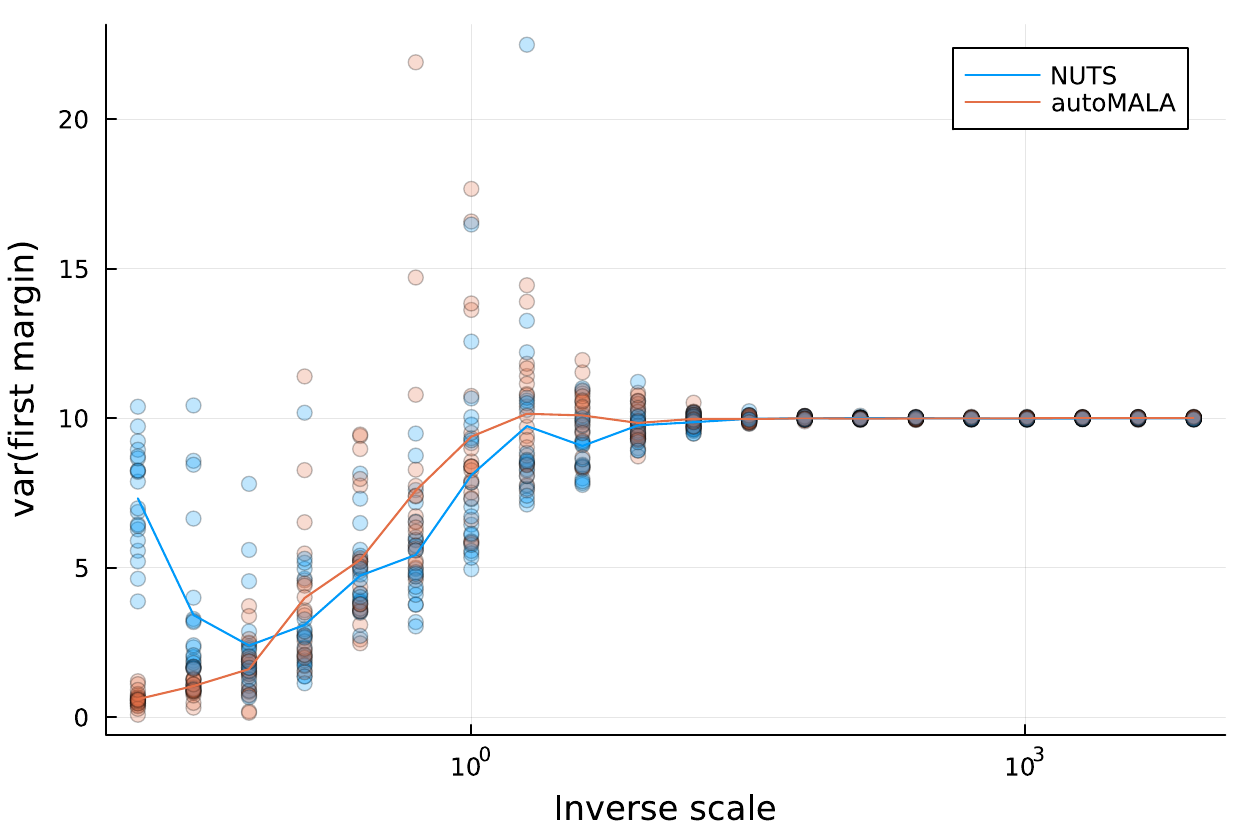}
    \end{subfigure}
    \begin{subfigure}{0.32\textwidth}
        \centering 
        \includegraphics[width=\textwidth]{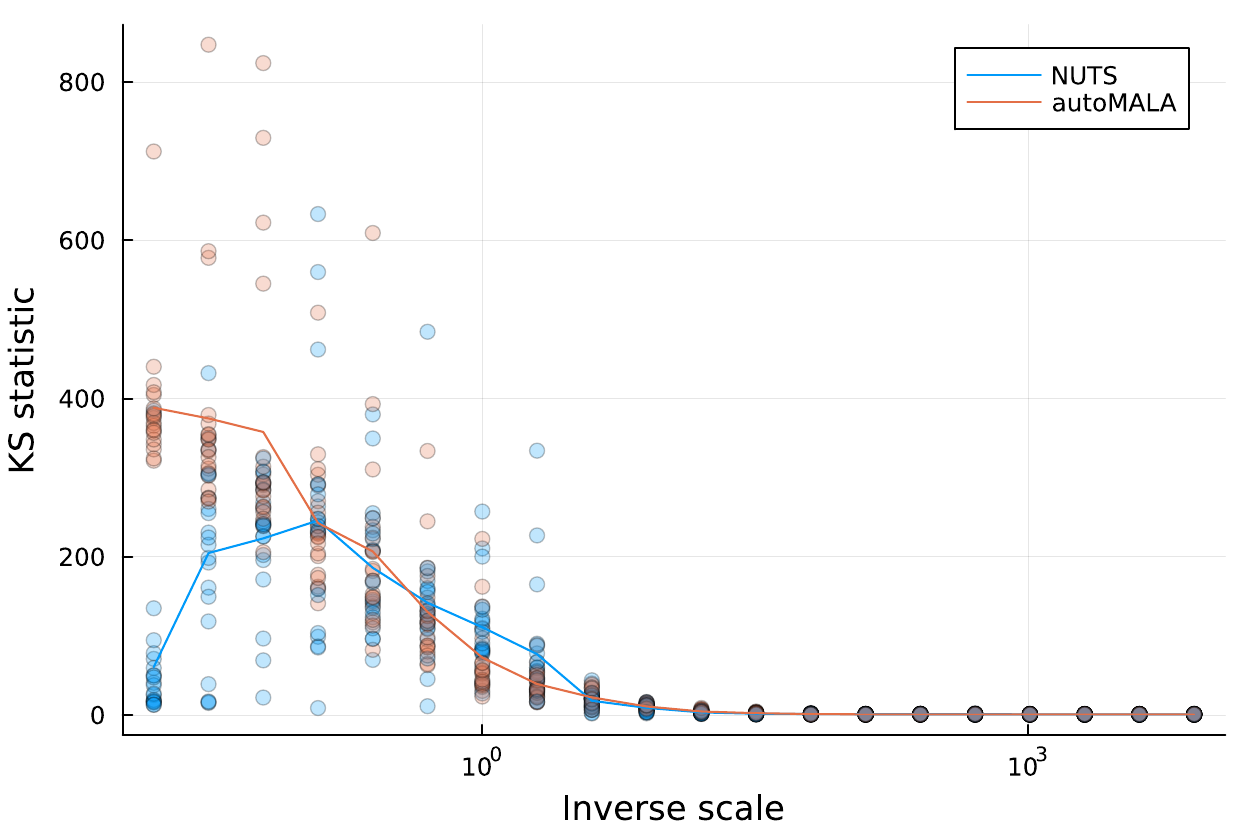}
    \end{subfigure}
    \caption{
        Various autoMALA and NUTS metrics for the scaled banana experiments. 
        \textbf{Top row:} Number of leapfrog steps per 1000 effective samples. 
        From left to right we consider the minESS, $\exactess$, and regular ESS. 
        \textbf{Middle row:} minESS, $\exactess$, and regular ESS. 
        \textbf{Bottom row:} mean, variance, and Kolomogorov-Smirnov test statistic 
        for the known first marginal of the distribution. 
    }
    \label{fig:banana_scale_all}
\end{figure*}

\begin{figure*}[!t]
    \centering 
    \includegraphics[width=0.7\textwidth]{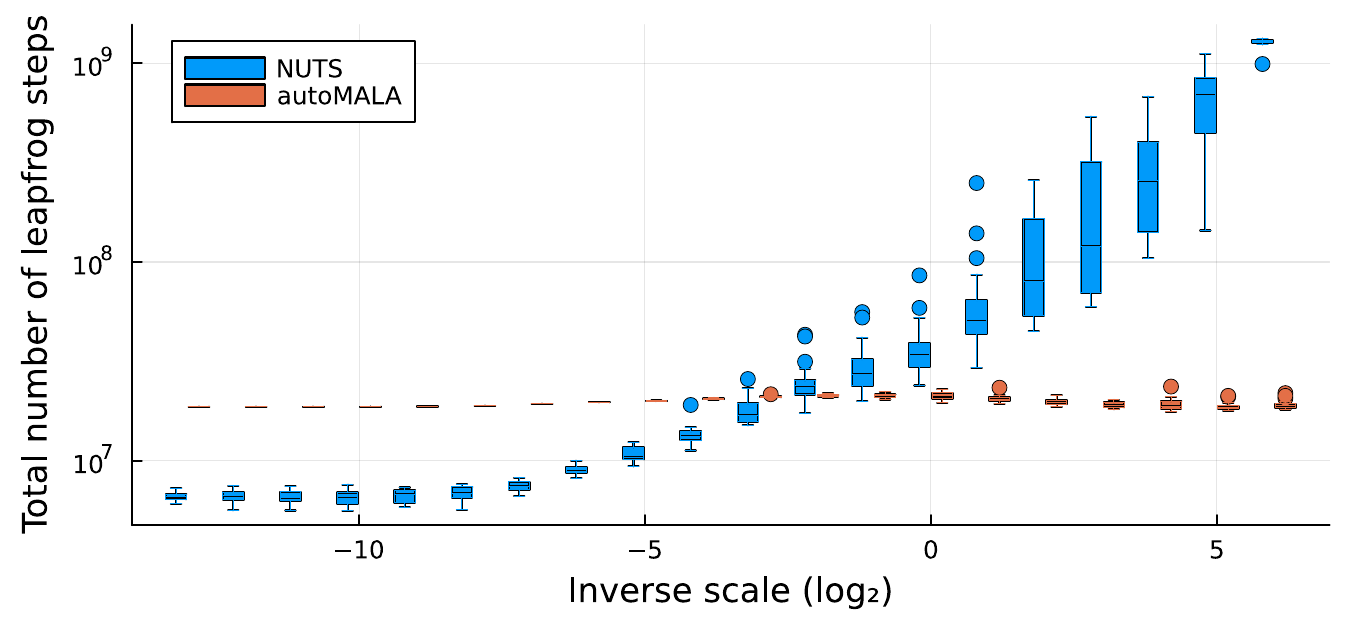}
    \caption{
       Number of leapfrog steps used by autoMALA and NUTS for each scale parameter
       of the banana distribution. 
    }
    \label{fig:banana_scale_leapfrogs}
\end{figure*}

\subsection{Dimensional scaling}
\begin{figure*}[!t]
  \centering
    \begin{subfigure}{0.32\textwidth}
        \centering 
        \includegraphics[width=\textwidth]{deliverables/AM_funnel_highdim/scaling-leapfrog_min-model-funnel.pdf}
    \end{subfigure}
    \begin{subfigure}{0.32\textwidth}
        \centering 
        \includegraphics[width=\textwidth]{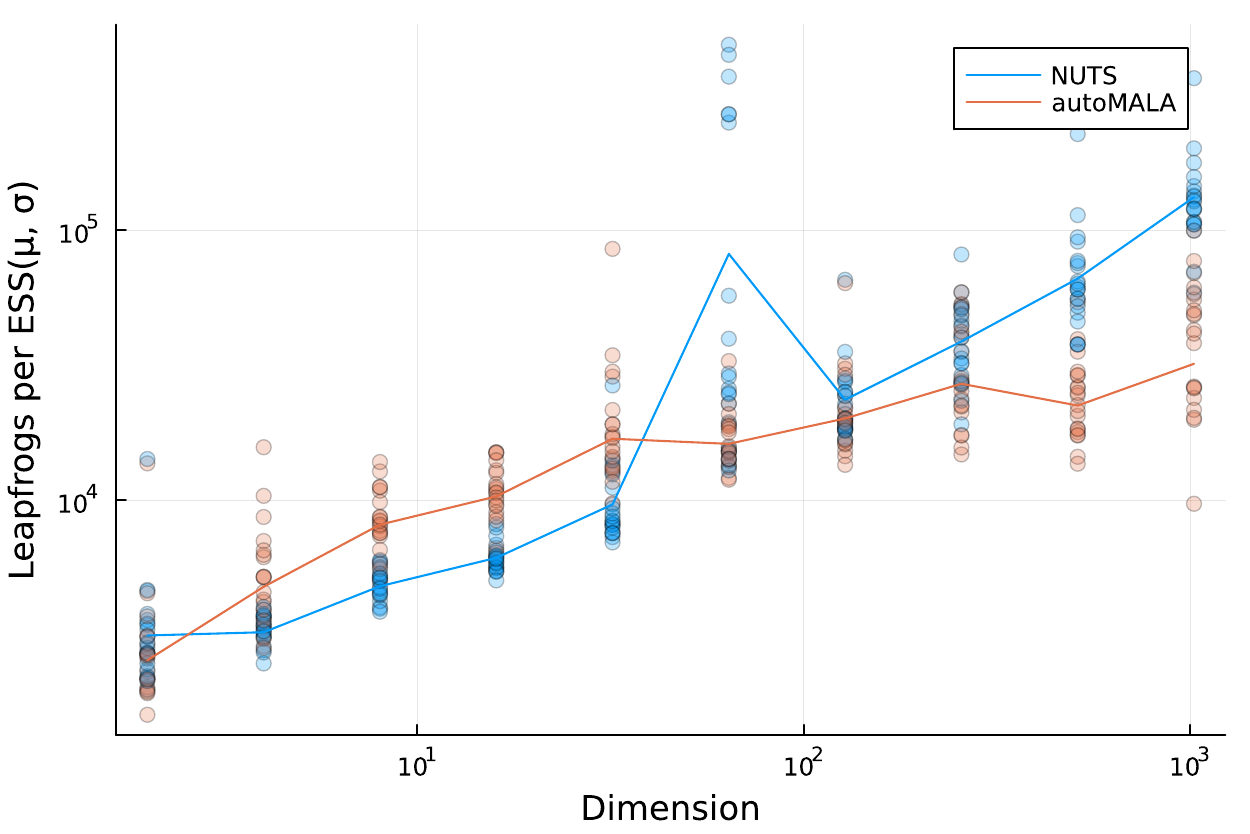}
    \end{subfigure}
    \begin{subfigure}{0.32\textwidth}
        \centering 
        \includegraphics[width=\textwidth]{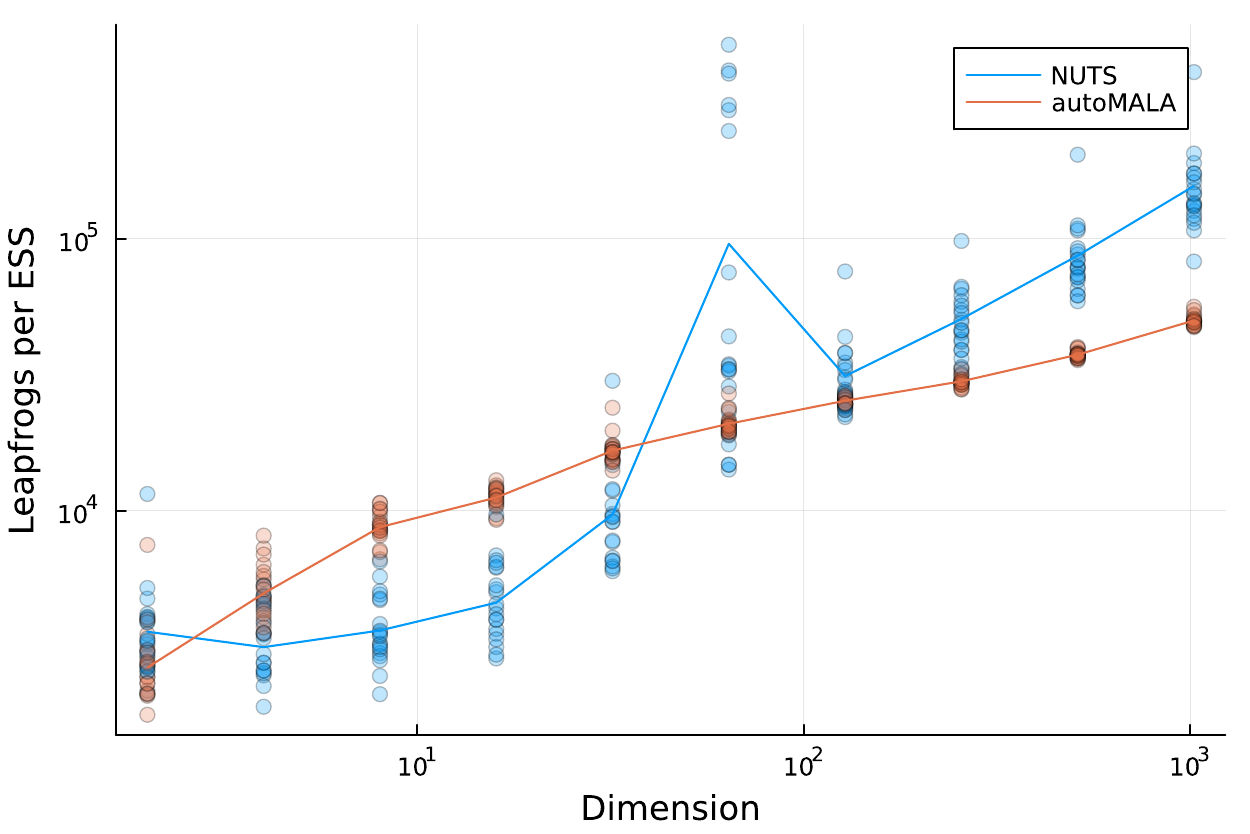} 
    \end{subfigure}
    \begin{subfigure}{0.32\textwidth}
        \centering 
        \includegraphics[width=\textwidth]{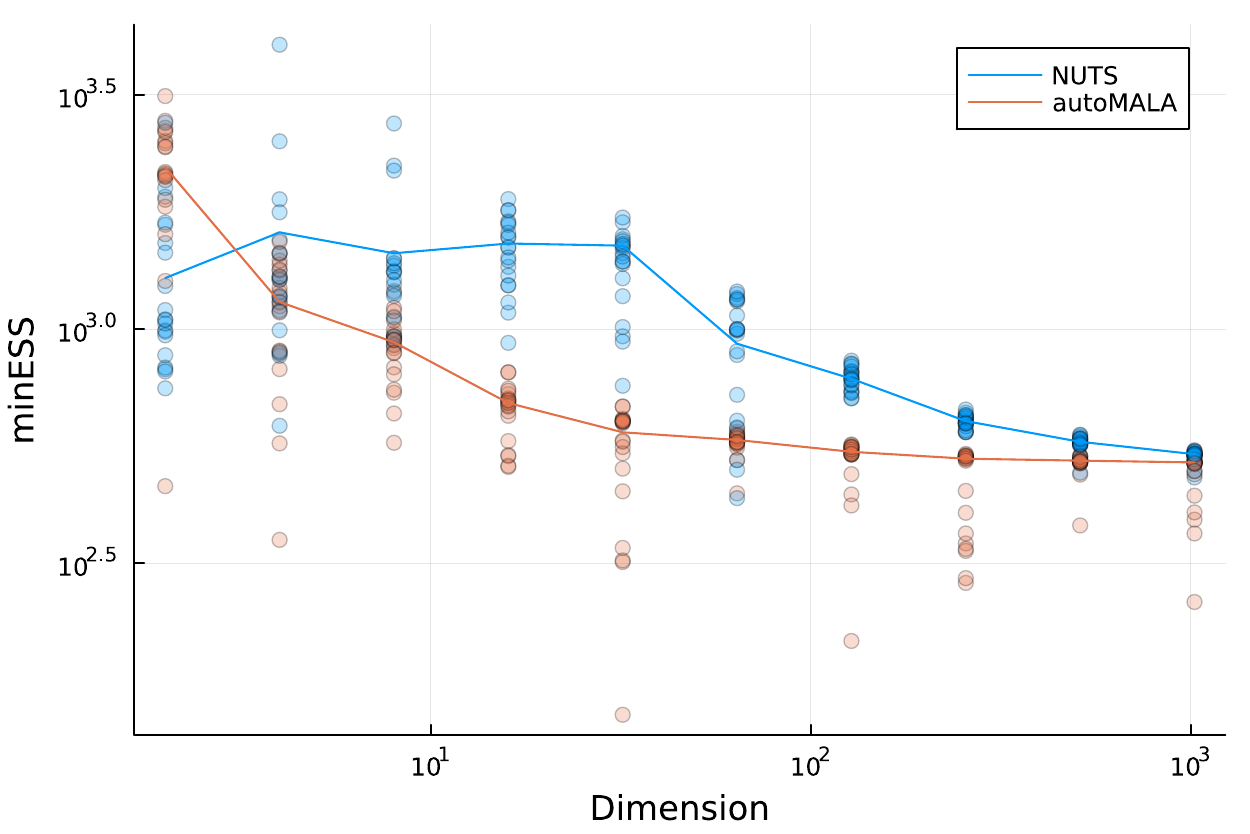}
    \end{subfigure}
    \begin{subfigure}{0.32\textwidth}
        \centering 
        \includegraphics[width=\textwidth]{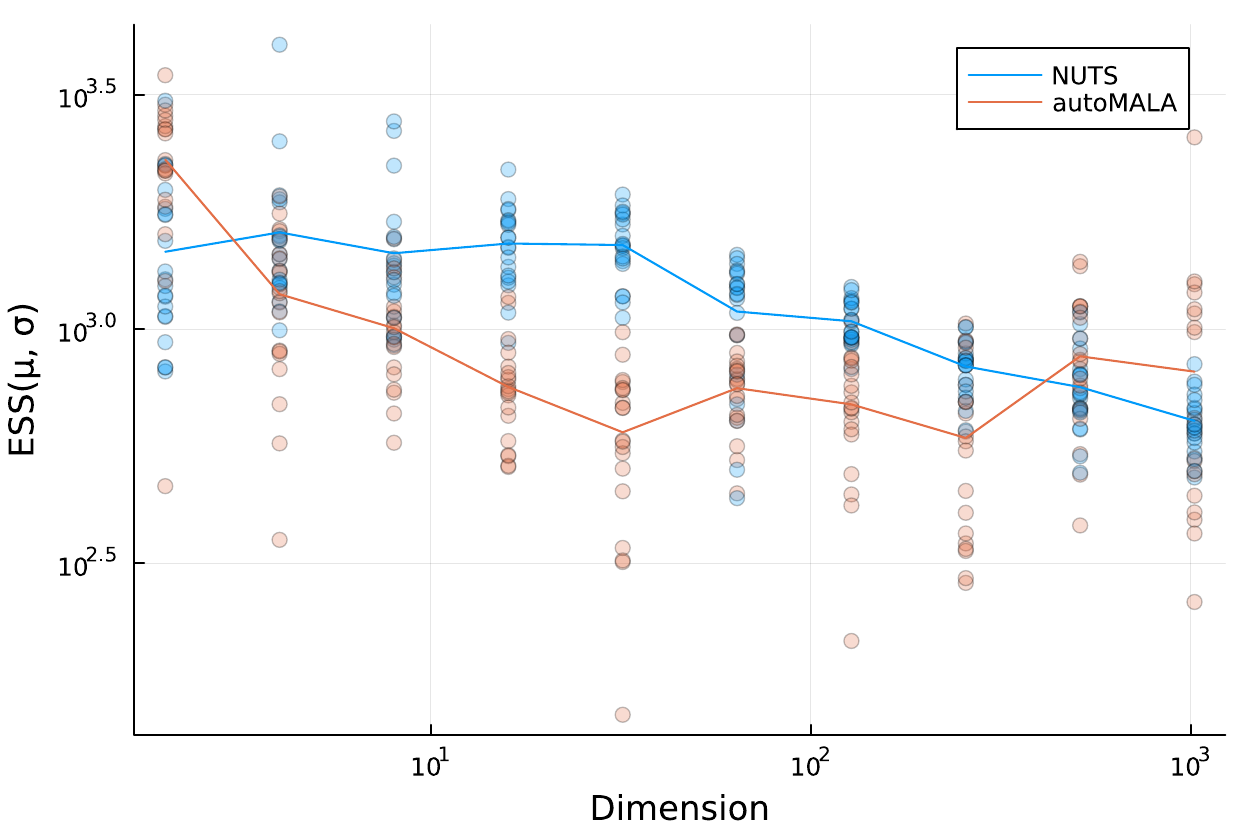}
    \end{subfigure}
    \begin{subfigure}{0.32\textwidth}
        \centering 
        \includegraphics[width=\textwidth]{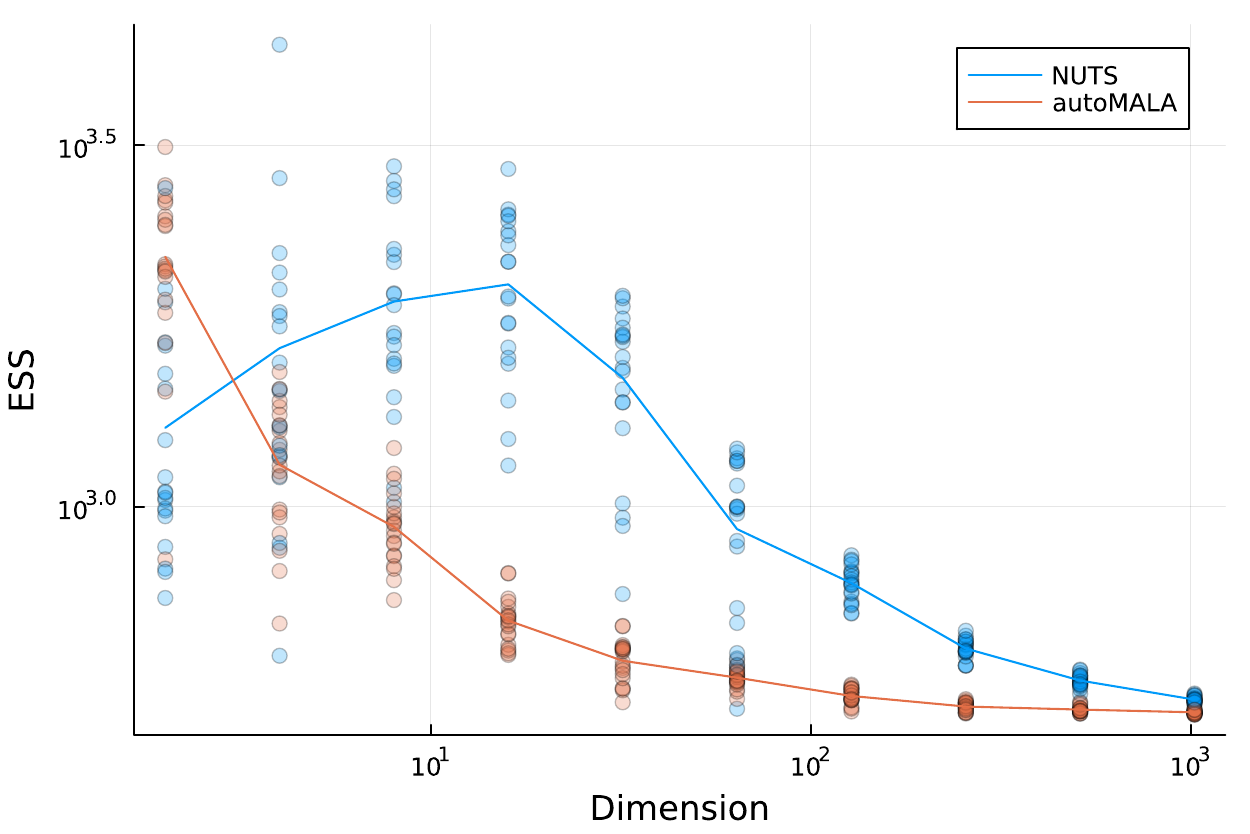}         
    \end{subfigure}
    \begin{subfigure}{0.32\textwidth}
        \centering 
        \includegraphics[width=\textwidth]{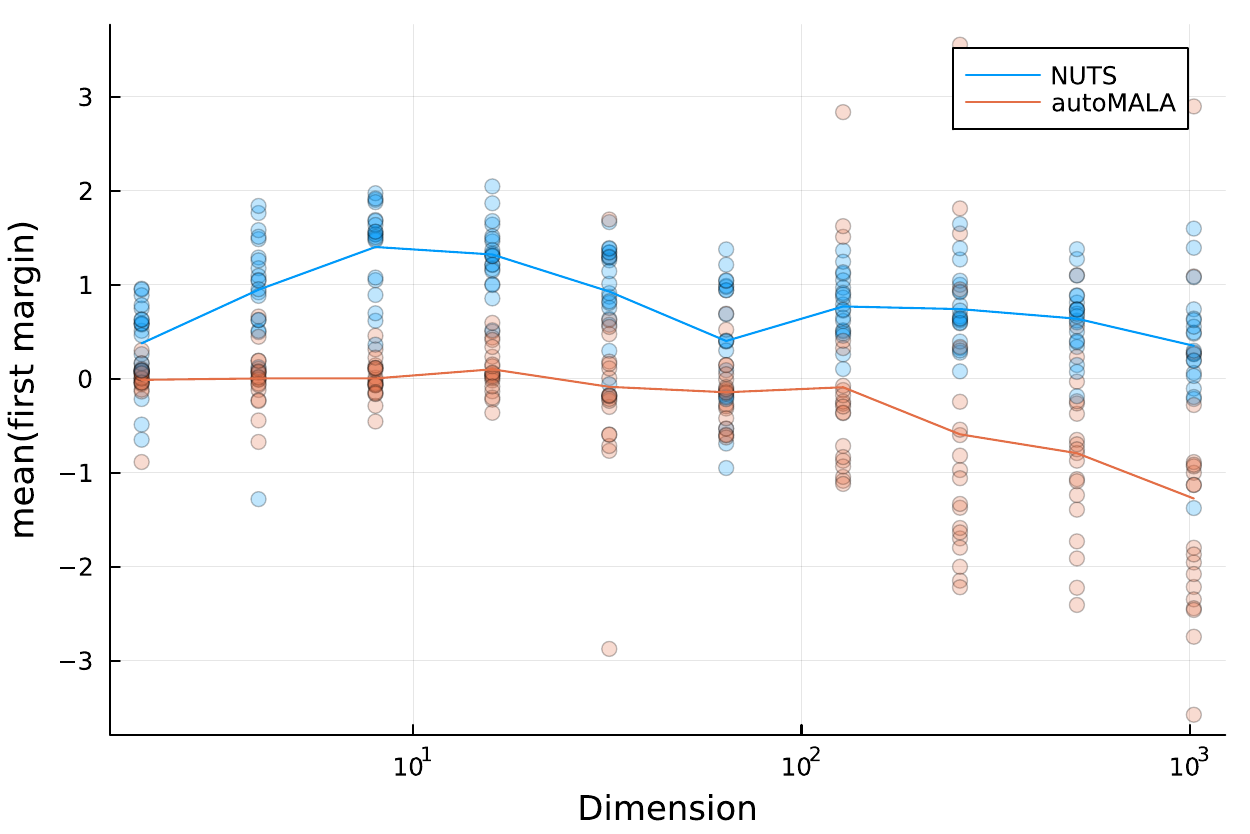}
    \end{subfigure}
    \begin{subfigure}{0.32\textwidth}
        \centering 
        \includegraphics[width=\textwidth]{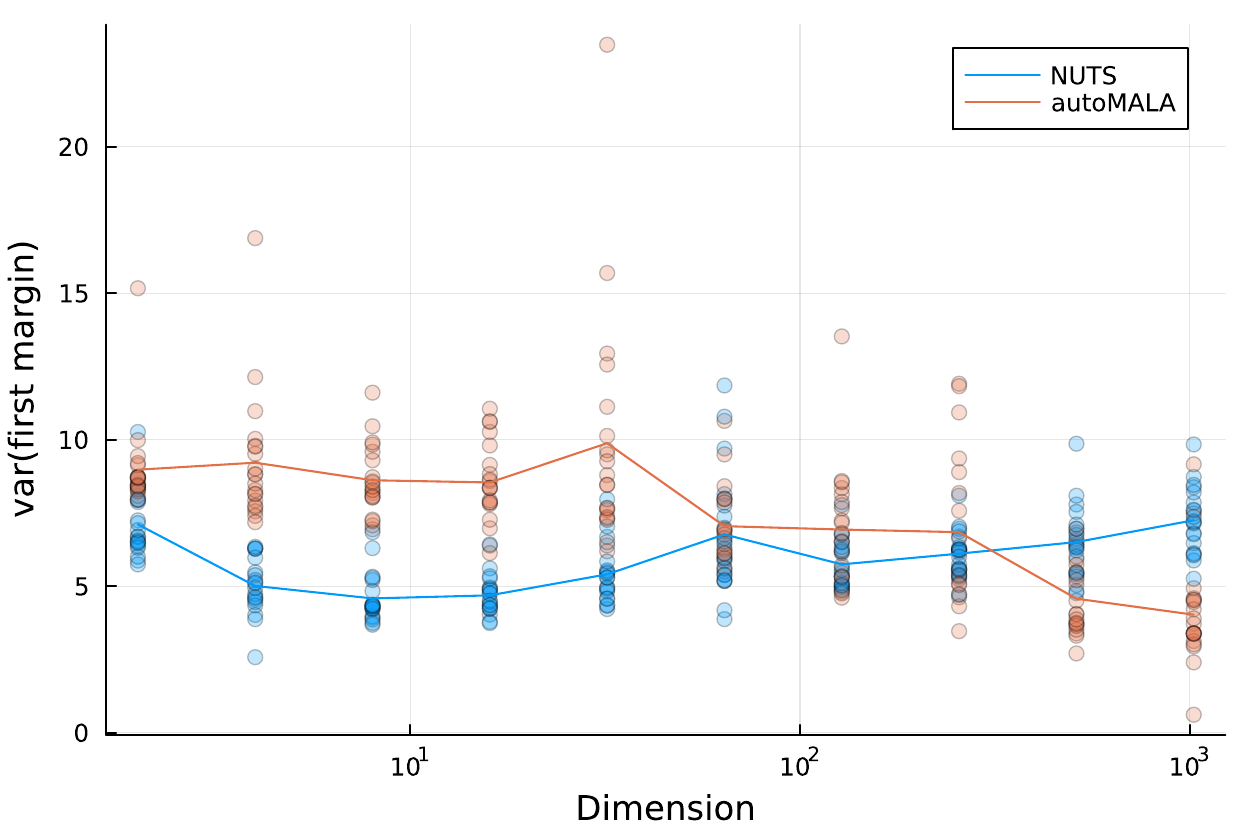}
    \end{subfigure}
    \begin{subfigure}{0.32\textwidth}
        \centering 
        \includegraphics[width=\textwidth]{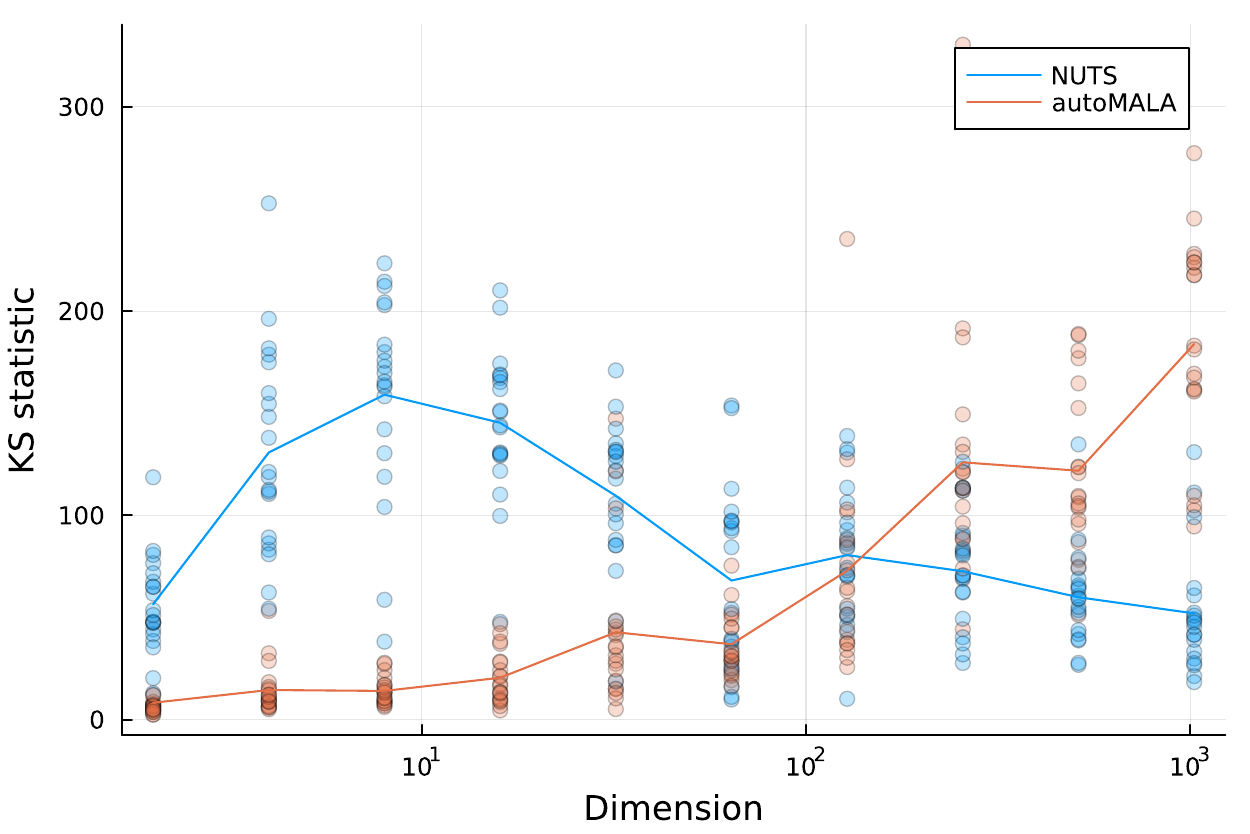}
    \end{subfigure}
    \caption{
        Various autoMALA and NUTS metrics for the high-dimensional funnel experiments. 
        \textbf{Top row:} Number of leapfrog steps per 1000 effective samples. 
        From left to right we consider the minESS, $\exactess$, and regular ESS. 
        \textbf{Middle row:} minESS, $\exactess$, and regular ESS. 
        \textbf{Bottom row:} mean, variance, and Kolomogorov-Smirnov test statistic 
        for the known first marginal of the distribution. 
    }
    \label{fig:funnel_highdim_all}
\end{figure*}

\begin{figure*}[!t]
  \centering
    \begin{subfigure}{0.32\textwidth}
        \centering 
        \includegraphics[width=\textwidth]{deliverables/AM_banana_highdim/scaling-leapfrog_min-model-banana.pdf}
    \end{subfigure}
    \begin{subfigure}{0.32\textwidth}
        \centering 
        \includegraphics[width=\textwidth]{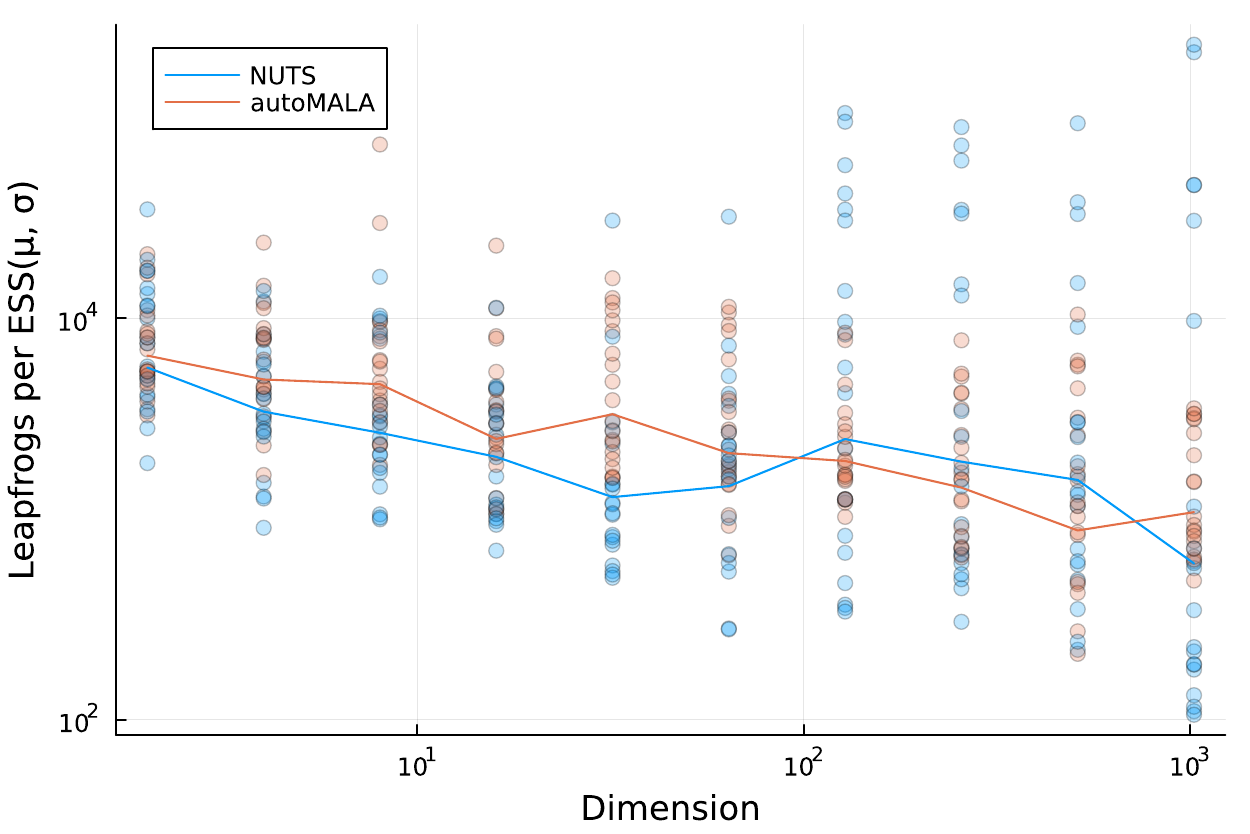}
    \end{subfigure}
    \begin{subfigure}{0.32\textwidth}
        \centering 
        \includegraphics[width=\textwidth]{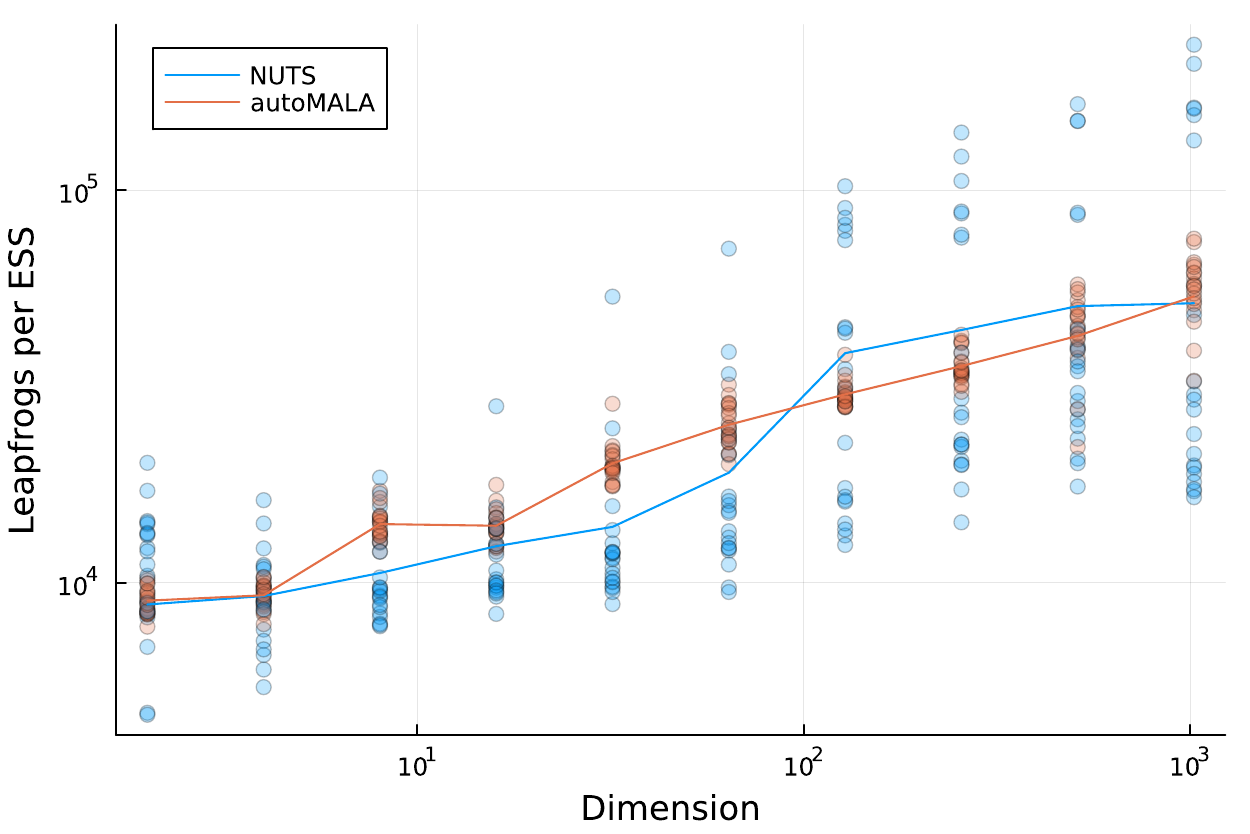} 
    \end{subfigure}
    \begin{subfigure}{0.32\textwidth}
        \centering 
        \includegraphics[width=\textwidth]{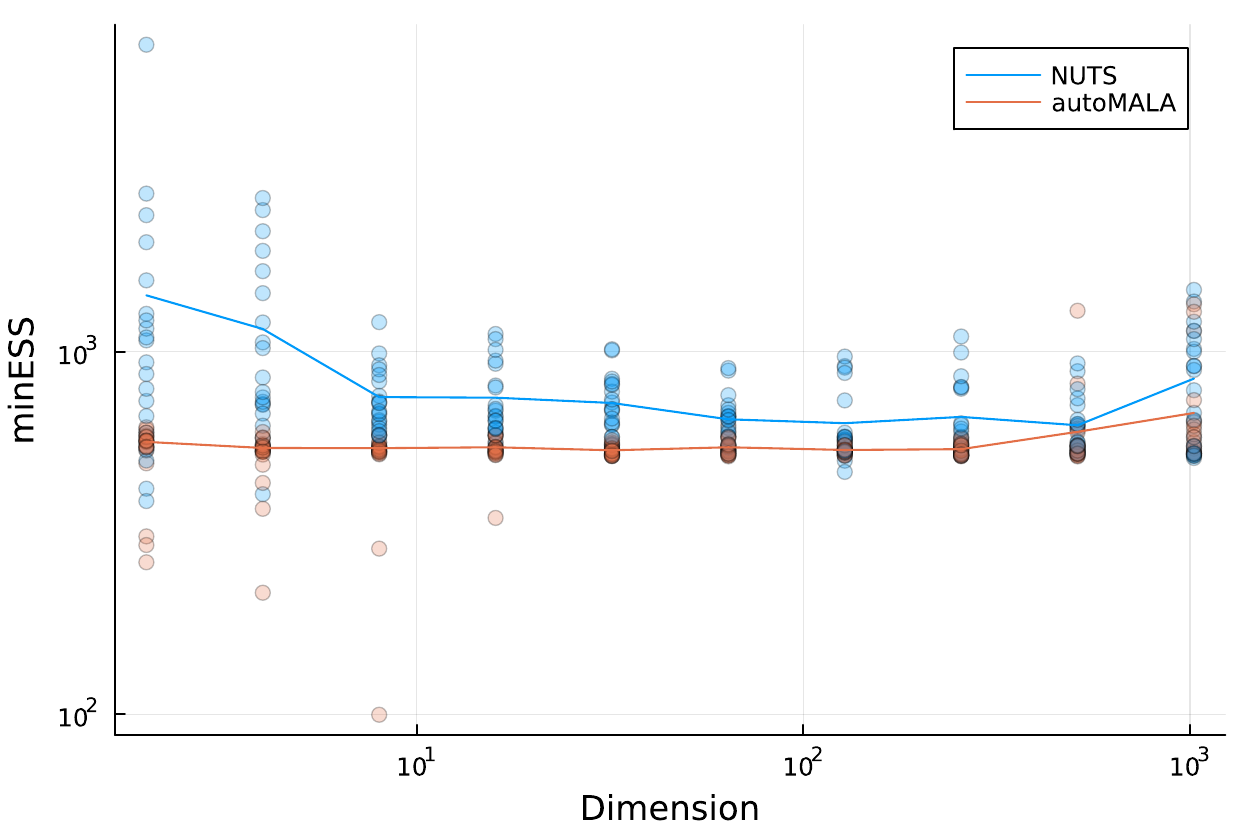}
    \end{subfigure}
    \begin{subfigure}{0.32\textwidth}
        \centering 
        \includegraphics[width=\textwidth]{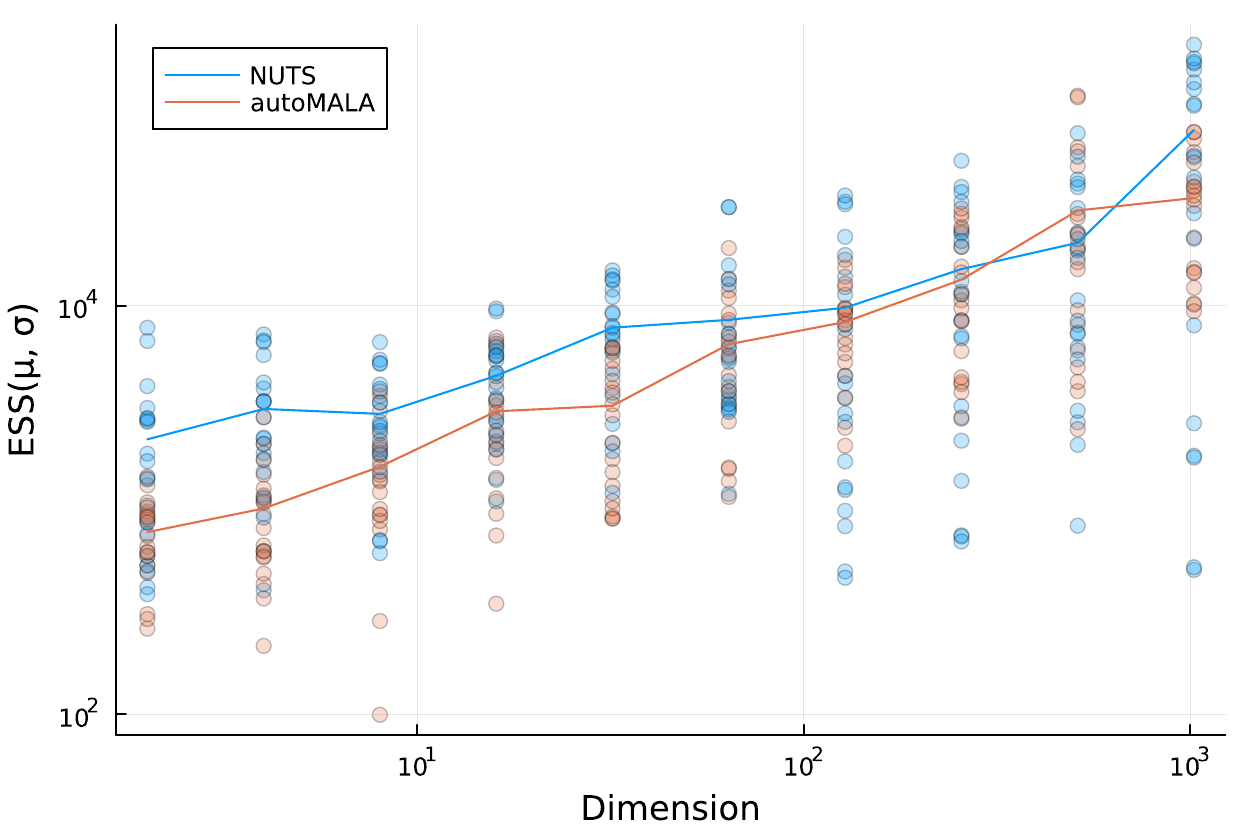}
    \end{subfigure}
    \begin{subfigure}{0.32\textwidth}
        \centering 
        \includegraphics[width=\textwidth]{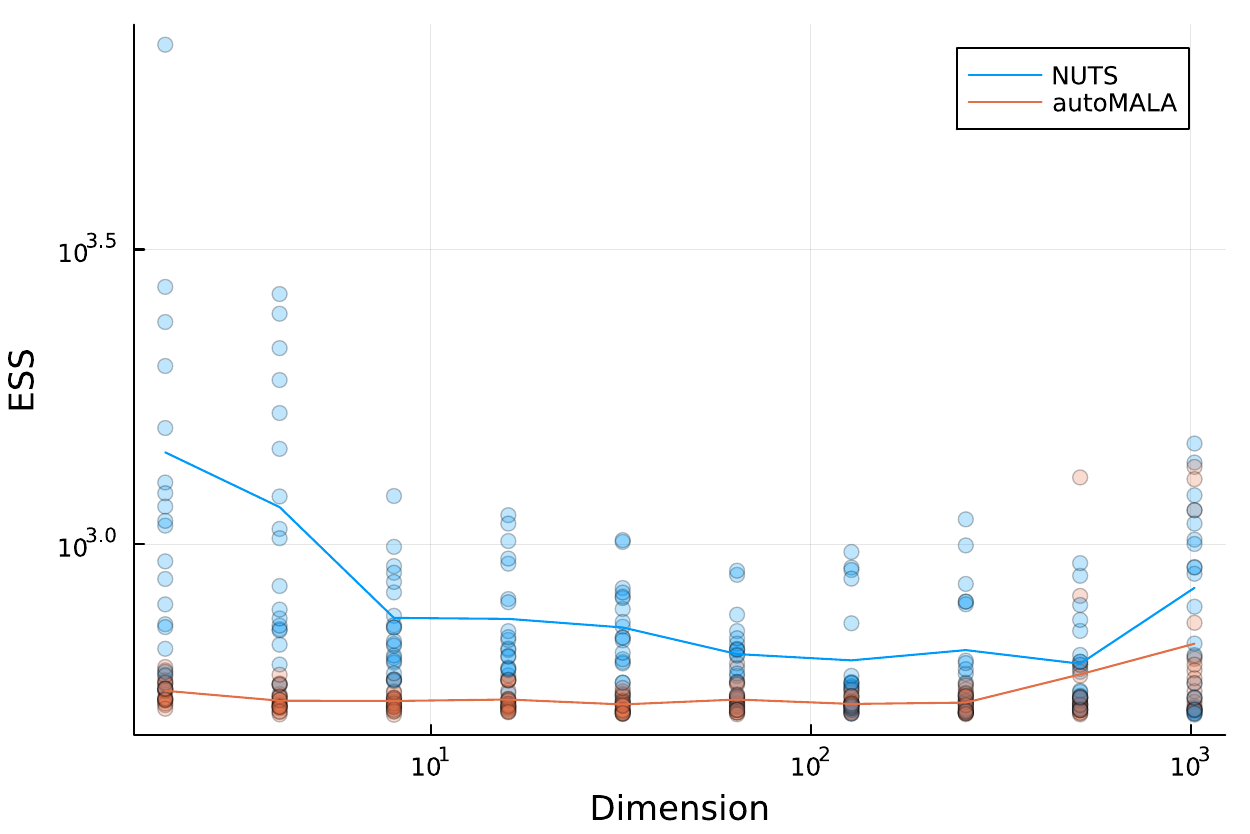}         
    \end{subfigure}
    \begin{subfigure}{0.32\textwidth}
        \centering 
        \includegraphics[width=\textwidth]{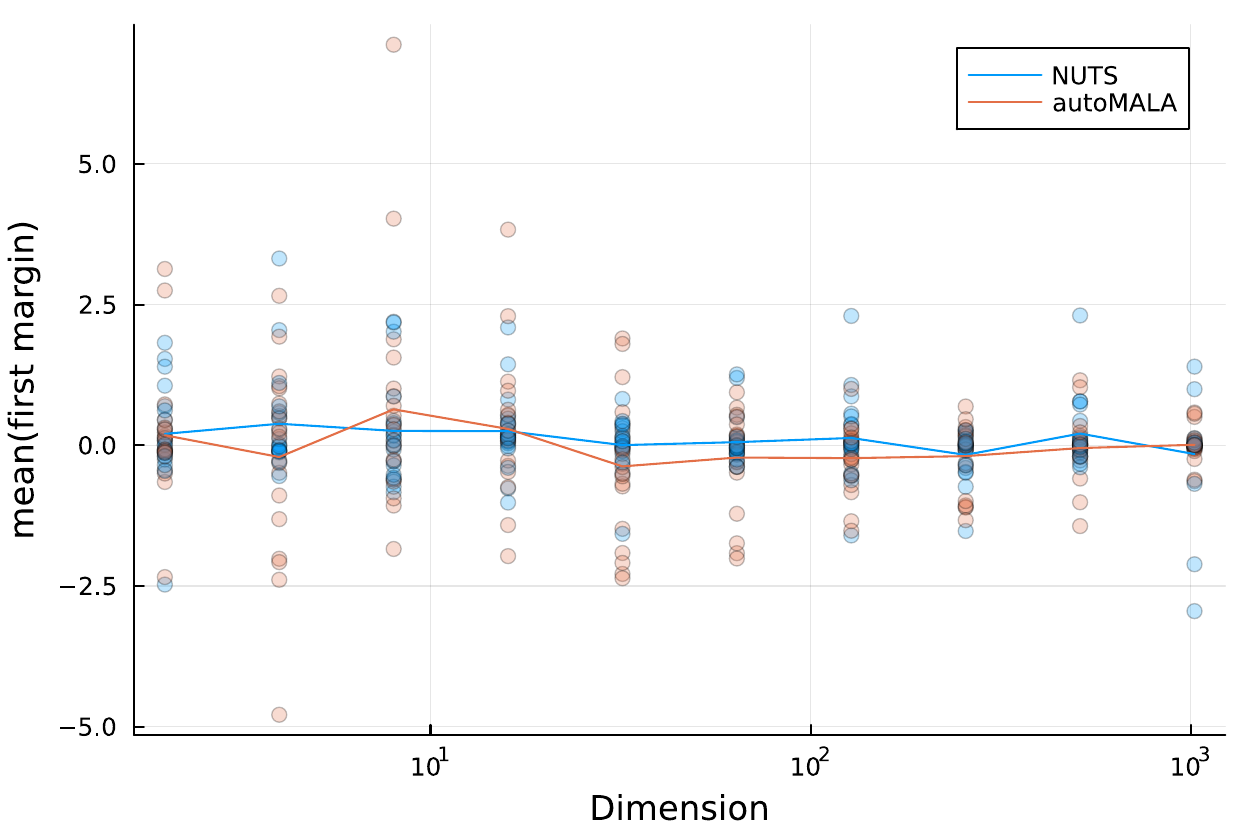}
    \end{subfigure}
    \begin{subfigure}{0.32\textwidth}
        \centering 
        \includegraphics[width=\textwidth]{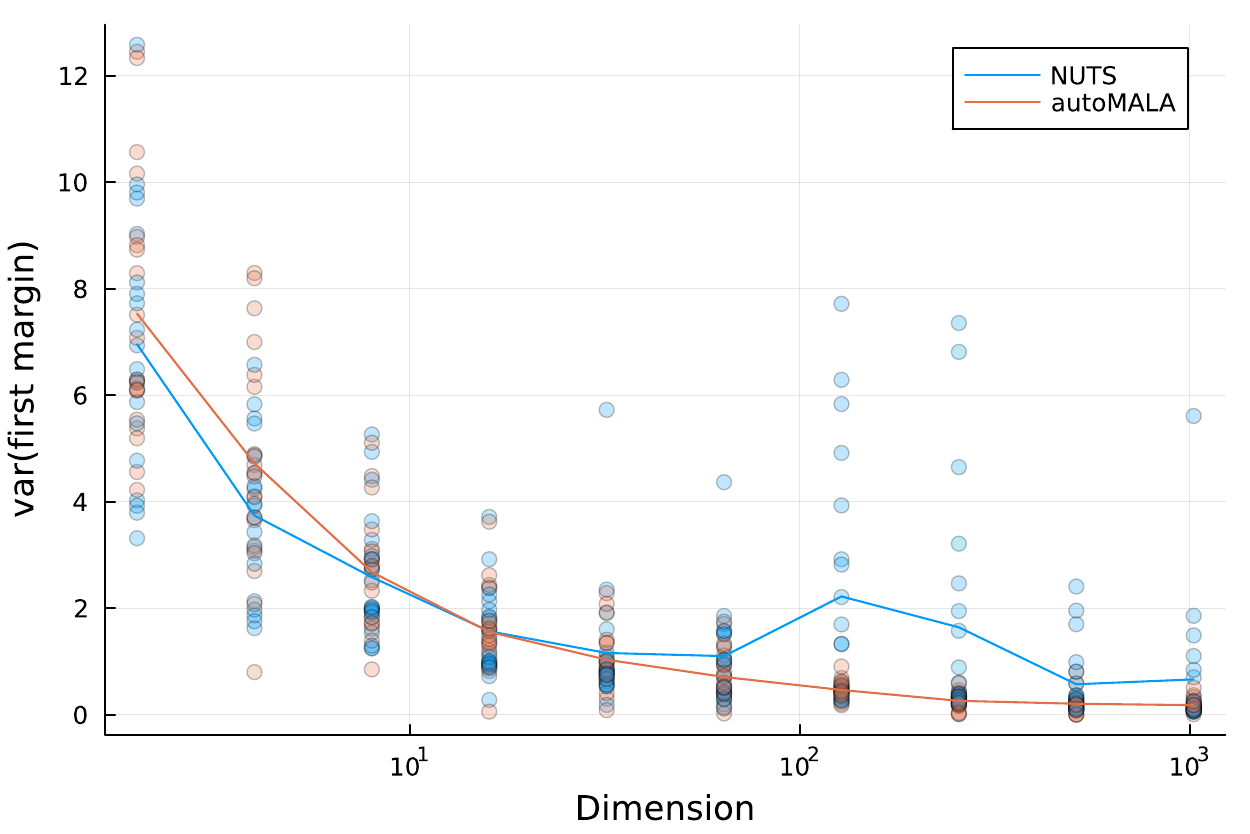}
    \end{subfigure}
    \begin{subfigure}{0.32\textwidth}
        \centering 
        \includegraphics[width=\textwidth]{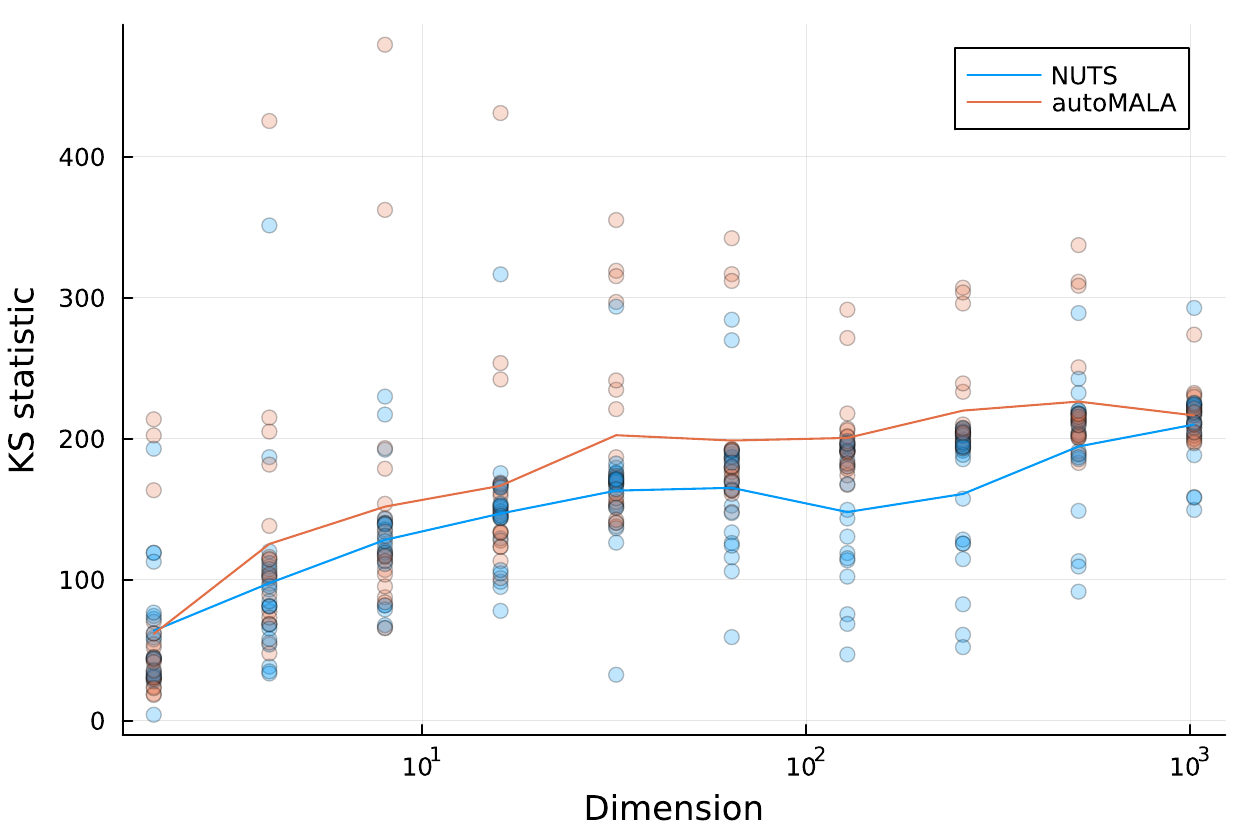}
    \end{subfigure}
    \caption{
        Various autoMALA and NUTS metrics for the high-dimensional banana experiments. 
        \textbf{Top row:} Number of leapfrog steps per 1000 effective samples. 
        From left to right we consider the minESS, $\exactess$, and regular ESS. 
        \textbf{Middle row:} minESS, $\exactess$, and regular ESS. 
        \textbf{Bottom row:} mean, variance, and Kolomogorov-Smirnov test statistic 
        for the known first marginal of the distribution. 
    }
    \label{fig:banana_highdim_all}
\end{figure*}

\begin{figure*}[!t]
  \centering
    \begin{subfigure}{0.32\textwidth}
        \centering 
        \includegraphics[width=\textwidth]{deliverables/AM_normal_highdim/scaling-leapfrog_min-model-normal.pdf}
    \end{subfigure}
    \begin{subfigure}{0.32\textwidth}
        \centering 
        \includegraphics[width=\textwidth]{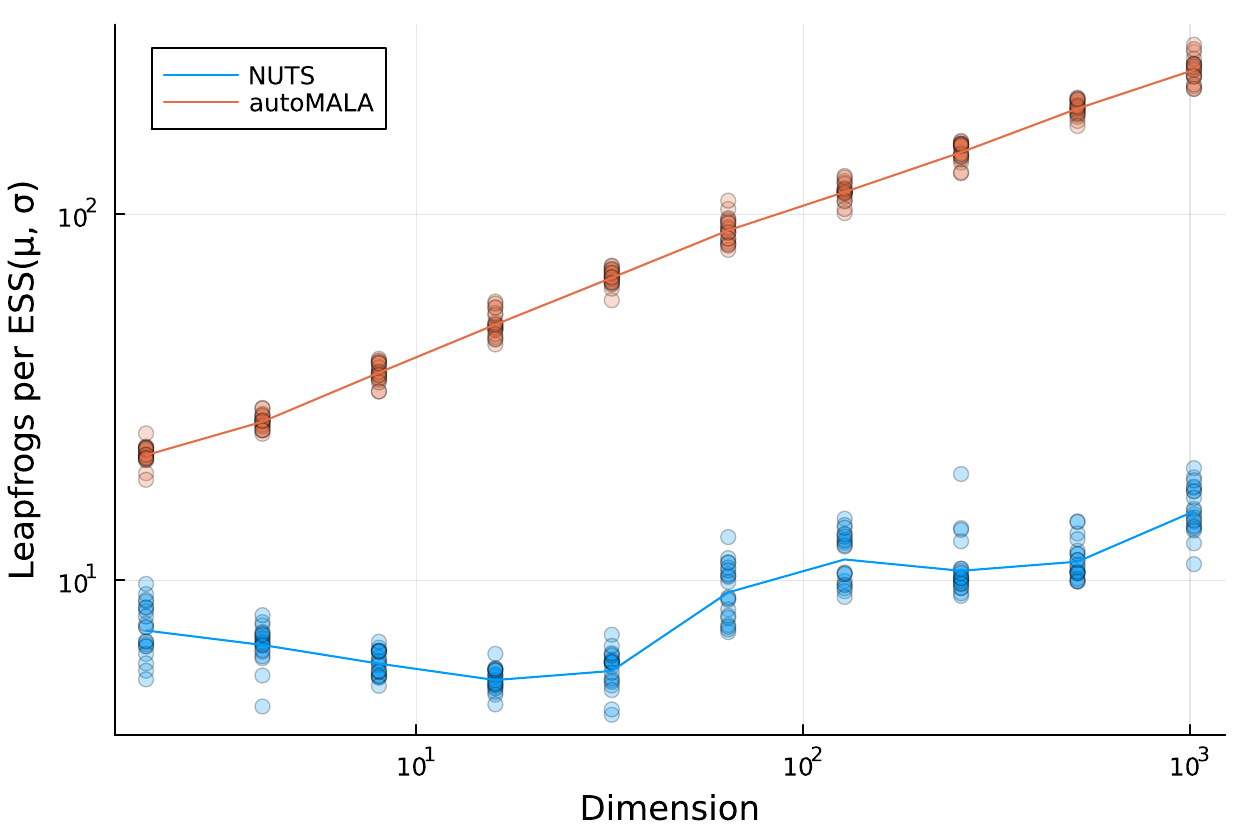}
    \end{subfigure}
    \begin{subfigure}{0.32\textwidth}
        \centering 
        \includegraphics[width=\textwidth]{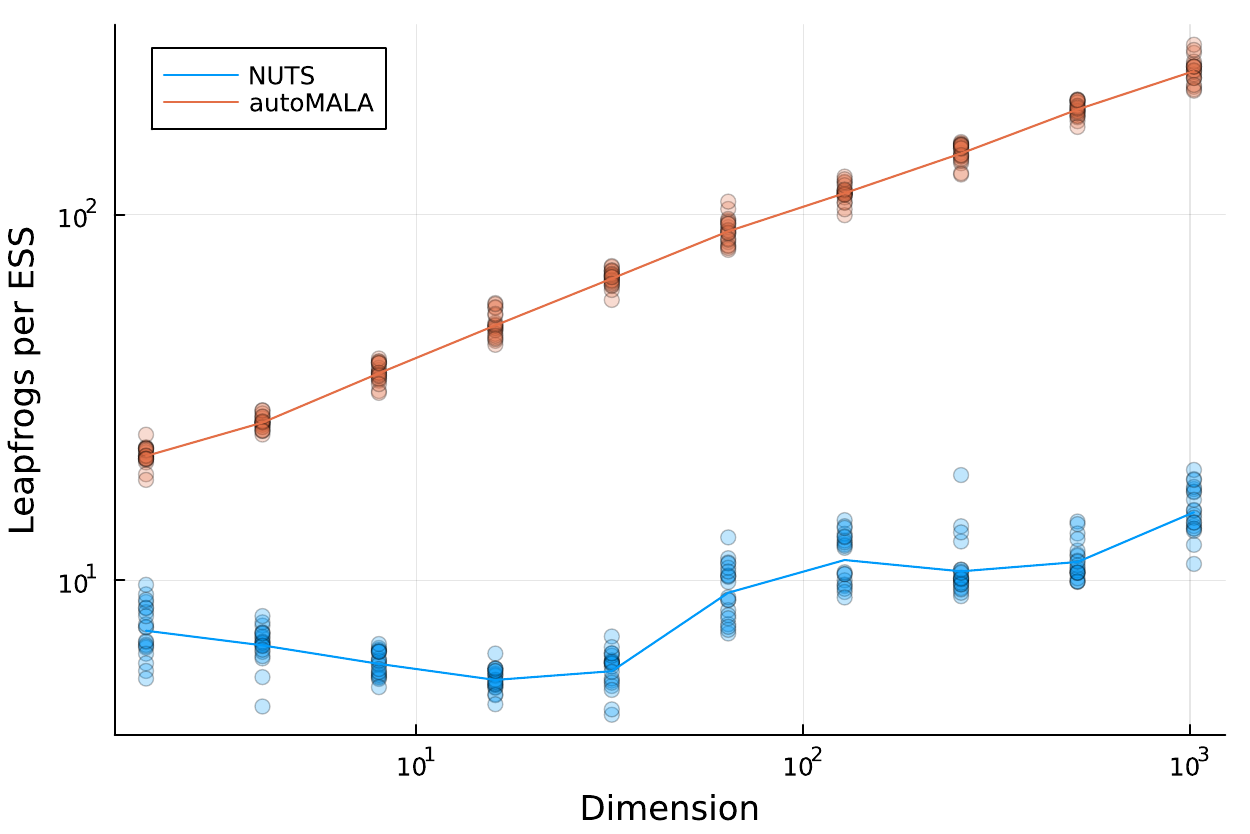} 
    \end{subfigure}
    \begin{subfigure}{0.32\textwidth}
        \centering 
        \includegraphics[width=\textwidth]{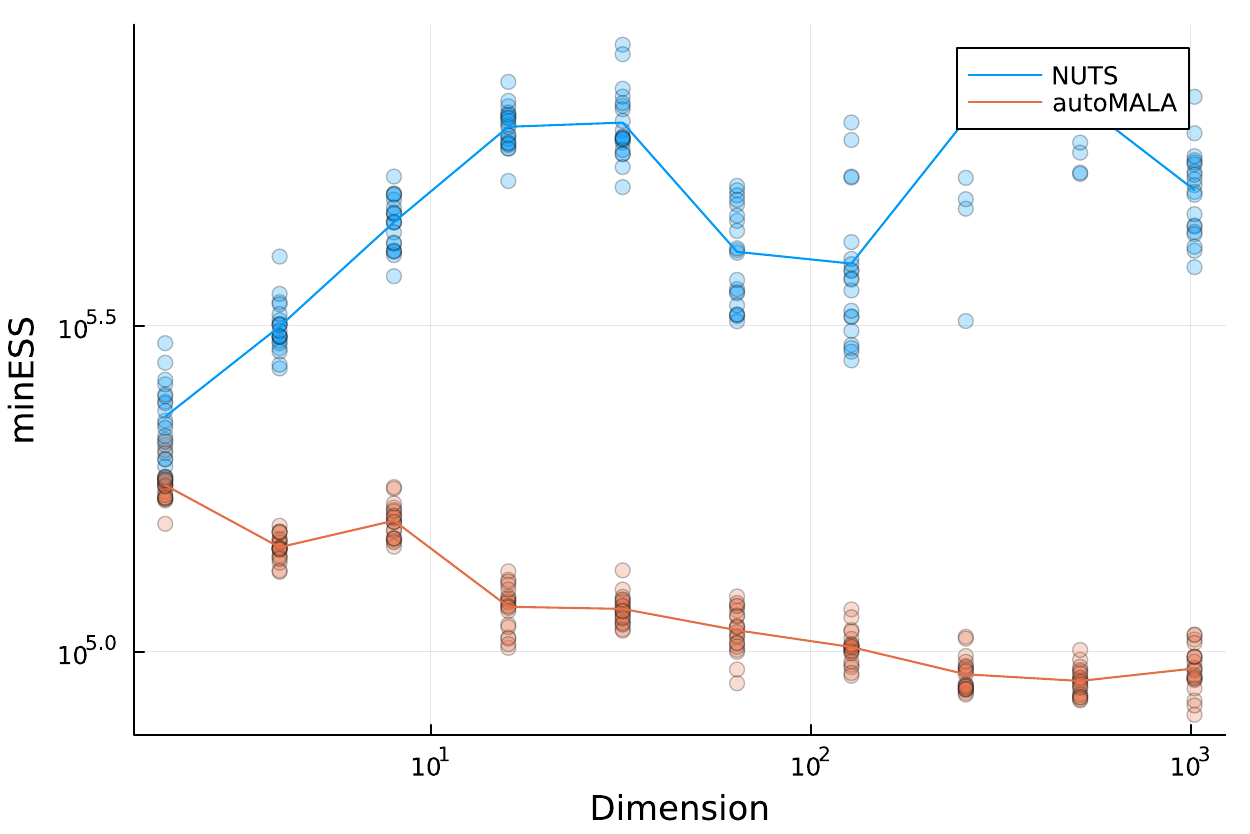}
    \end{subfigure}
    \begin{subfigure}{0.32\textwidth}
        \centering 
        \includegraphics[width=\textwidth]{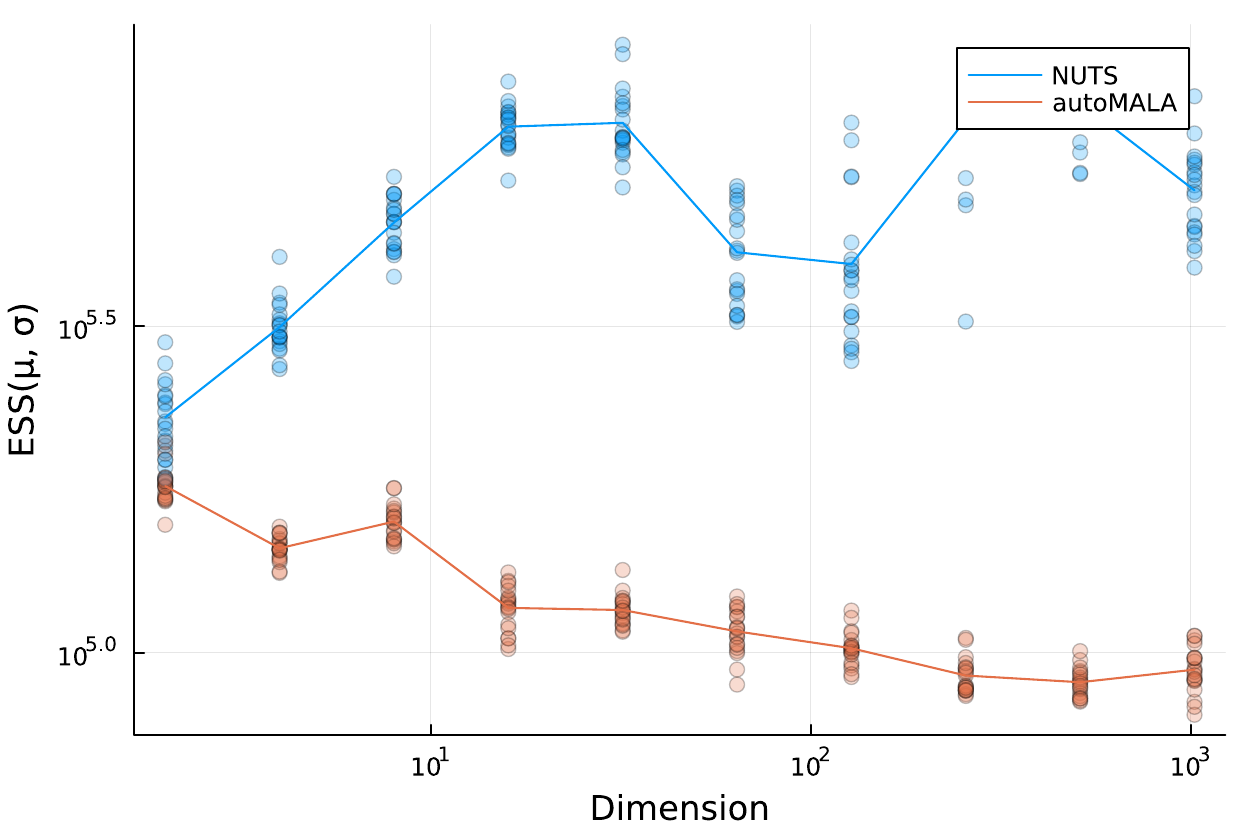}
    \end{subfigure}
    \begin{subfigure}{0.32\textwidth}
        \centering 
        \includegraphics[width=\textwidth]{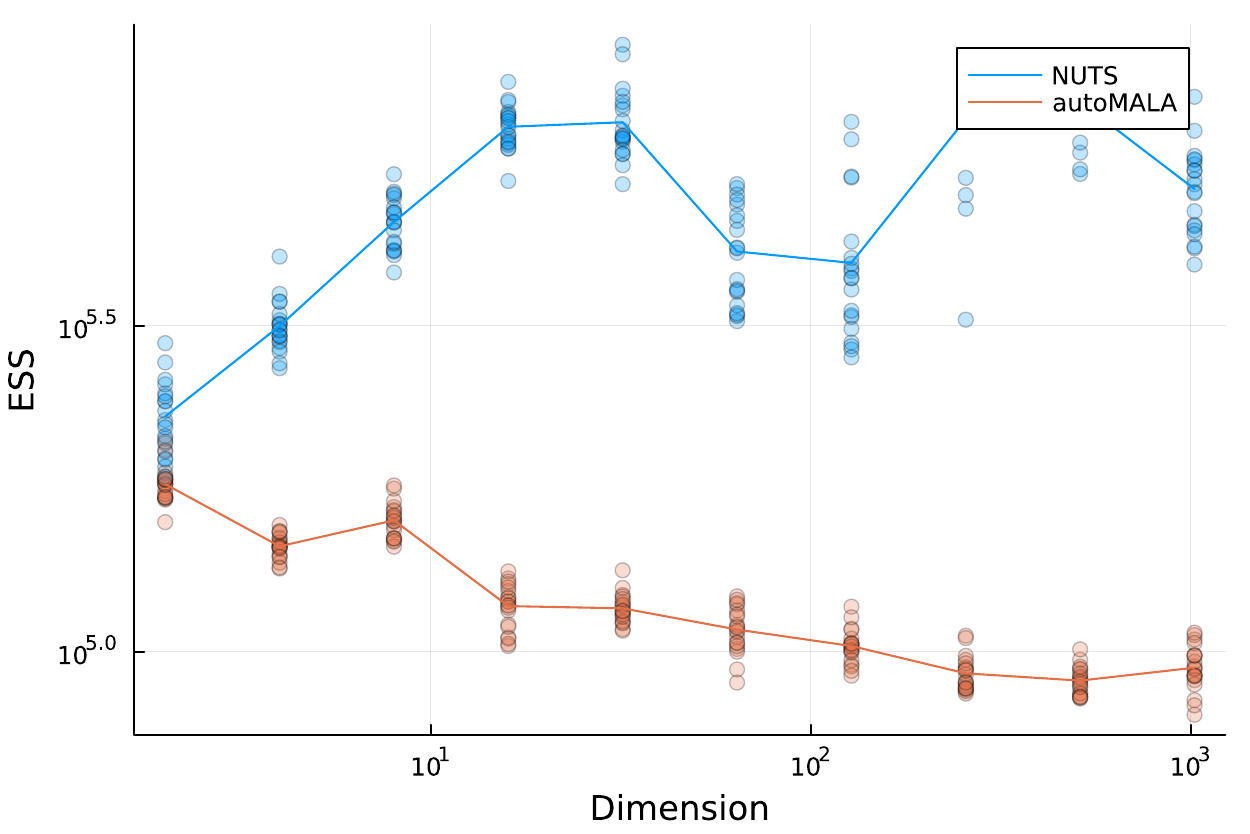}         
    \end{subfigure}
    \begin{subfigure}{0.32\textwidth}
        \centering 
        \includegraphics[width=\textwidth]{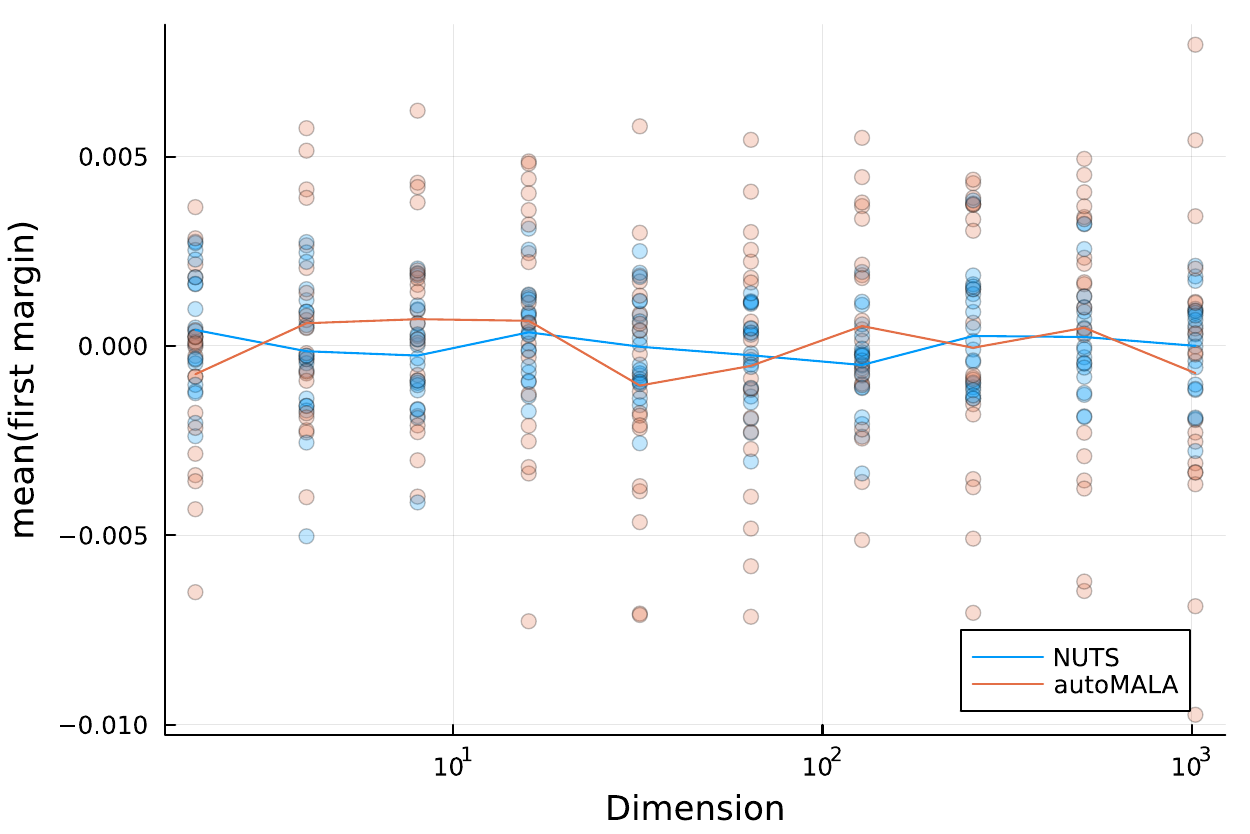}
    \end{subfigure}
    \begin{subfigure}{0.32\textwidth}
        \centering 
        \includegraphics[width=\textwidth]{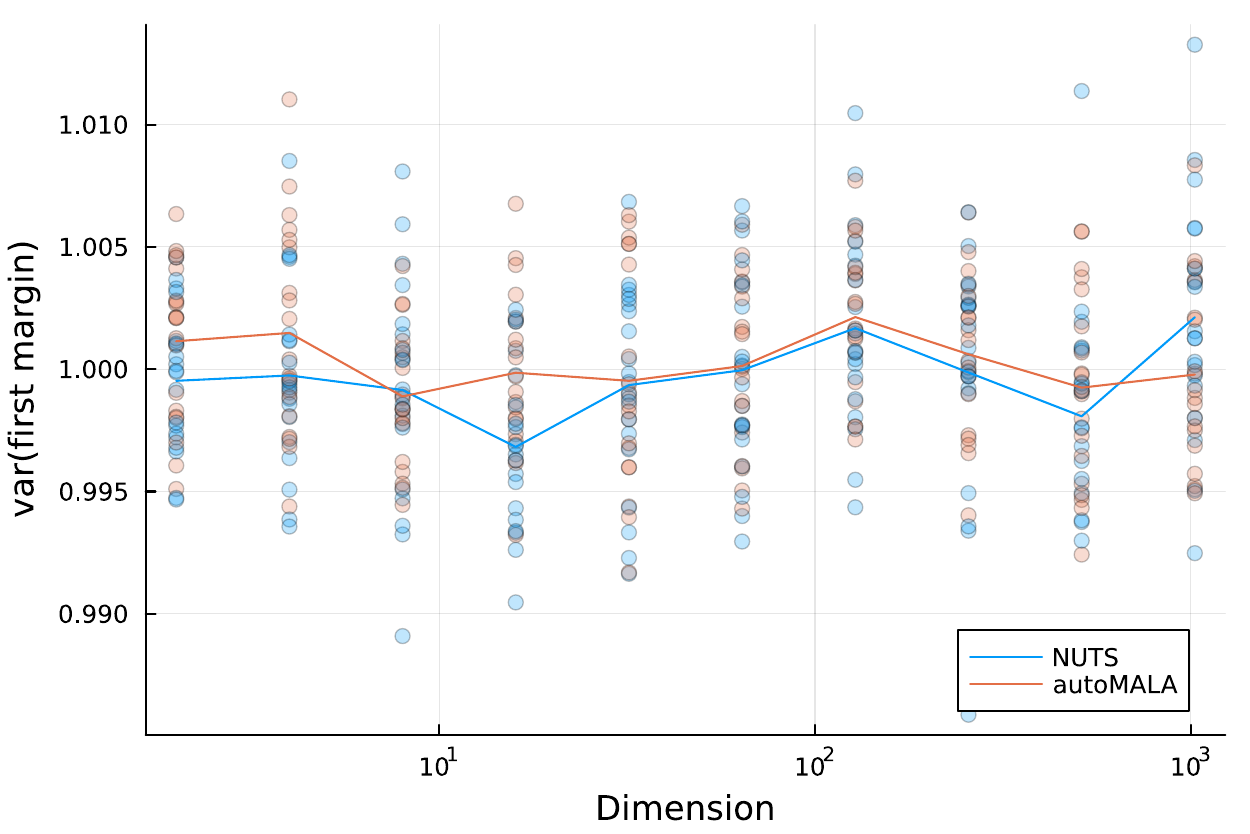}
    \end{subfigure}
    \begin{subfigure}{0.32\textwidth}
        \centering 
        \includegraphics[width=\textwidth]{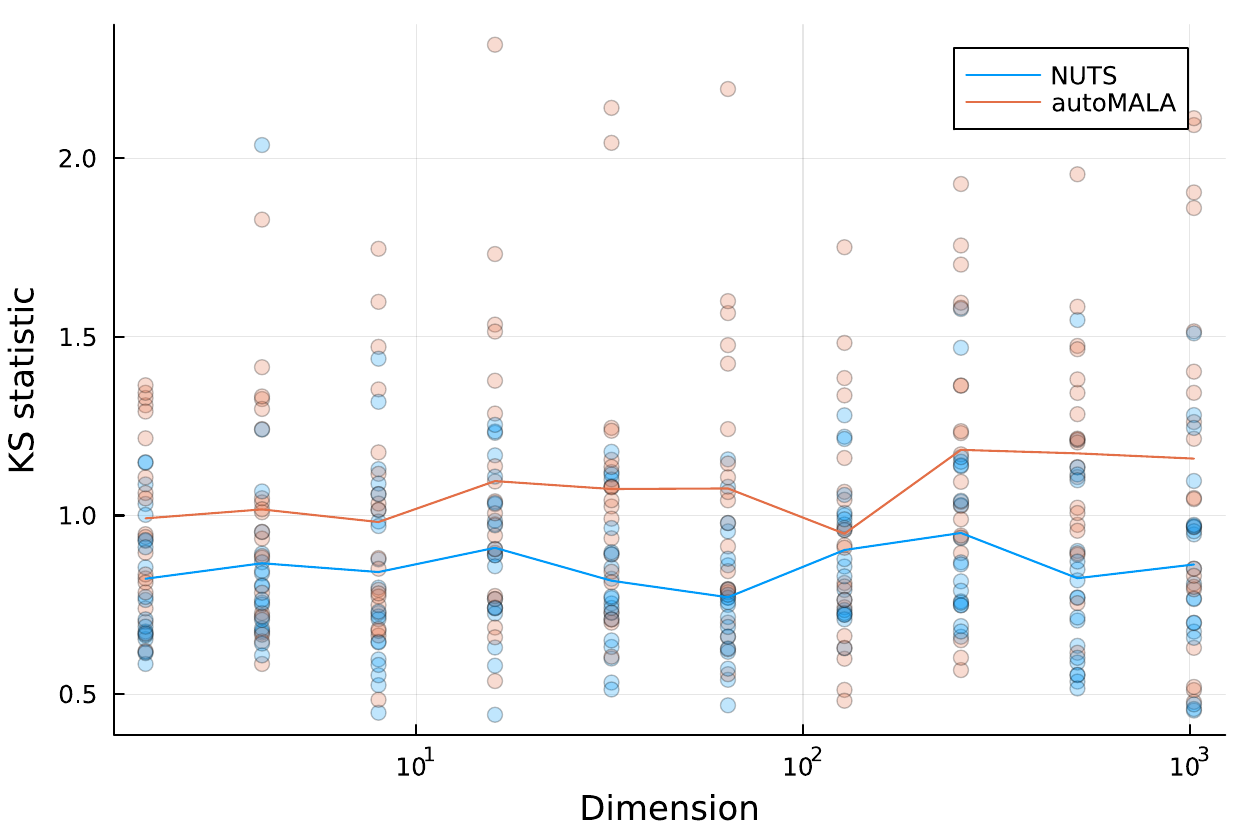}
    \end{subfigure}
    \caption{
        Various autoMALA and NUTS metrics for the high-dimensional normal experiments. 
        \textbf{Top row:} Number of leapfrog steps per 1000 effective samples. 
        From left to right we consider the minESS, $\exactess$, and regular ESS. 
        \textbf{Middle row:} minESS, $\exactess$, and regular ESS. 
        \textbf{Bottom row:} mean, variance, and Kolomogorov-Smirnov test statistic 
        for the known first marginal of the distribution. 
    }
    \label{fig:normal_highdim_all}
\end{figure*}

The simulation results for the comparison of autoMALA to NUTS on various high-dimensional 
target distributions (funnel, banana, and normal) are presented in 
\cref{fig:funnel_highdim_all}, \cref{fig:banana_highdim_all}, and 
\cref{fig:normal_highdim_all}.
For all three distributions we used $d \in \cbra{2^1, 2^2, \ldots, 2^{10}}$ with 
20 seeds and $2^{18}$ samples for each seed and $2^{18}$ samples for warmup for 
each seed. For Neal's funnel we set $\beta = 2$ and for the banana distribution 
we set $\beta = 1$.

\subsection{Step size convergence}

\begin{figure*}[!t]
    \begin{subfigure}{0.32\textwidth}
        \centering 
        \includegraphics[width=\textwidth]{deliverables/AM_stepsize_scaling/stepsize-scaling-dim-2.pdf}
    \end{subfigure}
    \begin{subfigure}{0.32\textwidth}
        \centering 
        \includegraphics[width=\textwidth]{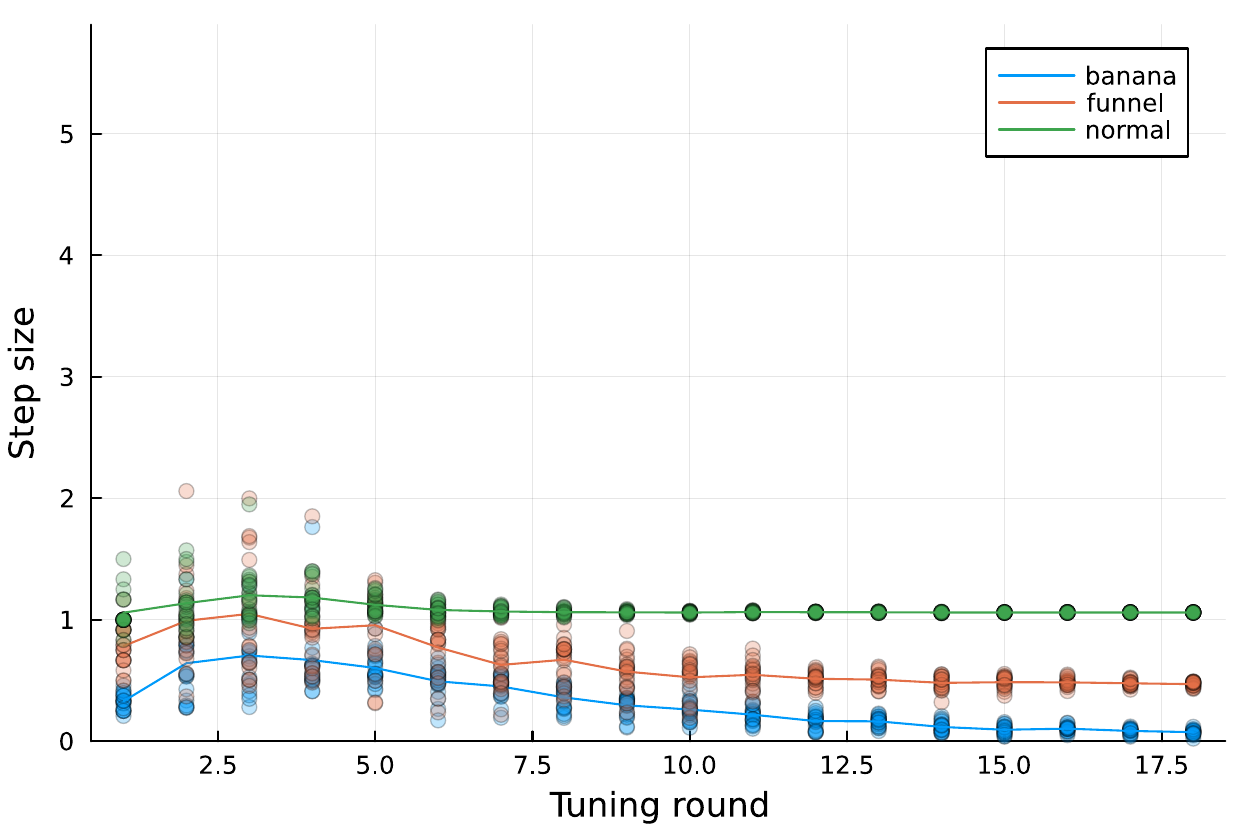}
    \end{subfigure}
    \begin{subfigure}{0.32\textwidth}
        \centering 
        \includegraphics[width=\textwidth]{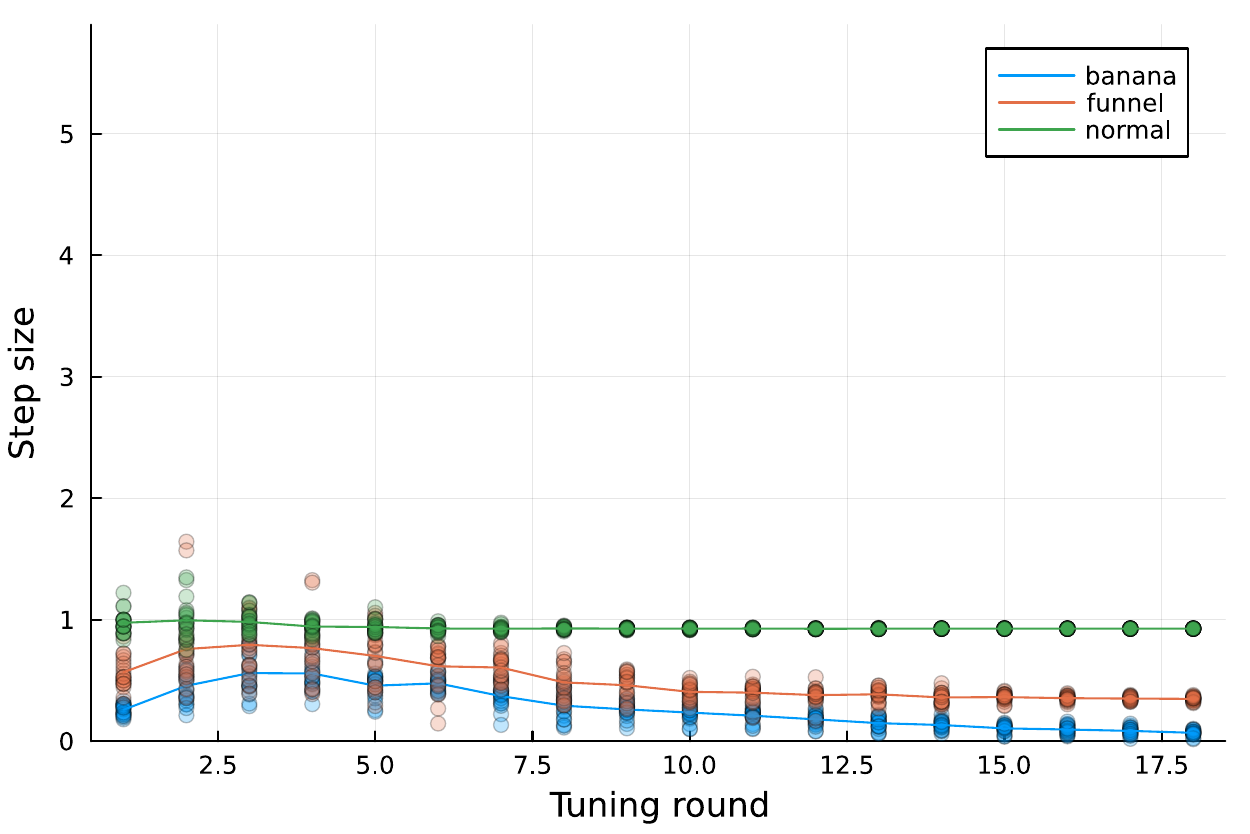}
    \end{subfigure}
    \begin{subfigure}{0.32\textwidth}
        \centering 
        \includegraphics[width=\textwidth]{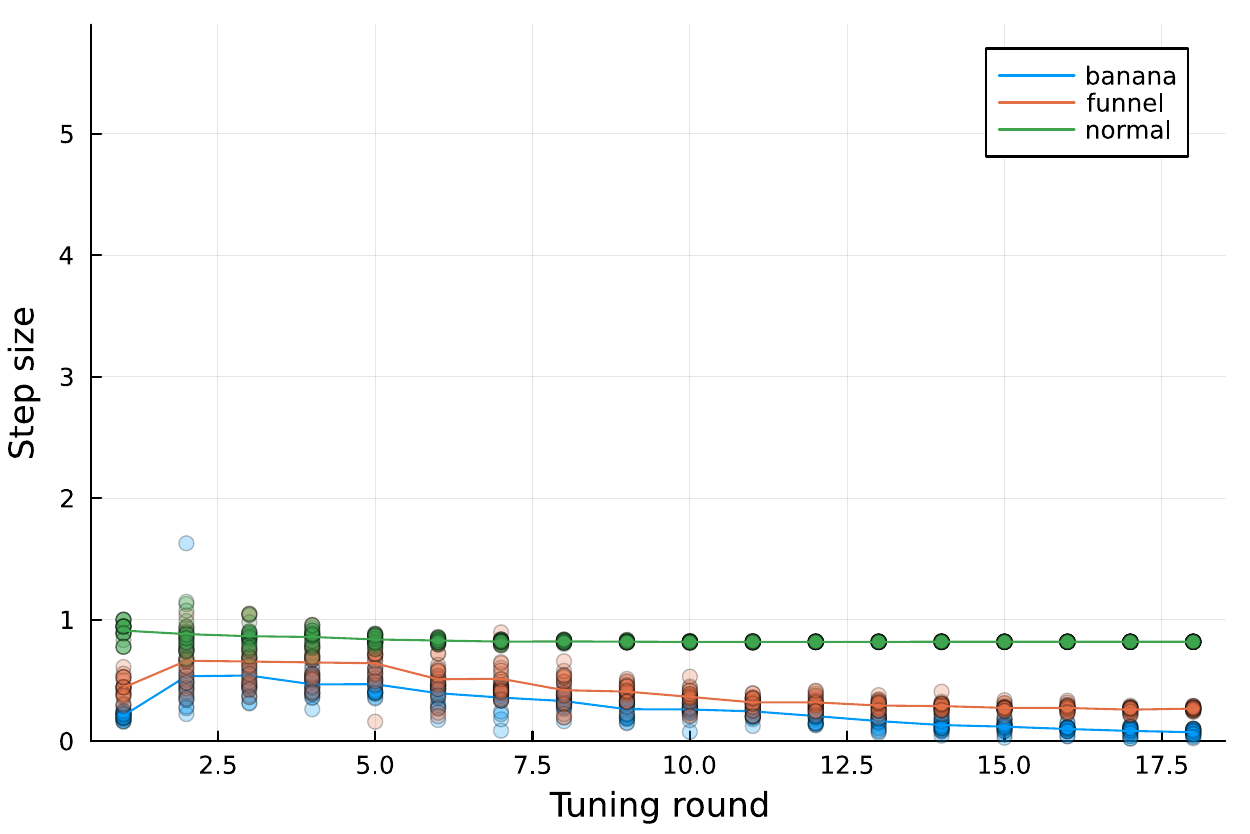}
    \end{subfigure}
    \begin{subfigure}{0.32\textwidth}
        \centering 
        \includegraphics[width=\textwidth]{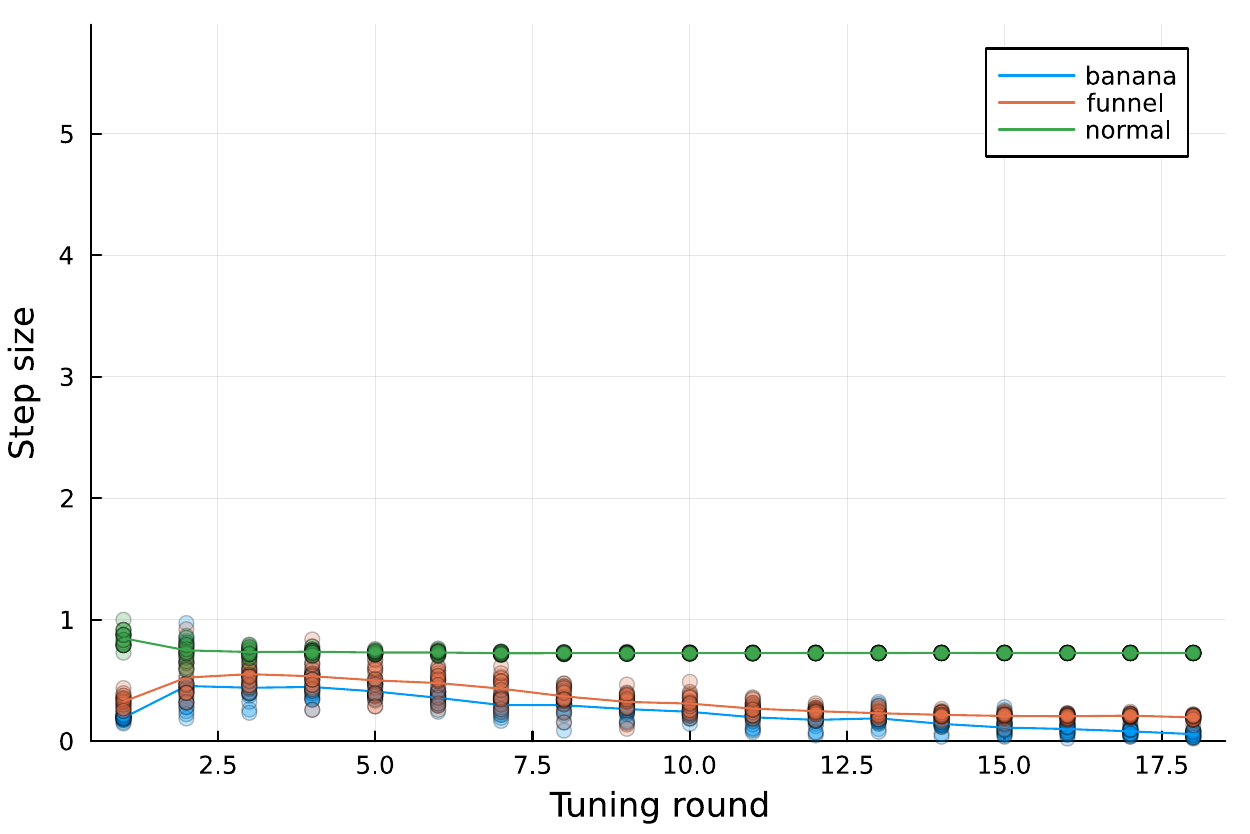}
    \end{subfigure}
    \begin{subfigure}{0.32\textwidth}
        \centering 
        \includegraphics[width=\textwidth]{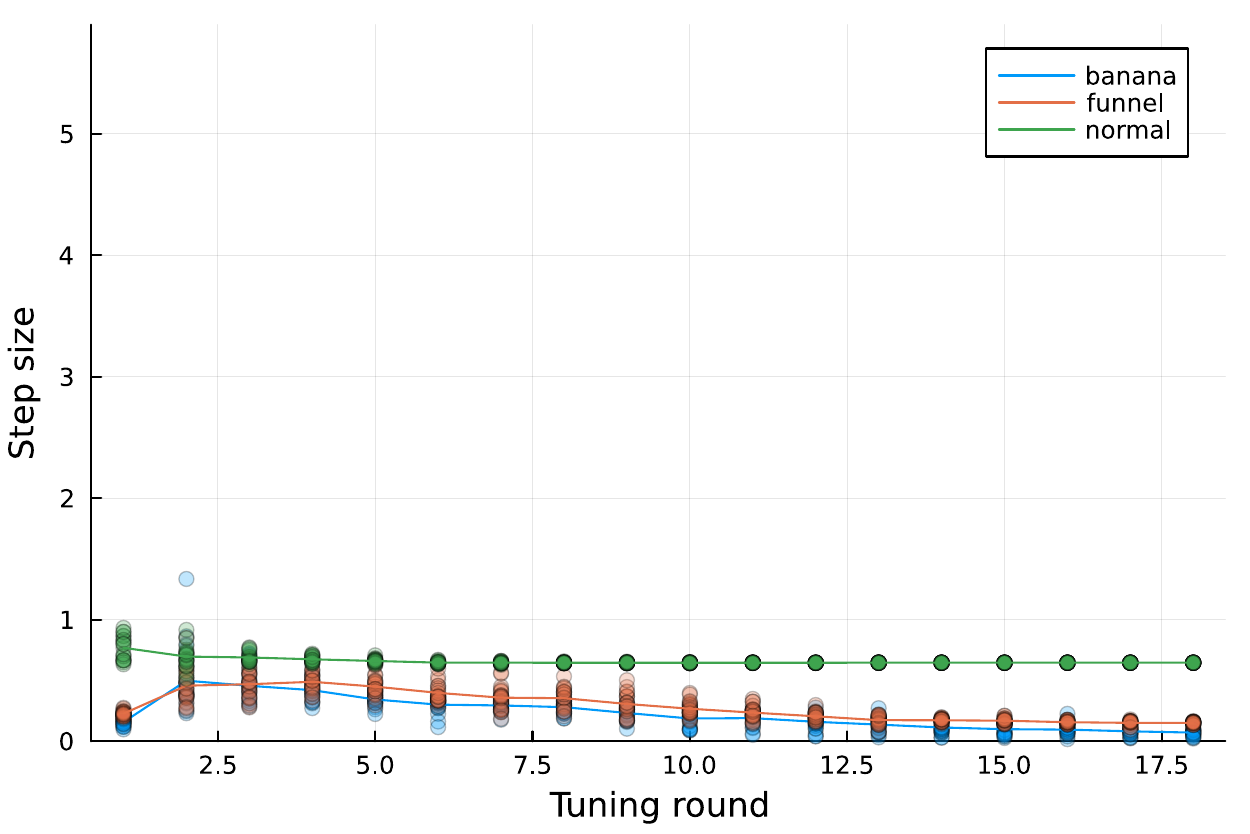}
    \end{subfigure}
    \begin{subfigure}{0.32\textwidth}
        \centering 
        \includegraphics[width=\textwidth]{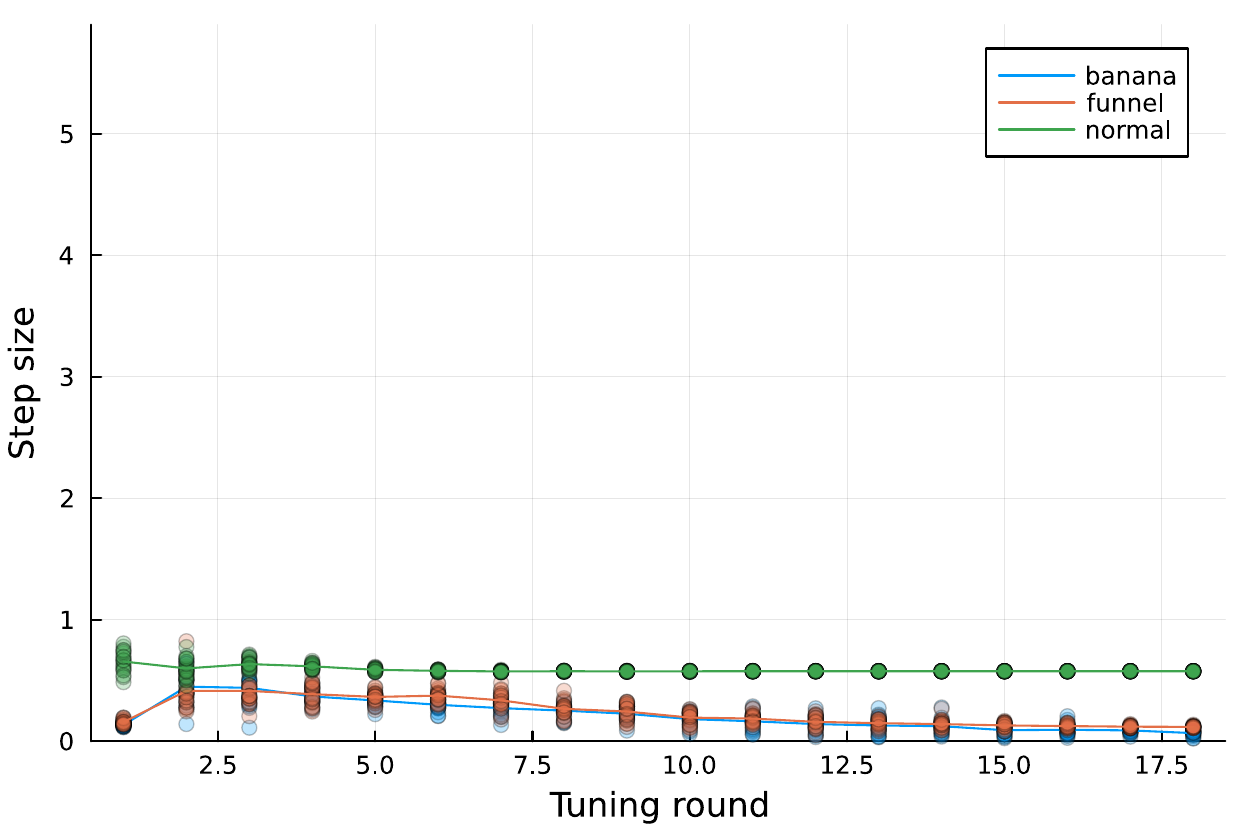}
    \end{subfigure}
    \begin{subfigure}{0.32\textwidth}
        \centering 
        \includegraphics[width=\textwidth]{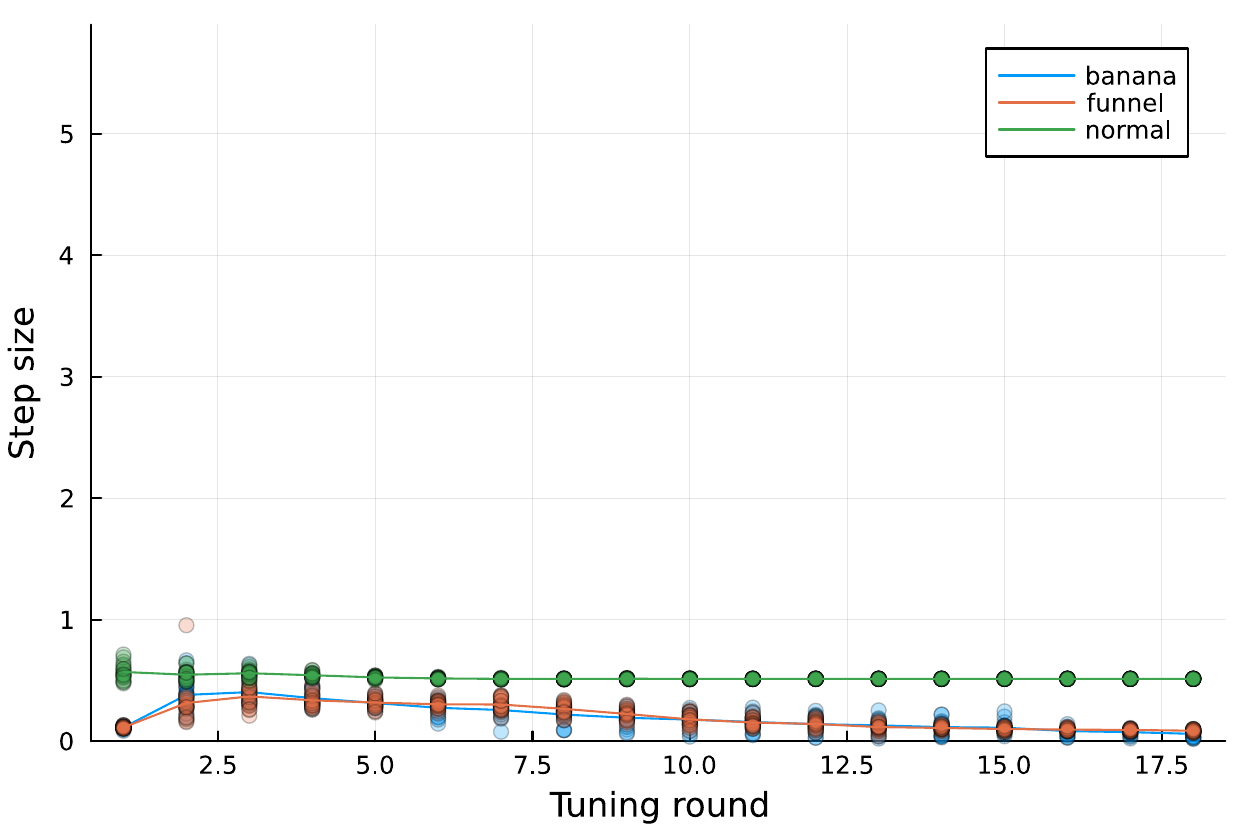}
    \end{subfigure}
    \begin{subfigure}{0.32\textwidth}
        \centering 
        \includegraphics[width=\textwidth]{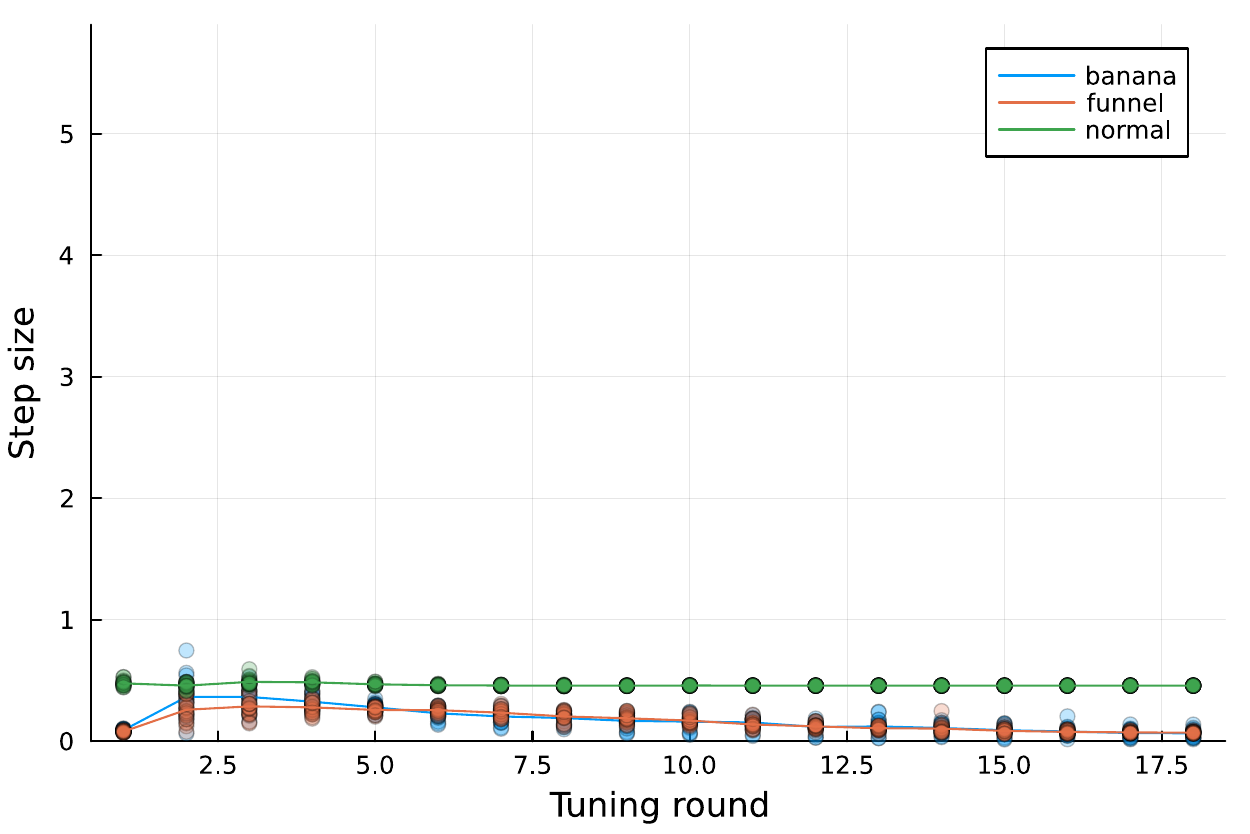}
    \end{subfigure}
    \caption{
        autoMALA default step sizes, $\epsi$, as a function of the tuning round, 
        where each tuning round doubles the number of samples used to estimate 
        $\epsi$. 
        \textbf{Top left to bottom right:} $2^j$-dimensional target distributions 
        for $j=1,2,\ldots, 9$.
    }
    \label{fig:stepsize_convergence_all}
\end{figure*}

We assessed whether the default step size, $\epsi$, converges as the number of tuning 
rounds increases. Each successive tuning round used twice the amount of MCMC iterations 
compared to the previous tuning round. 
The experimental results for the three synthetic target distributions 
for various dimensions $d$ are presented in \cref{fig:stepsize_convergence_all}.
In these experiments we used $d \in \cbra{2^1, 2^2, \ldots, 2^9}$, and 
20 different seeds for each setting. For each combination of simulation settings, we ran 
autoMALA for $2^{19}$ iterations ($2^{18}$ samples for warmup and $2^{18}$ final samples).

\subsection{Comparison to non-adaptive algorithms}
\begin{figure*}[!t]
    \begin{subfigure}{0.32\textwidth}
        \centering 
        \includegraphics[width=\textwidth]{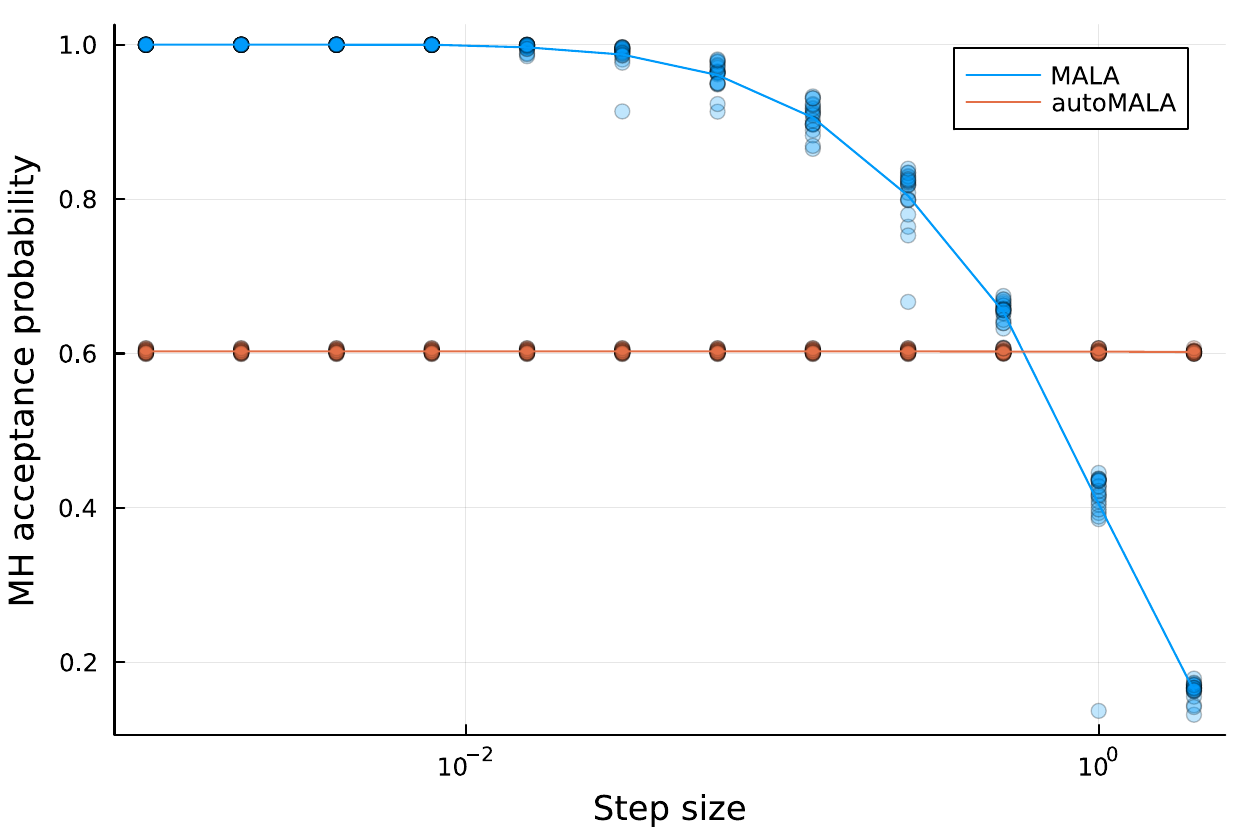}
    \end{subfigure}
    \begin{subfigure}{0.32\textwidth}
        \centering 
        \includegraphics[width=\textwidth]{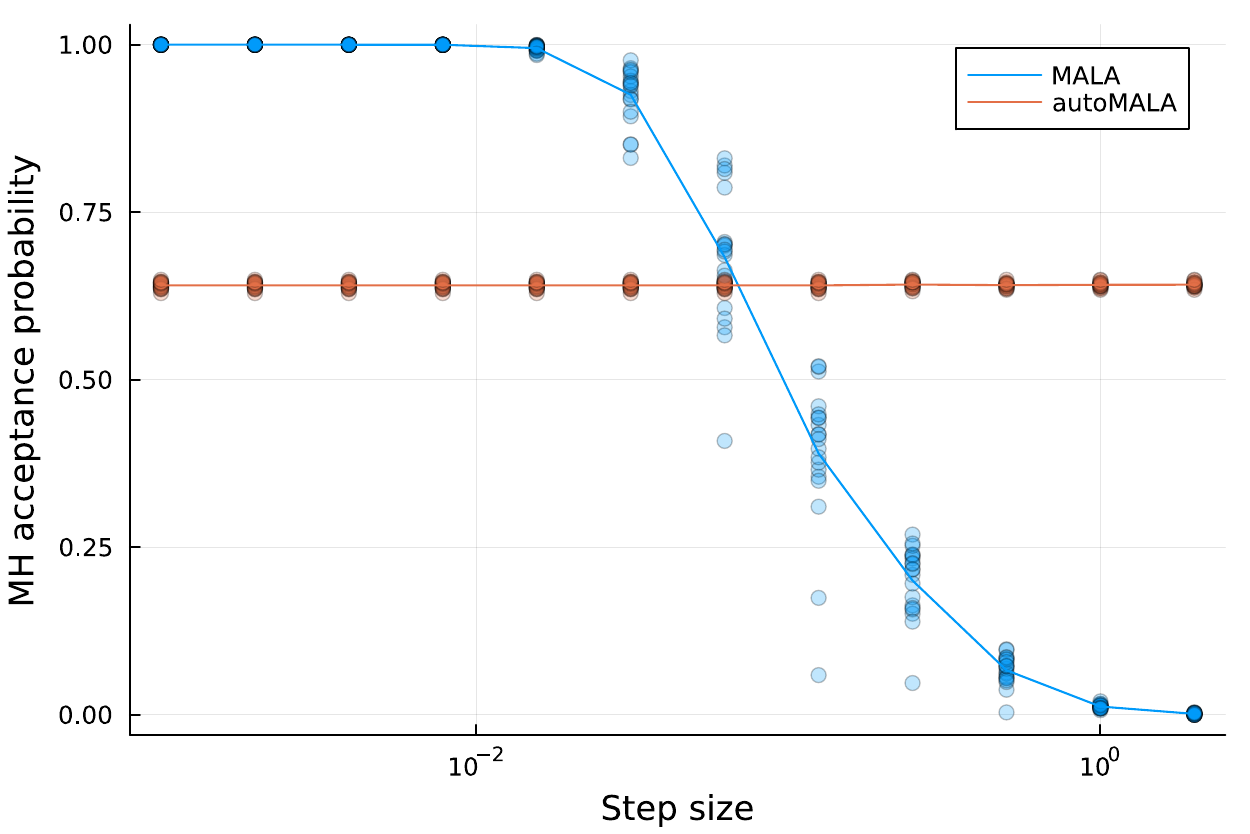}
    \end{subfigure}
    \begin{subfigure}{0.32\textwidth}
        \centering 
        \includegraphics[width=\textwidth]{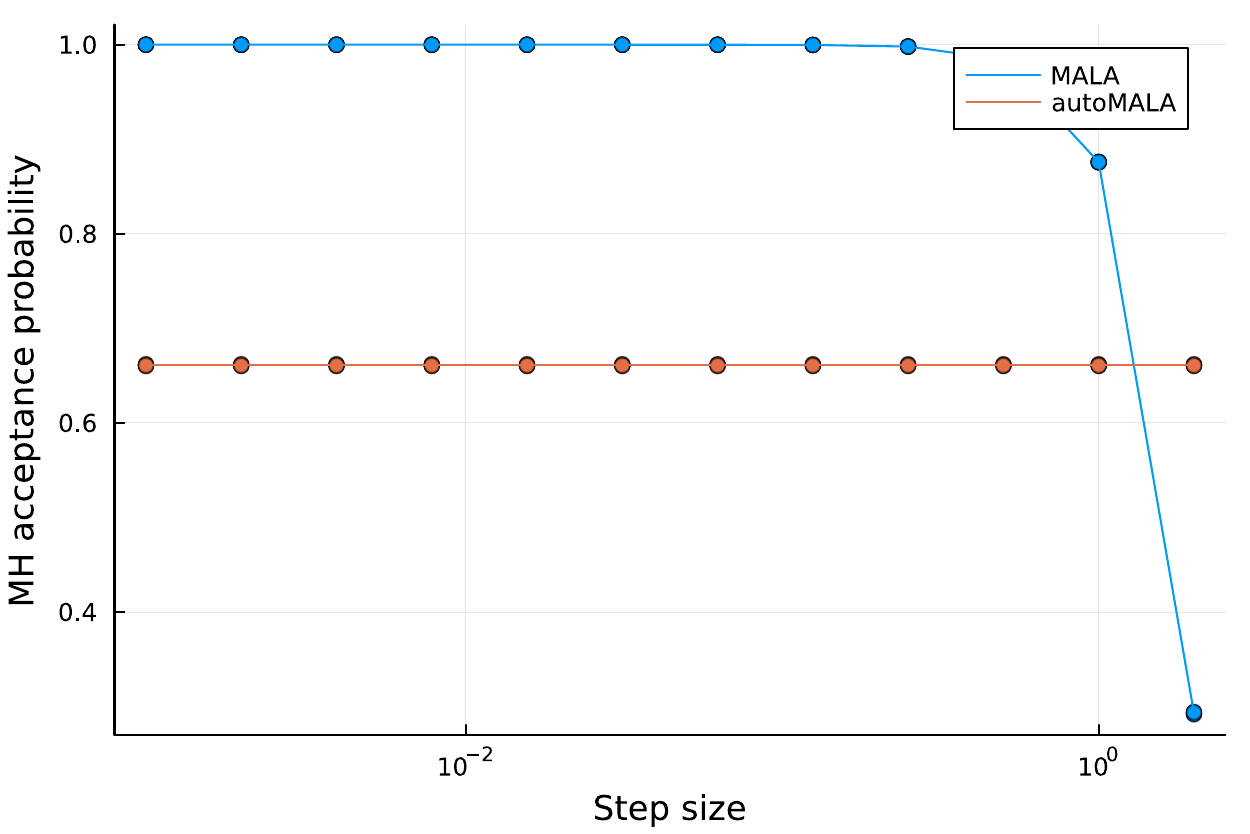}
    \end{subfigure}
    \caption{
        autoMALA and MALA acceptance probabilities as a function of initial step size 
        for a fixed grid. 
        \textbf{Left to right:} funnel, banana, and normal distributions ($d=2$). 
    }
    \label{fig:MALA_fixed_grid_all}
\end{figure*}

We compared autoMALA to MALA by considering both a fixed grid of step sizes 
and a grid relative to the ``optimal'' choice selected after a long run of autoMALA. 
Our main statistic is the average Metropolis--Hastings acceptance probability. 
For the fixed step size grid, we used 20 seeds for each of the three synthetic models 
with $d=2$. The final acceptance probabilities were calculated after running both 
autoMALA and MALA for $2^{18}$ warmup iterations followed by $2^{18}$ MCMC samples.  
We considered $\eps \in \cbra{2^{-10}, 2^{-9}, \ldots, 2^1}$. 
These results are presented in \cref{fig:MALA_fixed_grid_all}. 

\begin{figure*}[!t]
    \begin{subfigure}{0.32\textwidth}
        \centering 
        \includegraphics[width=\textwidth]{deliverables/mala_stepsize_performance/mala-stepsize-acceptance-model-funnel-dim-2.pdf}
    \end{subfigure}
    \begin{subfigure}{0.32\textwidth}
        \centering 
        \includegraphics[width=\textwidth]{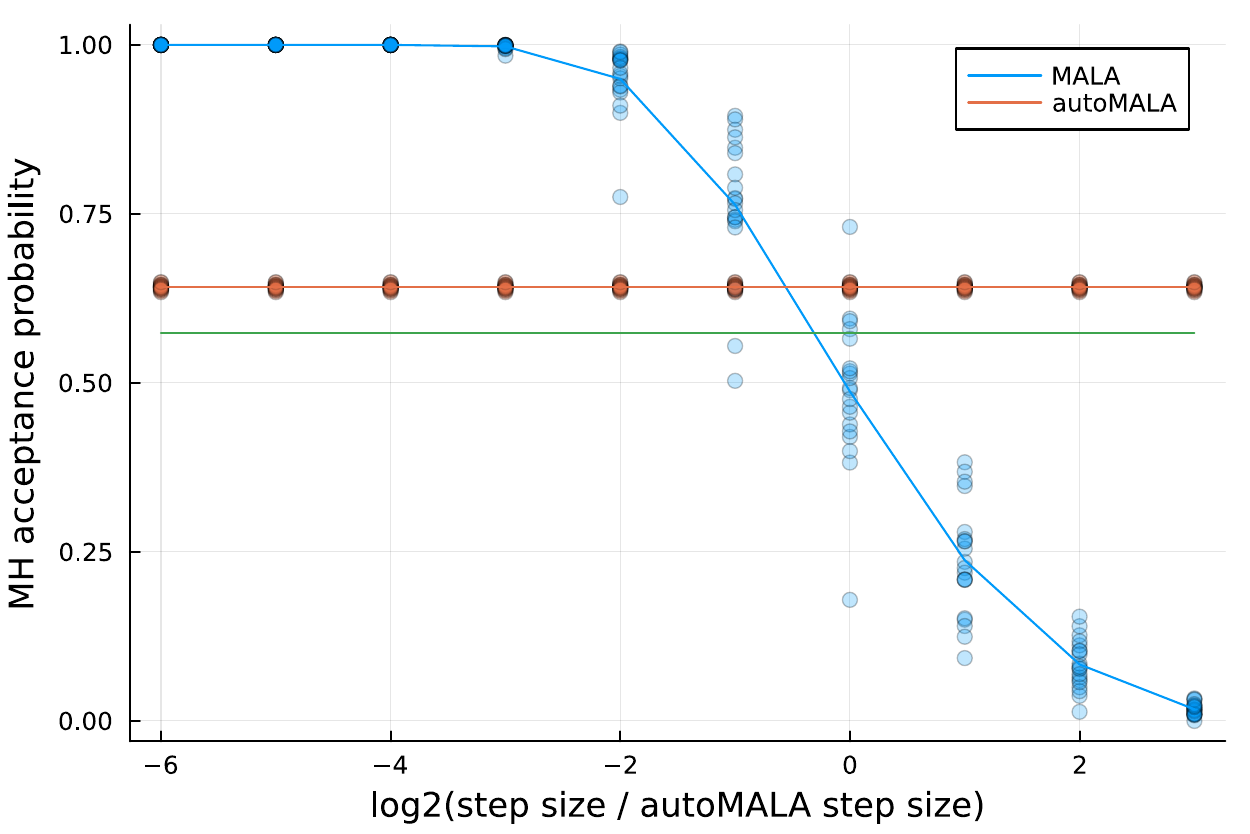}
    \end{subfigure}
    \begin{subfigure}{0.32\textwidth}
        \centering 
        \includegraphics[width=\textwidth]{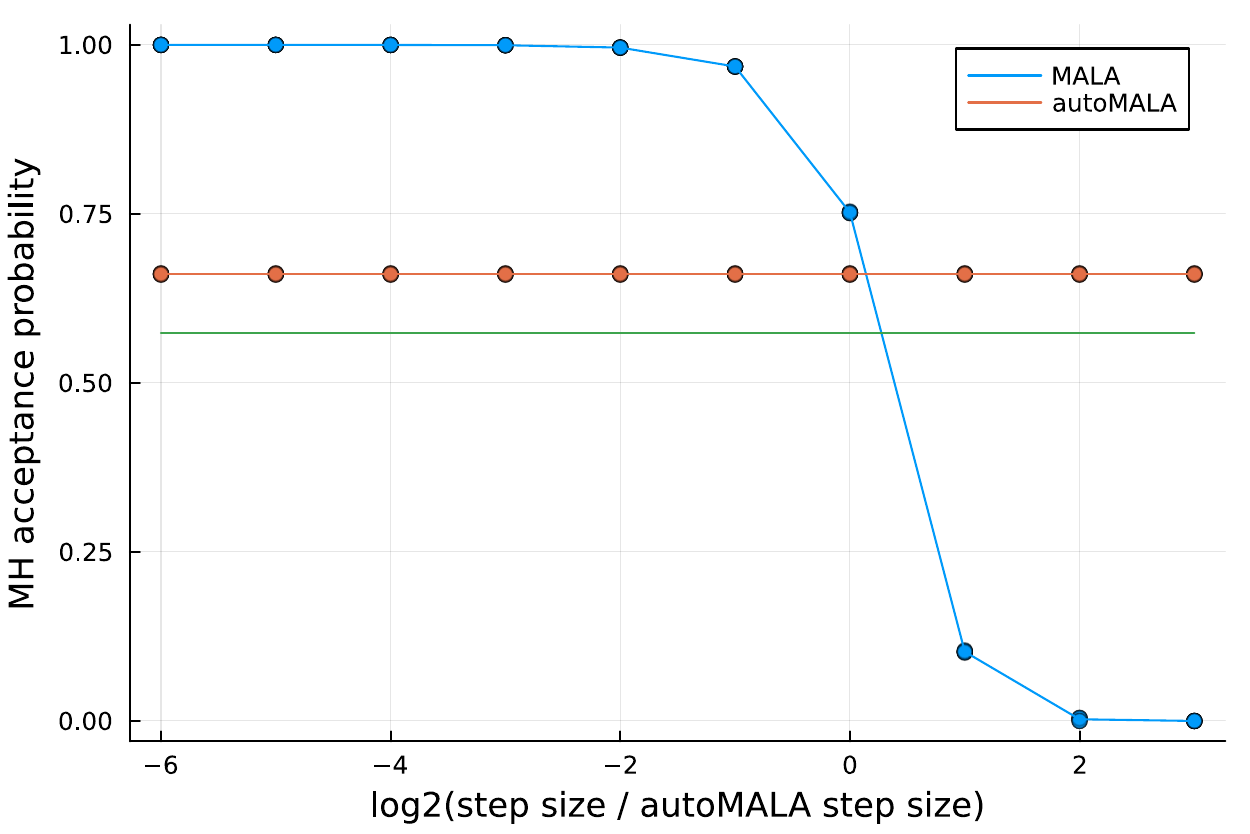}
    \end{subfigure}
    \caption{
        autoMALA and MALA acceptance probabilities as a function of initial step size 
        for a grid relative to the optimal choice for autoMALA. 
        \textbf{Left to right:} funnel, banana, and normal distributions ($d=2$). 
    }
    \label{fig:MALA_relative_grid_all}
\end{figure*}

For the step size grid relative to the ``optimal'' autoMALA choice, we also used 
20 seeds applied to the three synthetic models. 
After running autoMALA for $2^{18}$ warmup iterations and $2^{18}$ final samples, 
we extracted the selected default step size (denoted $\eps$ here). 
Then, we ran MALA with step sizes in the range 
$\cbra{\eps \cdot 2^{-6}, \eps \cdot 2^{-5}, \ldots, \eps \cdot 2^3}$. 
We ran MALA for the same number of warmup and final samples as autoMALA.
These results are presented in \cref{fig:MALA_relative_grid_all}. 

\subsection{autoMALA Metropolis--Hastings acceptance probability}
\begin{figure*}
  \centering 
  \includegraphics[width=0.5\textwidth]{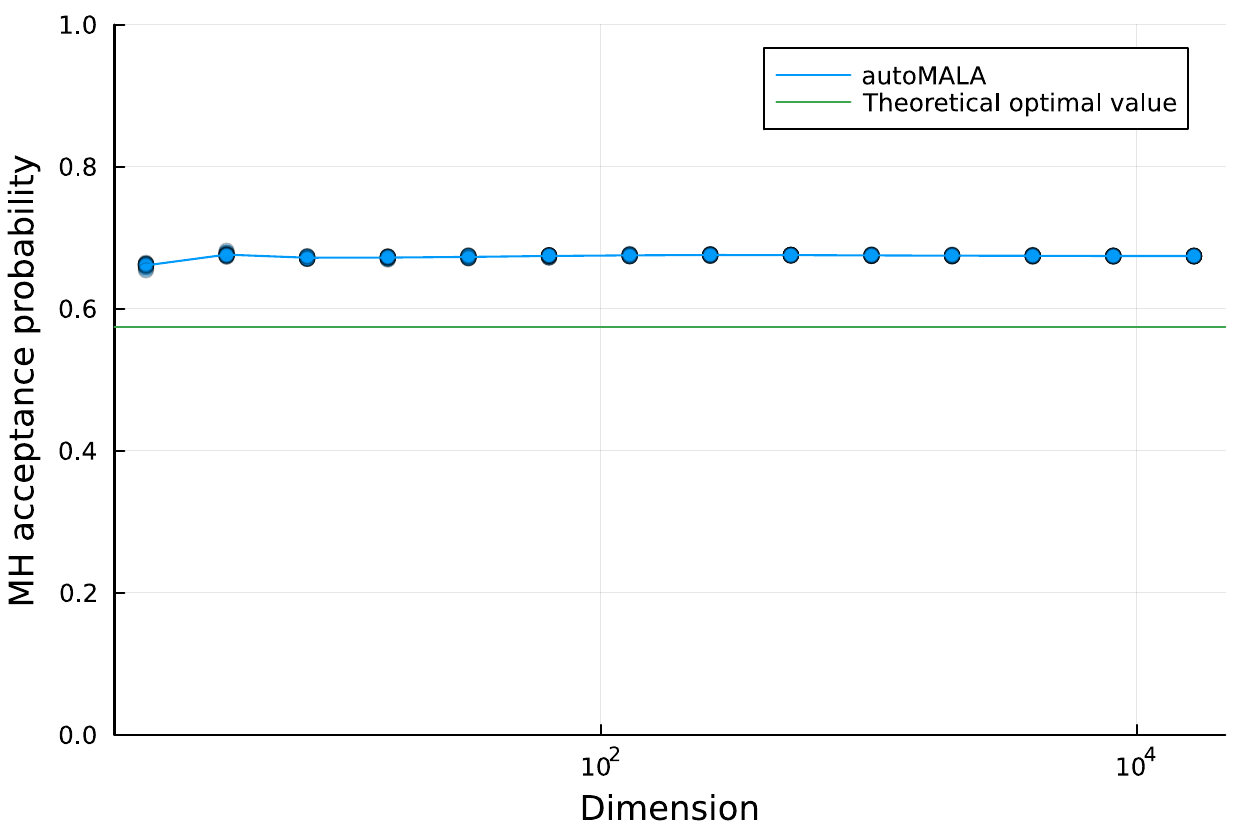}
  \caption{Metropolis--Hastings acceptance probability of autoMALA as a function of 
  the dimension of a multivariate standard normal target distribution for 
  various simulation seeds. 
  The green line indicates the theoretical optimal MALA acceptance probability. 
  Note that there is no guarantee that autoMALA should converge to the 
  MALA acceptance probability because the algorithm is inherently different 
  from MALA.}
  \label{fig:AM_normal_acceptance}
\end{figure*}

The optimal acceptance probability for MALA on a certain class of target distributions 
as the dimension tends to infinity has been shown to be 0.574. 
autoMALA is a different algorithm than MALA, and hence its ``optimal'' 
acceptance probability might be different. 
We assess the MH acceptance probability of autoMALA on standard multivariate normal 
targets as the dimension tends to infinity. The results of these simulations 
are shown in \cref{fig:AM_normal_acceptance}.
We see that the asymptotic autoMALA acceptance probability on the product 
distribution is close to the optimal MALA acceptance probability, but not exactly equal. 
In this experiment we used 20 different seeds with $2^{15}$ samples in the final round 
to estimate the step size. The simulations were run on distributions of dimension 
$d \in \cbra{2^1, 2^2, \ldots, 2^{14}}$.  

\subsection{Fixed point of autoMALA step size objective}

\begin{figure}[t]
	\centering
	\includegraphics[width=0.4\textwidth]{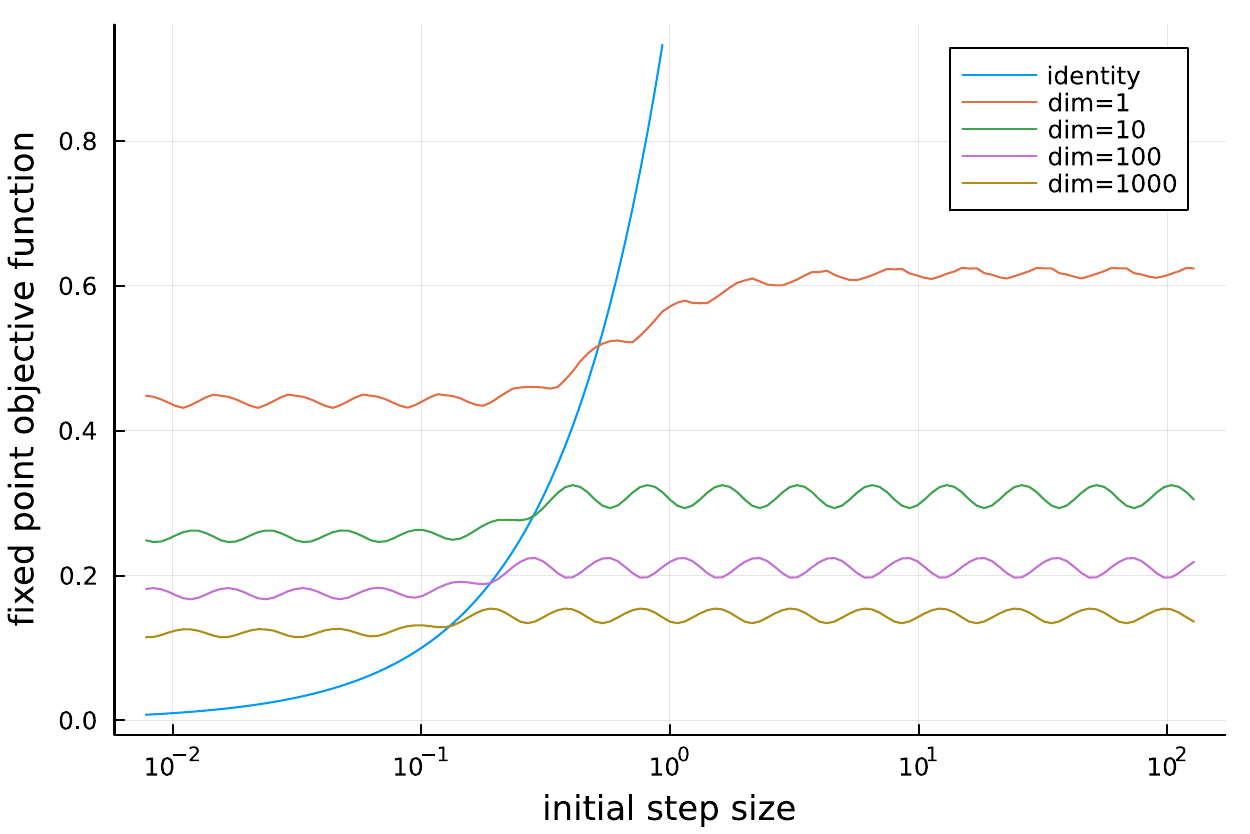}
	\caption{The objective function 
  $g(\epsi) = \ex_{\bar\pi, \epsi}[(\eps(S, \epsi) + \eps(S', \epsi))/2]$ 
	for isotropic normal targets of varying dimension. The x-axis shows a grid over $\epsi$ 
	in log scale. The y-axis shows the function $g$ approximated using $2^{15}$ samples. 
	The round-based scheme approaches the fixed point $\epsi = g(\epsi)$. 
  The identity function is shown 
  (sharply increasing function, due to the x-axis being in log-scale).}
	\label{fig:ideal-obj}
\end{figure}

We show in \cref{fig:ideal-obj} the objective function 
$g(\epsi) = \ex_{\bar\pi, \epsi}[(\eps(S, \epsi) + \eps(S', \epsi))/2]$ 
as a function of $\epsi$, approximated using $2^{15}$ iterations 
of autoMALA for each grid point $\epsi = 2^j$ for $j \in \{-7, -6.9, -6.8, \dots, 6.9, 7\}$. 
This was repeated for isotropic normal targets of dimension $d \in \{1, 10, 100, 1000\}$. 
In all cases, a unique fixed point is present upon inspection. 
Moreover, the shape of the objective 
appears to approximately converge to a constant function as $d$ increases, 
which suggests that the greater number of rounds required to converge in high 
dimensions is not necessarily due to the complexity of the idealized fixed 
point equation $\epsi = g(\epsi)$, but rather due to the increased difficulty in 
approximating $g$ from Monte Carlo samples.

\subsection{Preconditioning strategy}\label{app:preconditioning}

In \cref{sec:autoMALA} we described a simple strategy to obtain a robust
diagonal preconditioner by taking a random mixture of the form
\[
(\hat{\Sigma}_\text{AM})_{i,i}^{-1/2}=\eta \hat \Sigma_{i, i}^{-1/2}+(1-\eta),
\] 
with $\eta \sim \distBeta(\tilde \alpha, \tilde \beta)$ sampled independently 
for some values of $\tilde\alpha, \tilde\beta > 0$. The justification of this
strategy is that it adds robustness to the sampler when 1) the estimated standard
deviations are far from their true values under the target distribution, or 2)
the local geometry varies considerably so that a fixed preconditioner can fail
in particulars regions of the space. On the
other hand, the approach can cause issues when there are dimensions that have
scales considerably smaller than $1$. Indeed, consider the bivariate distribution
$\distNorm(0,\diag(10^{-8},10^{8})^2)$. When $\eta \sim \distUnif(0,1)$,
\[
\ex[(\hat{\Sigma}_\text{AM})_{2,2}^{-1/2}|\hat\Sigma] = 0.5 \hat \Sigma_{2,2}^{-1/2}+0.5 \approx 0.5 \cdot 10^{-8}+0.5 \approx 0.5.
\]
Since $p\sim \distNorm(0,\hat{\Sigma}_\text{AM}^{-1})$ (see \cref{alg:autoMALA}),
this means that $p_2$ will on average be eight orders of magnitude larger than
its optimal value \citep[see e.g.][\S 4.1]{neal2011mcmc}. 
In turn, this forces autoMALA to heavily shrink the step size at each iteration in 
order to reach the right scale, which results in slow performance.

\begin{figure}[ht]
	\centering
	\includegraphics[width=0.5\textwidth]{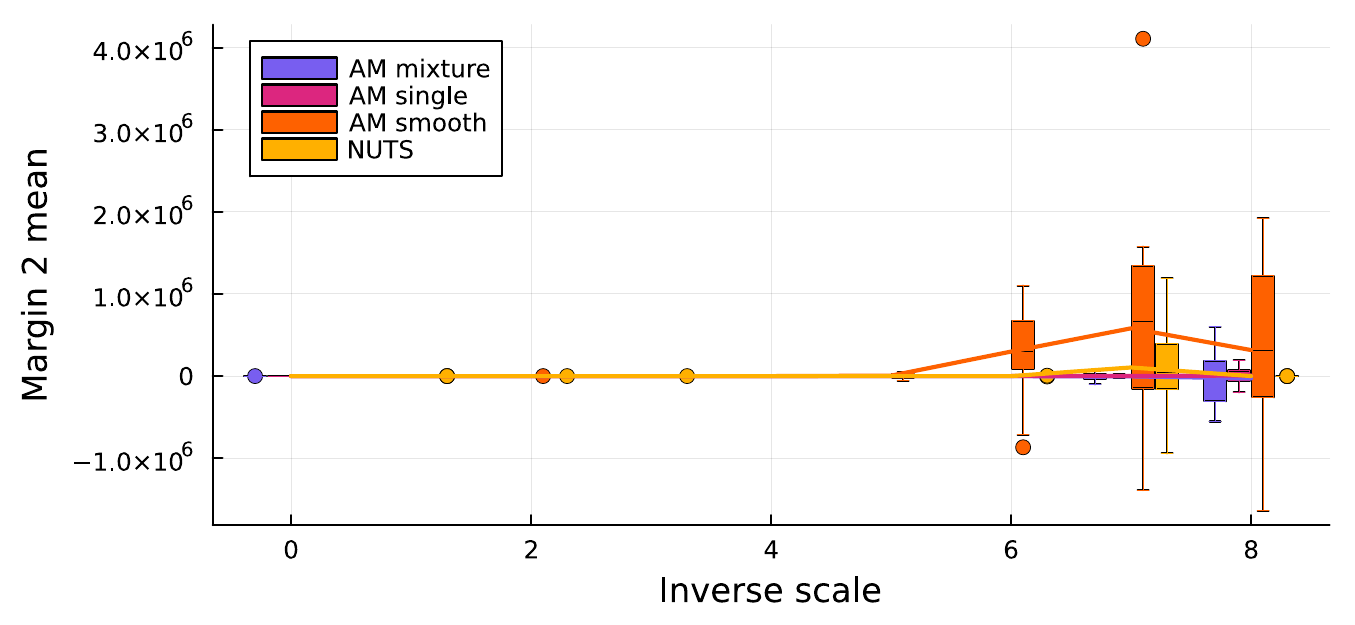}%
    \includegraphics[width=0.5\textwidth]{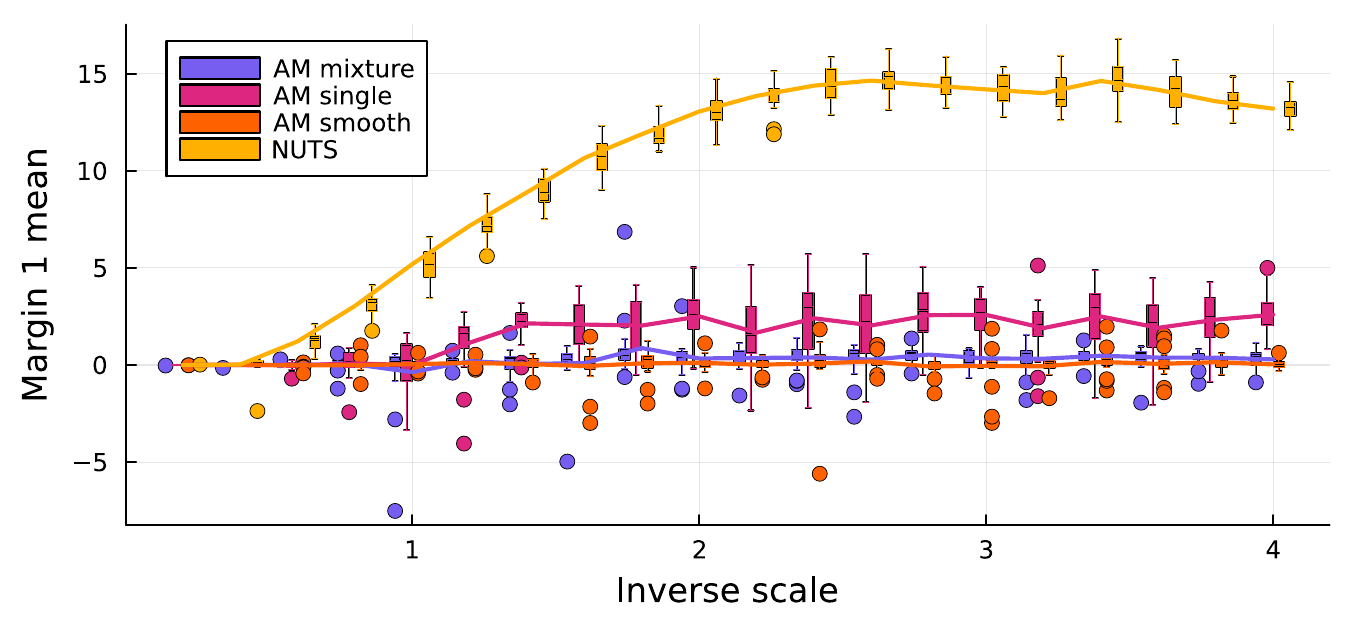} \\
	\includegraphics[width=0.5\textwidth]{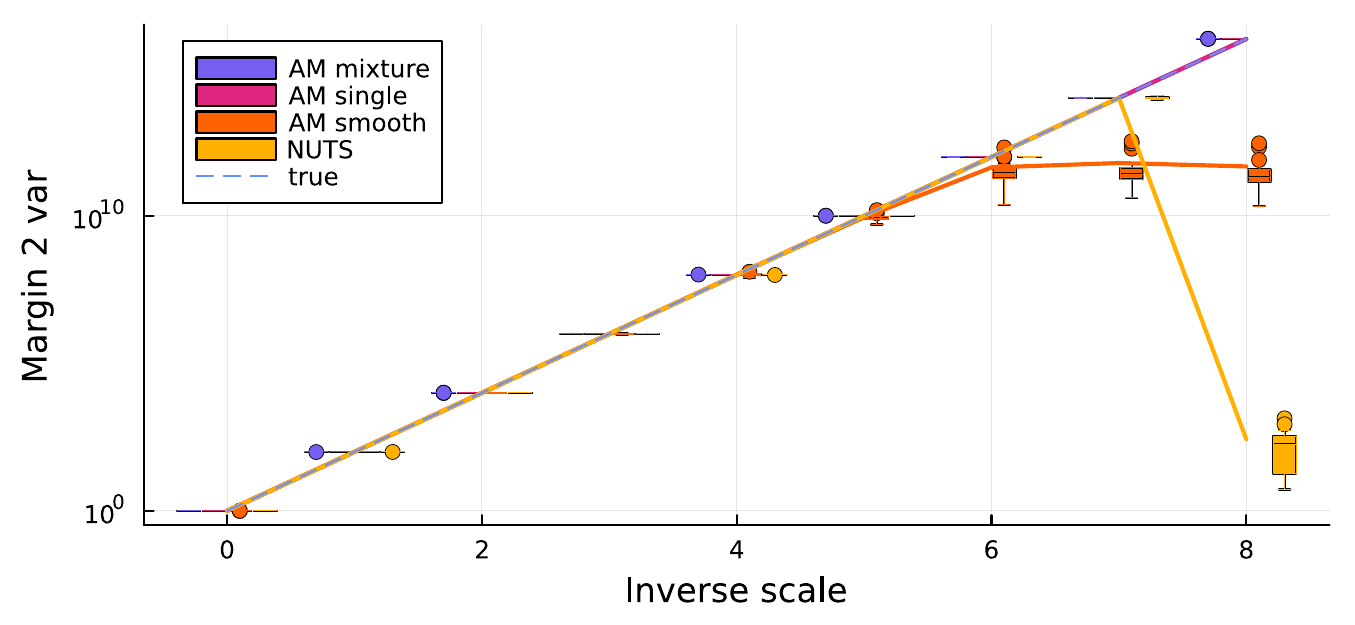}%
    \includegraphics[width=0.5\textwidth]{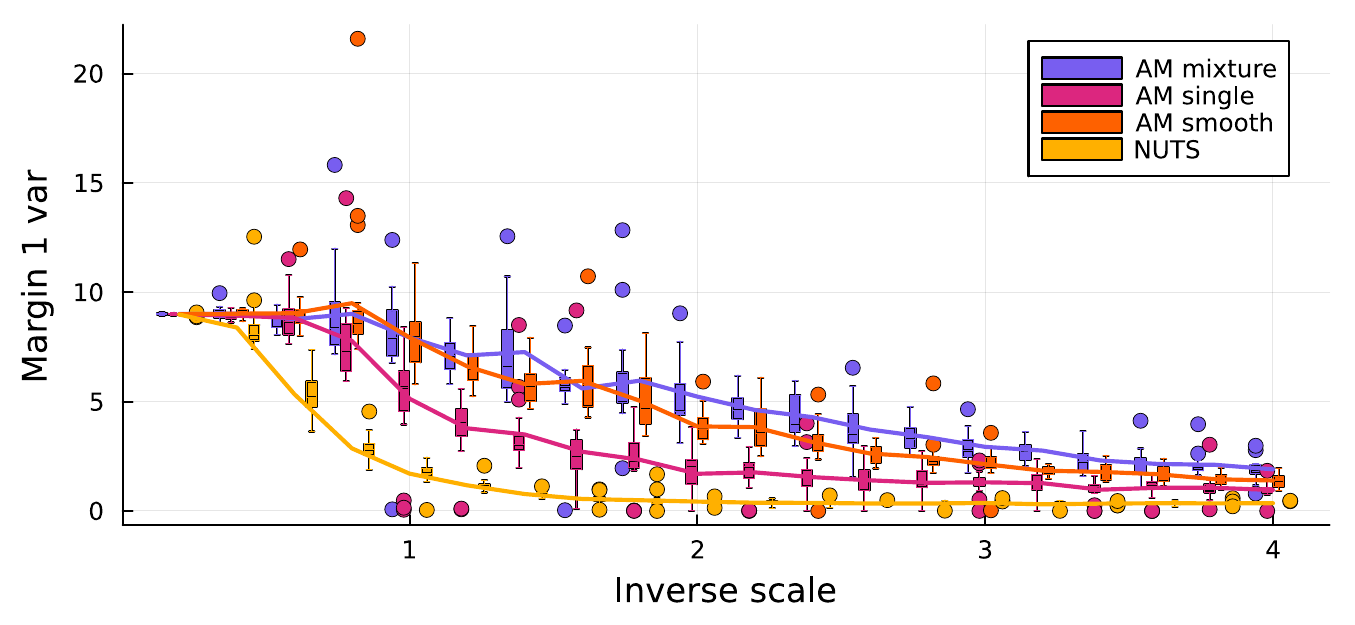} \\
	\includegraphics[width=0.5\textwidth]{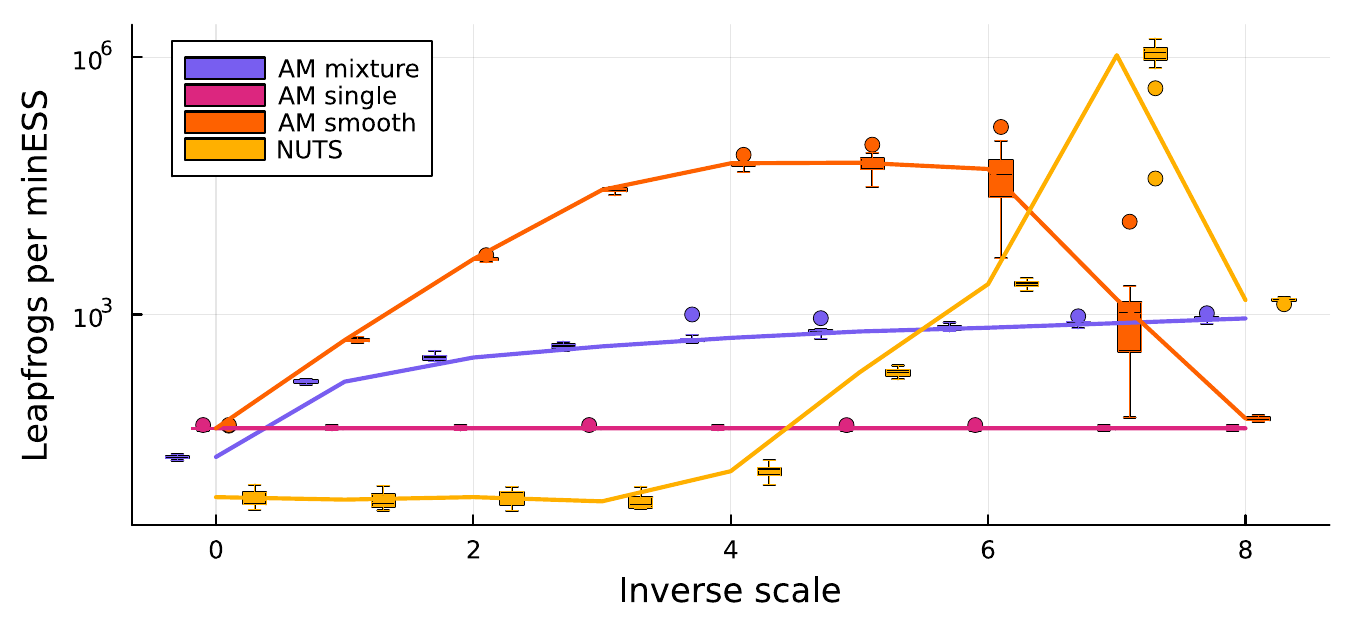}%
    \includegraphics[width=0.5\textwidth]{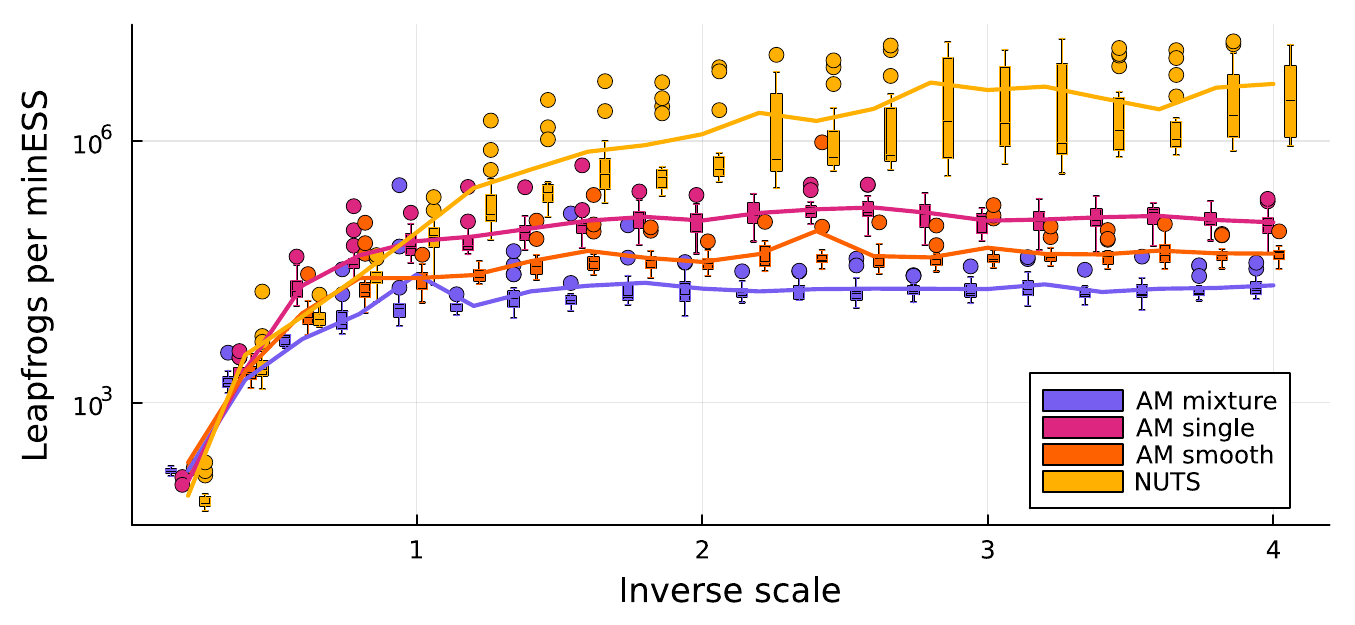}
	\caption{Comparison of three versions of autoMALA alongside NUTS on
	two synthetic targets ($d=2$). Boxplots denote repetition for $20$ seeds. Lines
	mark the mean across repetitions. \textbf{Left column:} anisotropic normal
    (inverse scale corresponds to exponent $c$). \textbf{Right column:} funnel. 
	\textbf{Top to bottom:} known margin mean, known margin variance, and leapfrog
	per minESS.
	}
	\label{fig:preconditioning}
\end{figure}

A slightly more complicated approach draws $\eta$ from a zero-one-inflated Beta 
distribution, which is a mixture between a Bernoulli and a Beta distribution
\[
\distBeta_{01}(\dee \eta; \tilde \alpha, \tilde \beta, p, m) = m\distBern(\dee \eta; p) + (1-m)\distBeta(\dee \eta;\tilde \alpha, \tilde \beta),
\]
for some $m,p\in[0,1]$. This approach has the benefit of letting the sampler use 
the exact adapted diagonal covariance matrix in some iterations. To investigate the 
benefits of this change, we ran NUTS alongside the following three versions of 
autoMALA:
\benum
\item Single preconditioner: $\eta\sim\distBeta_{01}(1, 1, 1, 1)$, so that $\diag(\hat{\Sigma}_\text{AM}) = \diag(\hat \Sigma)$.
\item Smooth preconditioner: $\eta\sim\distBeta_{01}(1, 1, 1, 0)$. This is the strategy described in \cref{sec:autoMALA}.
\item Mixture preconditioner: $\eta\sim\distBeta_{01}(1,1, 1/2, 2/3)$, so that the 
two endpoints $\{0,1\}$ and the interval $(0,1)$ have all equal chance ($1/3$) of
being picked. 
\eenum
We ran these $4$ samplers on the funnel scale experiment already presented, and also
on a simple $2$-dimensional anisotropic Gaussian parametrized so that the standard 
deviations are $(10^{-c}, 10^c)$ for $c\in\nats$. 
In this example we expect the single preconditioner approach to dominate, while the 
smooth approach should fail. This is exactly what is shown in 
\cref{fig:preconditioning}. Moreover, the mixture preconditioner is the second best
performing after the single preconditioner. In contrast, the mixture approach
dominates in the funnel target, although all three versions of autoMALA are
considerably more efficient than NUTS.